\documentclass[numberedappendix]{emulateapj}
\usepackage{amsmath}
\usepackage{amssymb}
\usepackage{aas_macros}
\usepackage[normalem]{ulem}
\usepackage[usenames]{color}
\usepackage{bm}
\usepackage{cancel}
\definecolor{DarkGreen}{rgb}{0.64,0.80,0.35}
\DeclareMathAlphabet\mathbfcal{OMS}{cmsy}{b}{n}
\newcommand{\calR}{\ensuremath{\bm\hat{\mathbfcal{R}}}}
\newcommand{\magcalR}{\ensuremath{\hat{\mathcal{R}}}}

\SetSymbolFont{symbols}{bold}{OMS}{cmsy}{b}{n}
\DeclareSymbolFont{bmisymbols}{OML}{cmm}{b}{it}
\DeclareMathSymbol{\balpha}{0}{bmisymbols}{"0B}
\DeclareMathSymbol{\bbeta}{0}{bmisymbols}{"0C}
\DeclareMathSymbol{\bgamma}{0}{bmisymbols}{"0D}
\DeclareMathSymbol{\bdelta}{0}{bmisymbols}{"0E}
\DeclareMathSymbol{\bepsilon}{0}{bmisymbols}{"0F}
\DeclareMathSymbol{\bzeta}{0}{bmisymbols}{"10}
\DeclareMathSymbol{\boldeta}{0}{bmisymbols}{"11}
\DeclareMathSymbol{\btheta}{0}{bmisymbols}{"12}
\DeclareMathSymbol{\biota}{0}{bmisymbols}{"13}
\DeclareMathSymbol{\bkappa}{0}{bmisymbols}{"14}
\DeclareMathSymbol{\blambda}{0}{bmisymbols}{"15}
\DeclareMathSymbol{\bmu}{0}{bmisymbols}{"16}
\DeclareMathSymbol{\bnu}{0}{bmisymbols}{"17}
\DeclareMathSymbol{\bxi}{0}{bmisymbols}{"18}
\DeclareMathSymbol{\bpi}{0}{bmisymbols}{"19}
\DeclareMathSymbol{\brho}{0}{bmisymbols}{"1A}
\DeclareMathSymbol{\bsigma}{0}{bmisymbols}{"1B}
\DeclareMathSymbol{\btau}{0}{bmisymbols}{"1C}
\DeclareMathSymbol{\bupsilon}{0}{bmisymbols}{"1D}
\DeclareMathSymbol{\bphi}{0}{bmisymbols}{"1E}
\DeclareMathSymbol{\bchi}{0}{bmisymbols}{"1F}
\DeclareMathSymbol{\bpsi}{0}{bmisymbols}{"20}
\DeclareMathSymbol{\bomega}{0}{bmisymbols}{"21}
\DeclareMathSymbol{\bvarepsilon}{0}{bmisymbols}{"22}
\DeclareMathSymbol{\bvartheta}{0}{bmisymbols}{"23}
\DeclareMathSymbol{\bvarpi}{0}{bmisymbols}{"24}
\DeclareMathSymbol{\bvarrho}{0}{bmisymbols}{"25}
\DeclareMathSymbol{\bvarsigma}{0}{bmisymbols}{"26}
\DeclareMathSymbol{\bvarphi}{0}{bmisymbols}{"27}

\begin{document}
\title{Patchy blazar heating: diversifying the thermal history of the intergalactic medium}
\author{Astrid Lamberts$^1$, Philip Chang$^1$,  Christoph Pfrommer$^2$, Ewald Puchwein$^3$, Avery E. Broderick$^{4,5}$ and Mohamad Shalaby$^{4,5,6}$}
\affil{$^1$ Department of Physics, University of Wisconsin-Milwaukee, 1900 East Kenwood Boulevard, Milwaukee WI 53211, USA}
\affil{$^2$ Heidelberg Institute for Theoretical Studies, Schloss-Wolfsbrunnenweg 35, D-69118 Heidelberg, Germany}
\affil{$^3$ Institute of Astronomy and Kavli Institute for Cosmology, University of Cambridge, Madingley Road, Cambridge, CB3 0HA, UK}
\affil{$^4$ Perimeter Institute for Theoretical Physics, 31 Caroline Street North, Waterloo, ON, N2L 2Y5, Canada}
\affil{$^5$ Department of Physics and Astronomy, University of Waterloo, 200 University Avenue West, Waterloo, ON, N2L 3G1, Canada}
\affil{$^6$ Department of Physics, Faculty of Science, Cairo University, Giza 12613, Egypt}
\email{lambera@uwm.edu}
\begin{abstract}
TeV-blazars potentially heat the intergalactic medium (IGM) as their gamma rays interact with photons of the extragalactic background light to produce electron-positron pairs, which lose their kinetic energy to the surrounding medium through plasma instabilities. This results in a heating mechanism that is only weakly sensitive to the local density, and therefore approximately spatially uniform, naturally  producing an inverted temperature-density relation in underdense regions. In this paper we go beyond the approximation of uniform heating and quantify the heating rate fluctuations due to the clustered distribution of blazars and how this impacts on the thermal history of the IGM. We analytically compute a filtering function that relates the heating rate fluctuations to the underlying dark matter density field. We implement it in the cosmological code GADGET-3 and perform large scale simulations to determine the impact of inhomogeneous heating. We show that, because of blazar clustering, blazar heating is inhomogeneous  for $z\gtrsim$ 2. At high redshift, the temperature-density relation shows an important scatter and presents a low temperature envelope of unheated regions, in particular at low densities and within voids. However, the median temperature of the IGM is close to that in the uniform case, albeit slightly lower at low redshift. We find that blazar heating is more complex than initially assumed and that the temperature-density relation is not unique. Our analytic model for the heating rate fluctuations couples well with large scale simulations and provides a cost-effective alternative to subgrid models.
\end{abstract}
\section{Introduction}
The intergalactic medium constitutes the bulk of the baryons forming the cosmic web \citep{1996Natur.380..603B} as well as the reservoir of baryons available for the formation of galaxies and clusters \citep{1997ApJ...489....7R}. Observations of metal-enriched gas link its evolution to the formation of galaxies and stars (see e.g. \citet{2009A&A...493..409S,2010MNRAS.407.2063W}). Therefore, the thermal history of the IGM plays a central role in determining the development of structure in the visible universe.

The temperature of the IGM is mostly set by photoionization of hydrogen and helium, competing with adiabatic cooling. As a result, the universe is slowly cooling once reionization is completed. The IGM temperature is expected to increase with density because denser regions experience a lower amount of adiabatic cooling, as well as more recombination-induced photoheating. Gas in the IGM that has not been shock heated provides a lower envelope to the temperature-density distribution.  Observations of Lyman $\alpha$ absorption lines with $4\leqslant z \leqslant 2$ show that lower column density gas exhibits the lowest line widths, which is commonly interpreted as an indication that lower density gas is colder than high density gas \citep{1997ApJ...484..672K,2000MNRAS.318..817S,2000ApJ...534...41R,2012ApJ...757L..30R}. A temporary flattening of the slope of the inferred temperature-density relation around $z\sim 3$ indicates an additional source of heating at that redshift.

The latter is expected due to He\,\textsc{II} reionization \citep[e.g.][]{2007MNRAS.380.1369T,2009ApJ...694..842M,2013MNRAS.435.3169C,2014arXiv1410.1531P}. Observations suggest that the tail end of He\,\textsc{II} reionization occurred between $z \simeq 3.2$ and 2.7 \citep{2001Sci...293.1112K,2014ApJ...784...42S}. However, the bulk of He\,\textsc{II} has likely ionized earlier, potentially as early as $z\ge 4$ \citep{2014arXiv1405.7405W}.  Its patchiness is probably related to the rarity of ionizing sources and the inhomogeneity of the IGM.  Including full radiative transfer, simulations show enhanced scatter in the temperature-density distribution around $z=3$ \citep{2009ApJ...694..842M}. Whether reionization may result in an inverted temperature-density distribution \citep{2012MNRAS.423....7M} is not firmly established yet \citep{2013MNRAS.435.3169C}. 

On top of that, recent measurements found that underdense regions may be warmer than predicted \citep{2009MNRAS.399L..39V,2008MNRAS.386.1131B}, as well as to exhibit unexpectedly high temperatures at $z<2$ \citep{2014MNRAS.441.1916B}, which is harder to explain solely by ionization of He\,\textsc{II}. Although an inverted temperature-density relation in underdense regions has not been firmly established yet \citep{2014MNRAS.438.2499B}, these observations suggest that the thermal history of the IGM may be more complex than initially assumed.

\citet{2012ApJ...752...22B} recently suggested a complementary heating mechanism, through TeV blazars. TeV blazars are active galactic nuclei (AGN) emitting very high energy gamma rays ($E\ge100$~GeV). They belong to the radio-loud subgroup of AGN, with the relativistic jet being pointed towards us. About 50 of these sources have been significantly discovered so far (http://tevcat.uchicago.edu/) by ground based Cerenkov telescopes such as MAGIC, H.E.S.S. and VERITAS. Those pointed observations have only skimmed the surface of a much larger population that manifest themselves as hard-spectra gamma-ray blazars as observed with the space-based \textit{Fermi}-LAT telescope \citep{2014ApJ...790..137B}. 

The universe is mainly opaque to very high energy gamma rays; they interact with the extragalactic background light (EBL) producing electron/positron pairs \citep{1967PhRv..155.1408G,1992ApJ...390L..49S}. It is commonly assumed that the electron/positron pairs inverse Compton scatter photons of the cosmic microwave background, resulting in a distribution of photons with energies between $0.1$ and $100$ GeV. Such an emission component towards TeV blazars has not been observed so far \citep{2010A&A...524A..77A}, despite extended searches \citep{2014A&A...562A.145H}. One solution would be pair deflection due to the intergalactic magnetic field, thus lowering the surface brightness of the formed pair halo at GeV energies \citep{2013A&ARv..21...62D,2012ApJ...747L..14V,2011ApJ...733L..21D}. 

However, the cascaded GeV emission would still contribute to the extragalactic gamma-ray background (EGRB). \textit{Fermi}-LAT was able to resolve more sources than its predecessor EGRET, thereby limiting the isotropic component of the unresolved EGRB and severely constraining the redshift evolution of hard blazars in the presence of inverse Compton cascades \citep[e.g.,][]{Vent:10,Murase:2012,Inoue:2012}. As a result of these, it is now well established that in this picture, the co-moving number density of gamma-ray blazars, and by extension the TeV blazars, must either be constant or decreasing with redshift \citep{Knei-Mann:08,Vent:10,Abazajian:2011,Inoue:2012}.  This lies in stark contrast to other classes of AGNs specifically, and all other tracers of the cosmological history of accretion onto galactic halos generally (e.g., star formation).  Furthermore, after careful modeling of gamma-ray and X-ray survey selection effects there appears to be no evidence for a significantly different evolution of blazars in comparison to their radio-loud analogues such as radio galaxies \citep{Giommi:2012,Giommi:2013}. Together this casts doubt on the inverse Compton cascade picture and provides circumstantial evidence that pair beams instead transfer their energy directly to the IGM through plasma instabilities \citep{2012ApJ...752...22B, 2012ApJ...758..102S, 2013ApJ...777...49S,2014ApJ...797..110C}. This naturally explains the EGRB with a blazar population that exhibits a redshift evolution in agreement with that of quasars \citep{2014ApJ...790..137B,2014ApJ...796...12B}.

The pairs constitute a dilute, ultrarelativistic beam, which is subject to several plasma instabilities, from which the ``oblique'' instability \citep{PhysRevE.70.046401} is the most powerful. Assuming its efficiency in the linear regime extends to the non-linear regime, \citet{2012ApJ...752...23C} show it is responsible for increasing the temperature of the IGM by almost a factor 10 in low density regions. While this assumption is still debated (see \citet{2013ApJ...770...54M,2014ApJ...787...49S} but also \citet{2013ApJ...777...49S,2012ApJ...758..102S,2014ApJ...797..110C}), throughout all this paper we assume plasma instabilities are the dominant mechanism for cooling of the pair beams.

Including TeV blazar heating in the thermal history of the IGM, \citet{2012ApJ...752...23C} were able to reproduce the inverted temperature-density relation for low density regions. \citet{2012ApJ...752...24P} found that TeV blazar heating  is capable of creating a redshift dependent entropy floor in clusters and galaxies, thus suppressing the formation of dwarf galaxies after the peak of blazar activity at redshift $z\simeq2$ and potentially providing an explanation to the ``missing satellite problem'' and the ``missing void dwarf problem'' \citep{2010AdAst2010E...8K}. Implementing uniform blazar heating, i.e. with a redshift-dependent but spatially homogeneous energy deposition rate per unit volume, in a cosmological hydrodynamical simulation of galaxy formation, \citet{2012MNRAS.423..149P} find excellent agreement with the one and two-point statistics of the Lyman $\alpha$ forest, which is the main observational tracer of low density regions in the universe.

However, the low thermal broadening of certain Lyman $\alpha$ lines indicates the presence of cold gas at $z =2.4$ \citep{2012ApJ...757L..30R}, which suggests TeV blazar heating does not uniformly heat the whole universe. It is natural to expect a larger TeV flux close to higher density regions where visible structures form. Conversely, in large underdense regions, far from massive black holes, heating is probably much lower. The goal of this paper is to go beyond the hypothesis of uniform heating and to link TeV blazar heating to the underlying clustered density field and take into account the bias of sources. This will lead to a more heterogeneous heating pattern and account for unheated regions while keeping the overall impact of blazar heating.

Self-consistently studying the evolution of the IGM from first principles involves modeling both the formation and evolution of galaxies at the largest scales of the universe. As this is still far beyond reach of current computers, we have determined a filter function which relates the heating fluctuations to the dark matter (DM) structure similarly to \citet{2007MNRAS.376.1680P,2005ApJ...626....1B,2014PhRvD..89h3010P}. Based on the hierarchical structure formation in a $\Lambda$CDM universe, it naturally selects the relevant length scales for TeV blazar heating ($\S$2). We have implemented it in large scale cosmological simulations ($\S$3) in order to focus on the equation of state and thermal evolution of the IGM ($\S$4). We then discuss how inhomogeneous heating could reconcile different observations ($\S$5) and conclude ($\S$6).

\section {Determining the window function}\label{window}
\subsection{Intuitive understanding}
One zone models \citep{2012ApJ...752...23C,2012ApJ...752...24P} and numerical simulations \citep{2012MNRAS.423..149P} of blazar heating on the IGM assume that the heating is uniform. Because the heating rate depends on the local density of EBL and TeV photons, the assumption of uniform heating implies that the distributions of EBL and TeV photons are uniform. The EBL photon density in voids is only 2$\%$ less than the average value \citep{2015MNRAS.446.2267F}. However, for TeV photons the mean free path compares with the separation between TeV sources for $z\geqslant$ 1, so the {\it spatial} fluctuations in the heating rate are likely nontrivial. Moreover, the sources of TeV photons tend to be clustered and so the IGM near these clustered regions will get an increased flux of photons in comparison to low-density regions because of the increased number of sources and the $1/r^2$ flux dilution with increasing distance $r$ from the source.

Our goal is to include a more realistic model for heating due to TeV blazars in numerical simulations.
To properly calculate the heating fluctuations due to TeV blazar heating, the formation and evolution of accreting supermassive black holes must be modeled in a full self-consistent cosmological simulation. In addition, the TeV radiation from these systems must be ray-traced through the simulation volume. Such a task is computationally intractable. As a result, we have elected to model this TeV blazar heating in a statistical manner.

We assume that TeV blazars are associated with galaxies or alternatively quasars and that they roughly emit over $4\pi$ steradian. The latter assumption remains valid as long as the duty cycle of blazars is small enough so that jets point in all directions over cosmological timescales and the number of TeV blazars is large enough such that every spot in the universe is illuminated by at least a few TeV blazars, which is the case for $z\lesssim 6$ \citep{2012ApJ...752...23C}.
The heating rate at a given point $\mathbf{x}$ is determined by the received TeV flux from all the sources 
\begin{equation}
\label{eq:heating_rate}
\dot{Q}(\mathbf{x},z)= \frac{1}{4\pi} \int_{\Omega} d\Omega\int_0^{\infty} dr'\frac{\mathcal{E}(\mathbf{r}'+\mathbf{x},z')}{D_{\mathrm{pp}}(z)} e^{-\tau},
\end{equation}
where $\dot{Q}$ and $\mathcal{E}$ are the heating rate density and emissivity of the sources (both in units of energy per unit time and per unit volume), $D_{\mathrm{pp}}$ is the mean free path for pair production on the EBL, $\tau$ is the associated optical depth along the line of sight and $z'$ is the redshift corresponding to the distance $|\mathbf{r}'|$.

One can then express the resulting heating rate as a mean value $\bar{\dot{Q}}$ and a first order correction $\delta_H$.
\begin{equation}
\label{eq:delta_h}
\dot{Q}(\mathbf{x},z)=\bar{\dot{Q}}(\mathbf{x},z)\left[1+\delta_H(\mathbf{x},z)\right].
\end{equation}
The method is based on a Taylor expansion of the quantities describing the TeV sources and keeping only the first order corrections. Transforming to Fourier space yields the fluctuation amplitude
\begin{equation}
\label{eq:use_window}
\tilde{\delta}_H(k,z)=\tilde{W}_H(k,z)\tilde{\delta}(k,z),
\end{equation}
where $\tilde{W}_H$ is defined as the window function and maps the Fourier transform of the dark matter overdensity, $\tilde{\delta}$, to the Fourier transform of the heating fluctuations, $\tilde{\delta}_H$. This naturally yields the relevant length scale for heating rate fluctuations. The detailed and pedagogical exposition of this method (in the Newtonian limit, in an expanding universe, and additionally accounting for a clustered source distribution) is given in the Appendices~\ref{sec:windon_newt} to \ref{sec:window_complete}. In the following section, we present the general method and highlight the underlying hypotheses of our work.

\subsection{Window function for TeV blazar heating}\label{sec:window}
To determine the heating rate fluctuations we express the TeV emissivity in Eq.~\eqref{eq:heating_rate} as a mean value and a first order correction. The heating rate at a given point is set by the received TeV flux from all the sources within a certain radius. We assume that the pairs lose their energy to the IGM at the point where they are created. As stated in \citet{2012ApJ...752...22B} and \citet{2014ApJ...797..110C}, this is a reasonable assumption as the plasma instability length scales are many orders of magnitude smaller than the mean free path of these TeV photons.

The TeV gamma rays are emitted by accreting supermassive black holes at centers of galaxies, which cluster in overdense regions. Matter is tightly coupled to the underlying dark matter, the evolution of which is straightforward to model analytically within the linear approximation. The linear approximation is valid as long as the overdensity is small, which is true in the early universe and then breaks down at small scales as very dense structures form. Our computation takes into account the bias between baryonic matter and dark matter \citep{1996MNRAS.282..347M}, as we detail below.

To model cosmic distances, Eq.~\eqref{eq:heating_rate} is integrated in redshift space and we take into account the resulting energy loss for the TeV photons as well as first order corrections due to a clustered blazar distribution and proper motions of the sources within the Hubble flow \citep{1987MNRAS.227....1K}. We integrate over the energy distribution of the TeV-emission. Following the detailed derivation in the Appendix, we obtain the window function $\tilde{W}_H(\hat{k},z)$ as a function of comoving wavelength $\hat{k}$,
\begin{eqnarray}
\label{eq:window}
\tilde{W}_H(\hat{k},z)&=&\frac{1}{N}\int_{E_{\mathrm{min}}}^{E_{\mathrm{max}}}\frac{dE}{D_{\mathrm{pp}}(E,z)}
\int_z^{\infty}dz' \frac{\mathcal{\bar{\hat E}}_{E'}(z')e^{-\tau}}{(1+z')\,H(z')} \nonumber\\
&\times& \frac{D(z')}{D(z)}\left[\left(b(z')+\frac{f}{3}\right)j_0(\hat k \hat r)-\frac{2f}{3}j_2(\hat k \hat r)\right]
\end{eqnarray}
with
\begin{eqnarray}
\label{eq:define_N}
N=\int_{E_{\mathrm{min}}}^{E_{\mathrm{max}}} \frac{dE}{{D_{\mathrm{pp}}}(E,z)}
\int_z^{\infty} dz'   \frac{\mathcal{\bar{\hat E}}_{E'}(z') e^{-\tau}}{(1+z')\,H(z')}.
\end{eqnarray}
Here, $\hat{r}$ is the comoving distance between the source and heated
region, $D$ denotes the linear growth factor defined in
Eq.~\eqref{eq:growth_1}, $f\equiv d\log \delta/d \log a$, $E$ is the
energy of the received TeV photon, $E'$ its initial energy, and
$\hat{\mathcal{E}}_E$ the comoving blazar spectral luminosity density, $b$
is the Eulerian bias and $j_0$ and $j_2$ are spherical Bessel
functions.

The blazar spectral luminosity density (in units of energy, per unit time, per volume, per energy) is determined by both the
intrinsic source spectra and the luminosity function of TeV blazars.
Here we adopt the model described in \citet{2014ApJ...790..137B}, in
which the spectra are assumed to be well described by a set of broken
power laws and a luminosity function that depends on redshift,
luminosity from 100~GeV to 10~TeV ($L_{\rm TeV}$), and the low energy
spectral index ($\Gamma_l$).  The assumed source spectral luminosity
density is given by 
$L_E(L_{\rm TeV},\Gamma_l)=\tilde{L}_E(\Gamma_l) L_{\rm TeV}$ where
\begin{equation}
\label{eq:intrinsic_spectrum}
\tilde{L}_E(\Gamma_l) = \frac{f_0 E}{(E/E_b)^{\Gamma_l} + (E/E_b)^{\Gamma_h}}\,,
\end{equation}
where the normalization $f_0$ is chosen such that 
$\int_{0.1~\mathrm{GeV}}^{10~\mathrm{TeV}} \tilde{L}_E dE = 1$, $E_b=1~{\rm TeV}$, and
$\Gamma_h=3$.  The luminosity function is based on the spectral
properties of the Fermi population of hard gamma-ray blazars and the
quasar luminosity function reported by \citet{2007ApJ...654..731H}.
In terms of these, the blazar spectral luminosity density is
\begin{equation}\label{eq:spec_lum_dens}
\hat{\mathcal{E}}_E(z)
=
\int d\log_{10}L_{\rm TeV}\,d\Gamma_l\, L_E(L_{\rm TeV},\Gamma_l)
\hat{\phi}(z,L_{\rm TeV},\Gamma_l)\,,
\end{equation}
where $\hat{\phi}(z,L_{\rm TeV},\Gamma_l)$ is the comoving analog of Eq.~6 in \citet{2014ApJ...790..137B}.  This may be simplified
further using the separability of $\hat{\phi}$ in $\Gamma_l$, i.e.,
\begin{equation}
\hat{\phi}(z,L_{\rm TeV},\Gamma_l)
=
\hat{\phi}(z,L_{\rm TeV})
\chi(\Gamma_l)\,,
\end{equation}
where $\int d\Gamma_l \chi(\Gamma_l) = 1$.  As a result,
\begin{equation}
\label{eq:mean_heat}
\hat{\mathcal{E}}_E(z)
=
\left< \tilde{L}_E \right>
\hat{\Lambda}(z)\,,
\end{equation}
where
\begin{equation}
\left< \tilde{L}_E \right>
=
\int d\Gamma_l\,\tilde{L}_E(\Gamma_l) \chi(\Gamma_l)
\end{equation}
which is now a function of energy alone, and
\begin{equation}
\hat{\Lambda}(z) = 
\int d\log_{10}L_{\rm TeV}\,\hat{\phi}(z,L_{\rm TeV})
\end{equation}
is the TeV luminosity density of TeV blazars.  

In practice, for the TeV blazar luminosity function in
\citet{2014ApJ...790..137B} $\left< \tilde{L}_E \right>$ is well fit
over the energy range of interest, 100~GeV to 10~TeV, by
Eq.~\eqref{eq:intrinsic_spectrum} with $\Gamma_l=1.80$ and $E_b=1.057$,
with an $L_1$ norm of 0.0048.  This spectral shape peaks near 1~TeV
and thus sets a characteristic $\gamma$-ray energy.

Since the TeV blazar luminosity function in
\citet{2014ApJ...790..137B} is simply the quasar luminosity function
in \citet{2007ApJ...654..731H} rescaled in energy and luminosity, the
comoving luminosity densities are also proportional.  That is,
\begin{equation}
\hat{\Lambda}(z) = \zeta \hat{\Lambda}_Q(z)\,,
\end{equation}
with $\zeta=2.1\times 10^{-3}$ and the comoving quasar luminosity
density $\hat{\Lambda}_Q$. The corresponding comoving TeV luminosity
density, found by \citet{2012ApJ...752...23C} based on a fit to the
comoving quasar luminosity density in \citet{2007ApJ...654..731H}, is
given by
\begin{equation}
\label{eq:phi_quasar}
\hat{\Lambda}(z)=\hat{\Lambda}_0 10^{1.18 z - 0.0812 z^2 - 0.182 z^3 + 0.0515 z^4 - 0.00418 z^5} 
\end{equation}
where 
$\hat{\Lambda}_0=(1.7-4.8)\times 10^{-36}$erg s$^{-1}$ cm$^{-3}$ the
blazar luminosity density at the current epoch.

The optical depth is set by the physical mean free path $D_\mathrm{pp}(E',z)$ of TeV photons before they interact with an EBL photon to produce an electron-positron pair. Its redshift evolution is set by the density of the EBL. Despite careful studies that aim at constraining it, there remain uncertainties in the star formation history of the universe as well as its metallicity and dust contents \citep[see, e.g.][]{2008A&A...487..837F,2006ApJ...648..774S}. Following \citet{2012ApJ...752...23C} we use a prescription
\begin{equation}
\label{eq:mean_free_path}
D_{\mathrm{pp}}(E',z')=35\left(\frac{E'}{1~\textrm{TeV}}\right)^{-1} \left(\frac{1+z}{2}\right)^{-\xi}~\textrm{Mpc,}
\end{equation}
where $\xi=4.5$ for $z<1$ and $\xi=0$ for $z>1$ \citep{2004A&A...413..807K,2009PhRvD..80l3012N} and the mean free path is expressed in physical units. The proper mean free path is constant for $z\geq 1$, however, the comoving mean free path increases for increasing redshift.

The fluctuations in TeV flux are related to the distribution of blazars, which is biased with respect to the distribution of dark matter halos. Luminous structures such as galaxies preferentially populate the high peaks of the dark matter density distribution. The square of the bias $b$ of a given structure is the ratio between its power spectrum to the power spectrum of the dark matter halos and it is stronger for more massive objects, such as the host galaxies of quasars \citep[see, e.g.][for a review]{2002PhR...372....1C}.

\begin{figure}[h]
\centering
\includegraphics[width = .4\textwidth ]{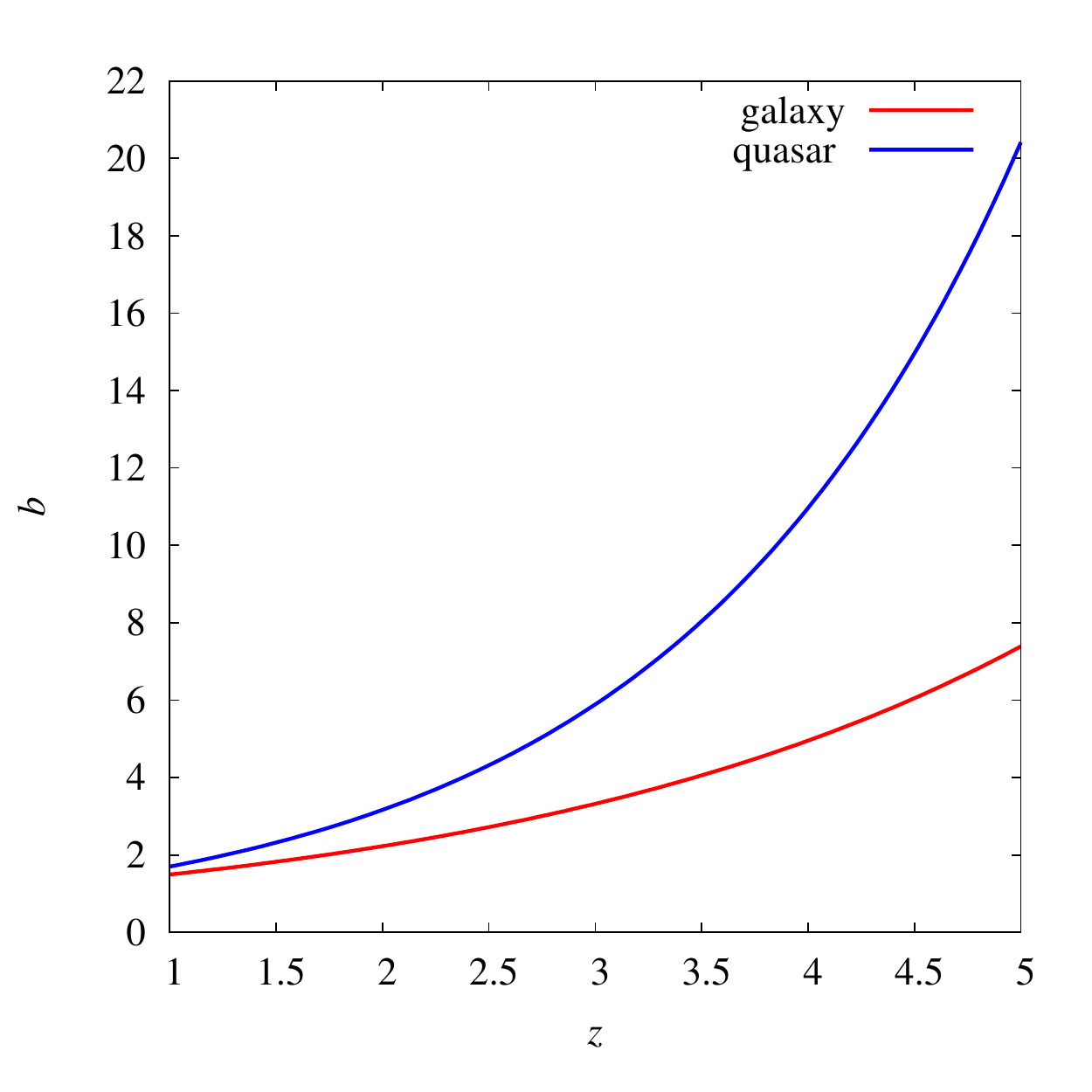}
\caption{Galaxy (red line) and quasar (blue line) bias models used in our simulations.}
\label{fig:bias}
\end{figure}

Due to the small number of sources and TeV photon absorption, there is no observation of TeV blazar bias.  Therefore, we  use a model for quasar bias as well as a model for galaxy bias to illustrate the impact of the uncertainty on the value of the bias. Since AGN are generally much more biased than galaxies the latter provides a robust lower bound. However, TeV blazars may have a stronger bias than quasars, as radio-loud AGN are generally found in a more clustered environment than quasars \citep{2009MNRAS.393..377M,2012MNRAS.421.3060S}. Our model for quasar  bias for $z>1$ is based on a fit of data by \citet{2005MNRAS.356..415C,2007ApJ...658...85M,2007AJ....133.2222S}. We use

  \begin{equation}
    \label{eq:gal_bias}
    b_{\mathrm{quasar}}(z)=10^{0.27z-0.04},
  \end{equation}
which is very similar to \citet{2012MNRAS.422..106P}. Our model for galaxy bias is based on a fit of data by \citet{2005A&A...442..801M,1998ApJ...492..428S,2006ApJ...637..631K}. We use
  \begin{equation}
    \label{eq:qso_bias}
    b_{\mathrm{galaxy}}(z)=10^{0.174z}.
  \end{equation}
Figure~\ref{fig:bias} shows the corresponding evolution of quasar and galaxy bias as a function of redshift. Using the same data, \citet{2008ApJ...678..627B} find that the quasar and galaxy bias can be fitted using a dark matter halo model of  $10^{13}h^{-1}M_{\odot}$ and $10^{12}h^{-1} M_{\odot}$ respectively.

\begin{figure}[h]
\centering
 \includegraphics[width = .45\textwidth ]{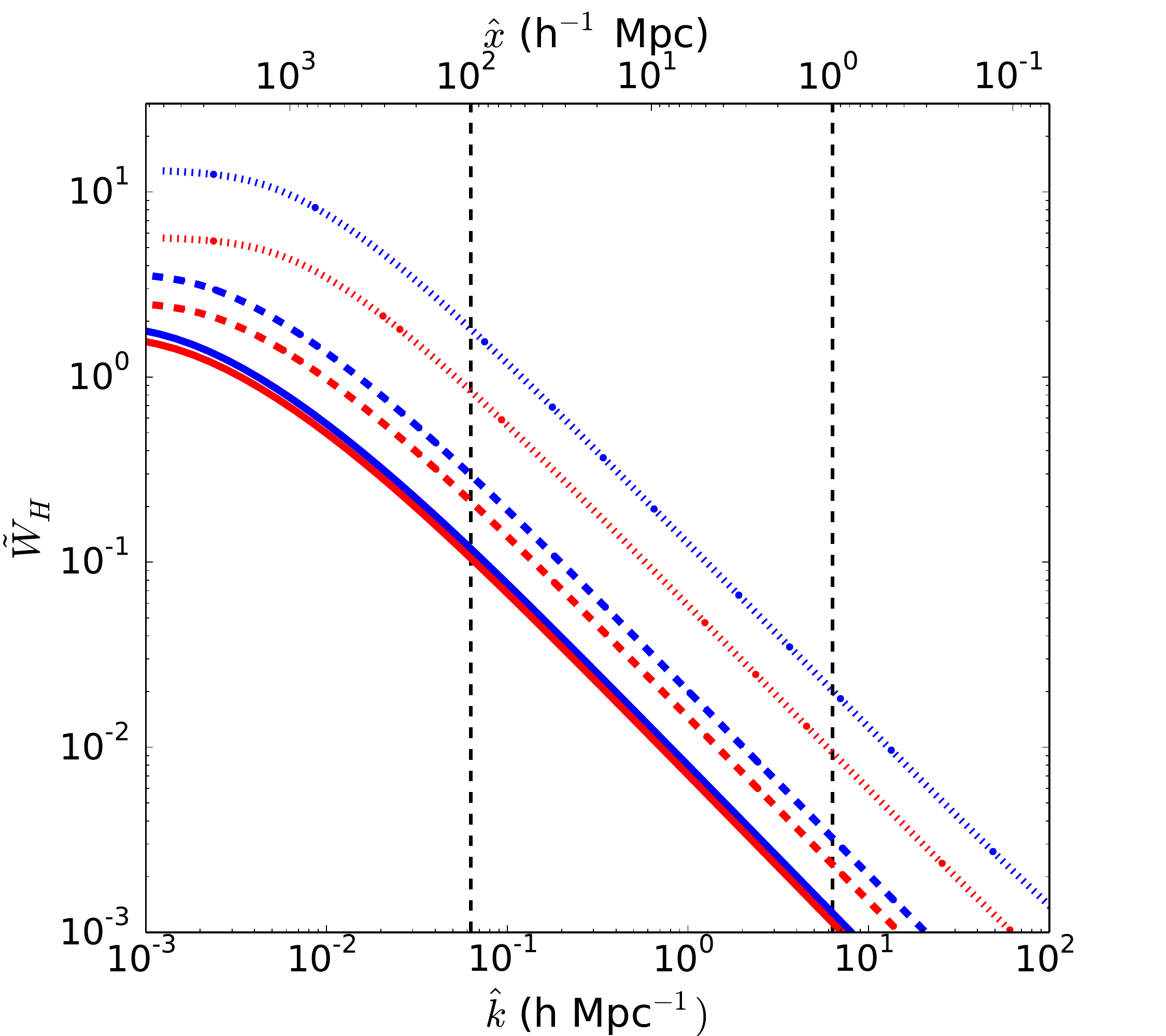}

\caption{Window function for TeV blazar heating from $z=1$ (solid lines), $z=2$ (dashed lines) and $z=4$ (dotted lines) for the galaxy bias model (red) and the quasar bias model (blue).  The vertical dashed lines indicate the minimal and maximal comoving wavenumber modeled in the simulation. }
\label{fig:window}
\end{figure}

Substituting Eqs.~\eqref{eq:mean_heat} and \eqref{eq:mean_free_path} into Eq.~\eqref{eq:window} then gives the complete window function for TeV blazar heating. Figure~\ref{fig:window} shows the filter for $z=1,2$ and 4 for a model with galaxy and quasar bias. We have computed the window function using an embedded Runge-Kutta method which is able to capture the fast variation of the Bessel functions at large wave numbers while decreasing computing time at smaller wave numbers. The window function is integrated up to $z=8$, which ensures that all the sources are taken into account. Changes in maximum redshift of the integration have no discernible impact on the window function.

The window function describes how density fluctuations translate into heating fluctuations.  The $\hat{k}^{-1}$ slope for large wavenumbers results from the $r^{-2}$ decrease in the TeV-flux. The position of the break is set by a combination between the mean free path and the redshift evolution of the blazar luminosity density. One might expect the comoving mean free path, which increases for increasing redshift at $z>1$, would move the break to smaller wavenumbers at higher $z$. However, by integrating over the energy of the photons and including lower energy photons that have longer mean free paths, the effect of the change in the mean free path is mitigated. This allows the rapidly decreasing blazar luminosity density for $z\gtrsim 2$ to move the break to larger wavenumber: the luminosity density drops too steeply so that more distant sources within the photon mean free path do not contribute as much as sources more close by. This latter effect moves the break to larger wavenumber for increasing redshift when lower energy photons with longer mean free paths are accounted for.  This implicitly assumes that these low energy photons would contribute to the local heating rate via pair production and plasma instabilities.  A more rigorous calculation that accounts for the efficiency of blazar heating would vary with photon energy and redshift (see for instance \citealt{2014ApJ...797..110C} for the effect of nonlinear Landau damping). However, this efficiency is not yet fully understood and so we assume in this paper that it as a function of photon energy is unity. As the normalization is set  by the bias, the quasar bias model has more power at all scales and density fluctuations will lead to more enhanced heating.  In  both models, at high redshift, most of the power resides on large scales, where blazar heating traces the density fluctuations. At smaller scales, density fluctuations have no impact and blazar heating is uniform. At the current epoch, TeV blazar heating is close to uniform because there is power only at the largest scales (above 100 Mpc) where the universe is essentially uniform (see \citet{2013MNRAS.429.2910C} and references therein).

In both models, the window function remains positive at all scales; thus, underdense regions are the only areas where a lower than average heating rate is expected. Heating at rates larger than the average is possible for large-scale overdensities but also on small scales for highly non-linear objects with $\delta\gg1$. We caution that our simplified approach with a linear bias description starts to break down there and would have to be augmented with a non-linear description of bias. However, the gas in these regions experiences shock heating in addition to photoheating such that the influence of blazar heating becomes negligible there. Moreover, those densities have little influence on the statistics of the high-redshift Lyman $\alpha$ forest and thus shall not be the subject of the present work. After these first analytic estimates, we include the window function to model TeV blazar heating in cosmological simulations.

\section{Numerical method}
\subsection{Cosmological simulations}
We perform simulations with the smoothed particle hydrodynamics (SPH) code \textsc{GADGET-3}, an upgraded version of the publicly available \textsc{GADGET-2} code \citep{2005MNRAS.364.1105S}. The code solves the gravitational evolution of both dark matter and gas particles following a TreePM N-body method. The hydrodynamical evolution of the gas is modeled using an entropy conserving scheme \citep{2002MNRAS.333..649S}.

The cosmological model is based on the \textit{WMAP} 7-year data \citep{2011ApJS..192...18K}: $\Omega_M=0.272$, $\Omega_{\Lambda}=0.728$, $\Omega_{B}= 0.0465$, $h=0.704$ and $\sigma_8=0.809$. The initial conditions were evolved from $z=100$ until $z=1$ in boxes with comoving side length of 100 $h^{-1}$ Mpc and periodic boundary conditions. We use $N= 2\times 512^3$ particles, which gives a mass of $m_\mathrm{gas}=3.8\times10^{6} h^{-1} M_{\odot}$ and $m_\mathrm{DM}=1.8\times 10^{7} h^{-1} M_{\odot}$ for baryonic and dark matter particles, respectively. We used a comoving gravitational softening length of 7.8 $h^{-1}$ kpc. We checked that the resolution has a barely discernible impact on the results of our simulations by performing test simulations with a comoving side length of 50 $h^{-1}$ Mpc at different resolutions, up to $2\times 512^3$. The spread of the temperature in the low density regions increases with increasing resolution. However, this effect is approaching saturation at a resolution of $N=2\times 256^3$. We find less than $4\%$ difference in the median temperature and less than $10\%$ difference in its root mean square in low density regions between $N=2\times 256^3$ and $2\times 512^3$ simulations, indicating that our chosen resolution is sufficient to accurately capture the temperature-density distribution. 

As we are only interested in the low density intergalactic medium, we use a simplified model for star formation which significantly speeds up the simulations. In this model, gas particles with $\delta_{\mathrm{gas}}\geq 1000$ and T $\leq 10^5$ are directly converted into stars \citep{2004MNRAS.354..684V}. Although it results in unrealistic galaxy properties, this approximation does not affect regions with $\delta_{\mathrm{gas}} \lesssim 10$. Local black hole feedback is not included. Photoheating is set by ionization equilibrium of H, He\,\textsc{I} and He\,\textsc{II} in the presence of an external UV field, which is parameterized according to \citet{2009ApJ...703.1416F}. As our version of the \textsc{GADGET-3} code assumes ionization equilibrium when computing photoheating rates, the heating is rather inefficient during reionization where this assumption is not well satisfied \citep[see e.g.][]{2014arXiv1410.1531P}. Following \citet{2012MNRAS.423..149P} we thus include the equivalent heat input by hand at redshift $z=10$. Our simulation does not include radiative transfer effects on the photoionization of He\,\textsc{II}.

The size of the box is set to model the heating perturbations on the scales determined by the window function in Figure \ref{fig:window}. We want to model a representative cosmic sample and probe distances beyond the mean free path of the TeV photons, which is of order of 70 (comoving) Mpc at $z=1$ (but 105 Mpc at $z=2$). To confirm that 100 $h^{-1}$ Mpc is a satisfactory size to model all the significant length scale, we performed a $L_\mathrm{box}=200 $ $h^{-1}$ Mpc simulation with $512^3$ particles. We compared the resulting temperature distribution function with the 100 $h^{-1}$ Mpc box, with the same mass resolution. The mean temperature varies by less than $2\%$ for regions with $\delta_{\mathrm{gas}}\leqslant 0$ at all redshifts. The standard deviation varies by 10$\%$ at $z=3$ and is below $5\%$ at $z=1$. This is consistent with studies showing that the one-and two-point statistics of the Lyman $\alpha$ forest are well captured with $\simeq 50$ $h^{-1}$ boxes \citep{2007MNRAS.374..196R,2009MNRAS.398L..26B}.
\subsection{Including the TeV blazar heating fluctuations}
We model the impact of the fluctuations in TeV blazar heating on the thermodynamics of the IGM. For every gas particle, the blazar heating is set by the mean value plus some correction depending on the local density field (Eq.~\eqref{eq:delta_h}). As in \citet{2012MNRAS.423..149P}, we adopt the mean heating rate computed by \citet{2012ApJ...752...23C} (Eq.~\eqref{eq:mean_heat}). We focus on the model with ``intermediate'' values for the blazar heating.

Modeling fluctuations by implementing a filtering function in a large scale simulation is a new method. The computation of the fluctuations is done in Fourier space and is inspired by the numerical particle-mesh (PM) algorithm used to solve the long-range gravitational force. The Fourier transforms are performed with the parallel extension of the Fast Fourier Transform Library. The first step is map the particles onto a mesh, which is done with a clouds-in-cells algorithm \citep{1981csup.book.....H}. We determine the Fourier transform of the DM density field. Then the density field is multiplied by the window function performing a bilinear interpolation of tabulated values for certain values of redshift and wavenumber. In this paper we use 21 equally spaced redshift bins from $z=5$, where blazar heating turns on in our model, until the end of the simulation. We use 128 logarithmically equally spaced wavenumber bins. Similarly to the long-range gravitational PM force, we deconvolve for the clouds-in-cells kernel by dividing by $\mathrm{sinc}^2(k_x L/2N)\mathrm{sinc}^2(k_y L/2N)\mathrm{sinc}^2(k_z L/2N)$. We then perform the inverse Fourier transform and renormalize. The last step is necessary to remap the results onto the gas particles. As our filtering technique may result, in rare cases, in unphysical cooling ($\dot{Q}< 0$), we set a lower limit of  $\delta_H =-1$.
\section{Results}
\begin{figure*}
\centering
\includegraphics[width = .45\textwidth ]{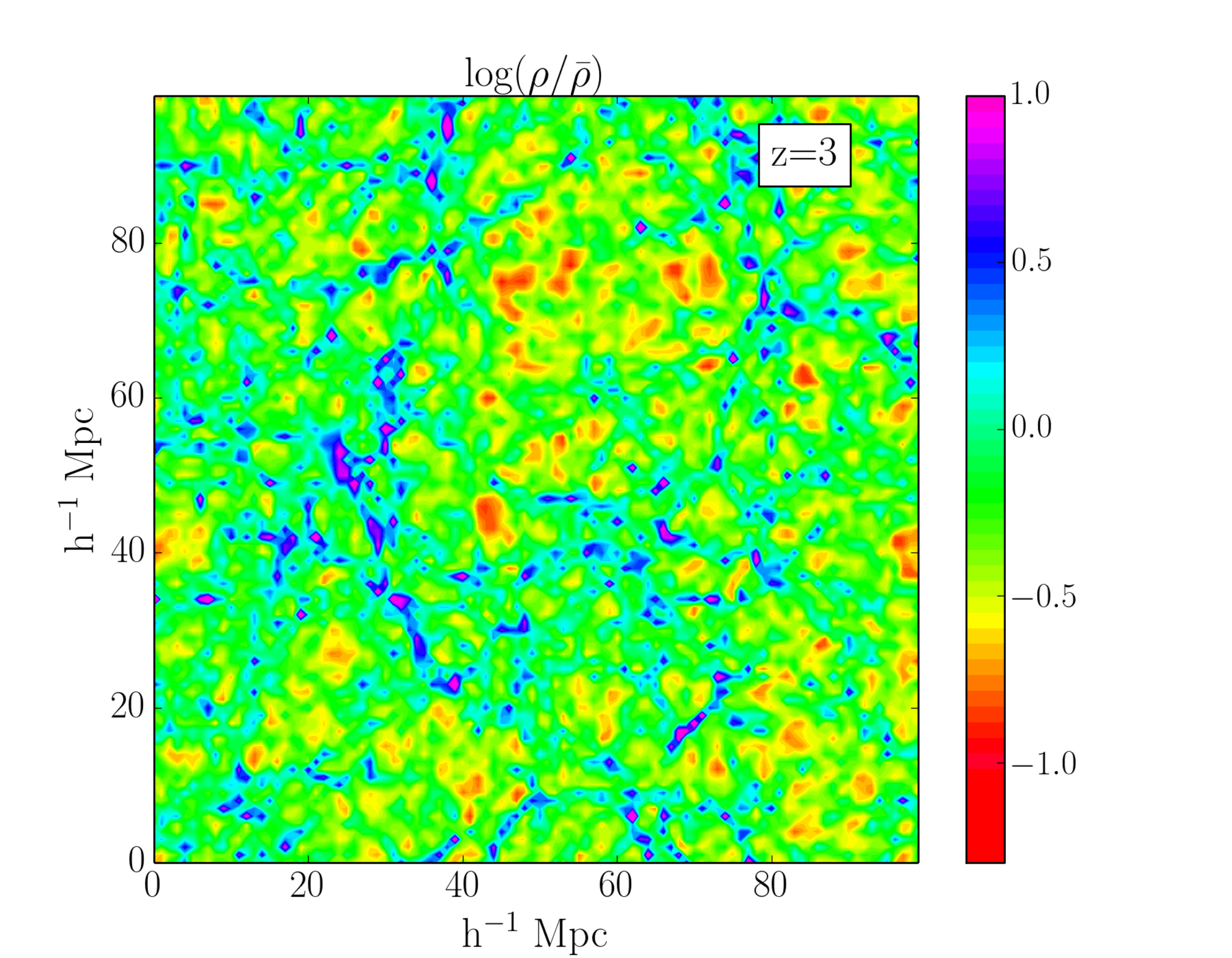}
\includegraphics[width = .45\textwidth ]{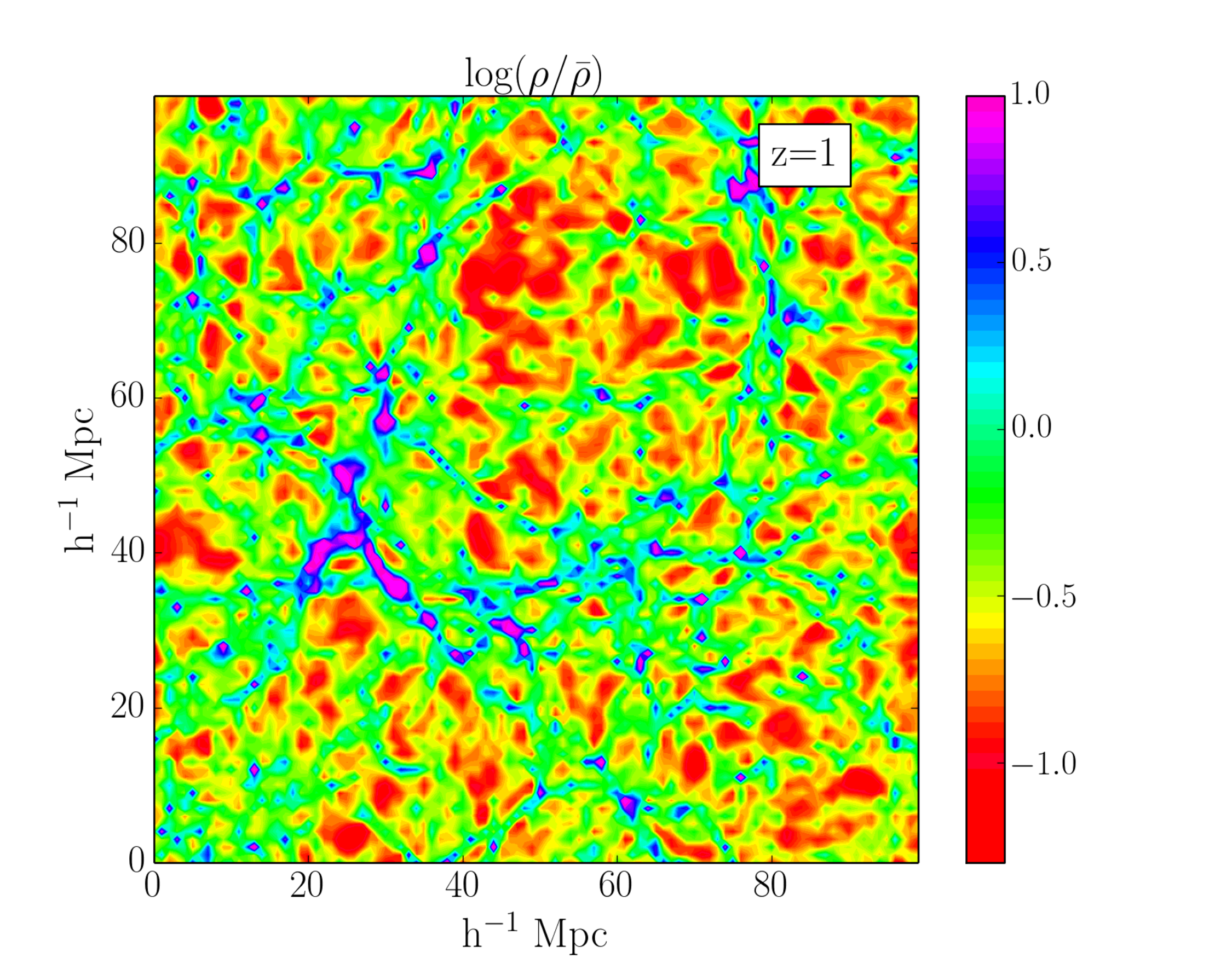}
\includegraphics[width = .45\textwidth ]{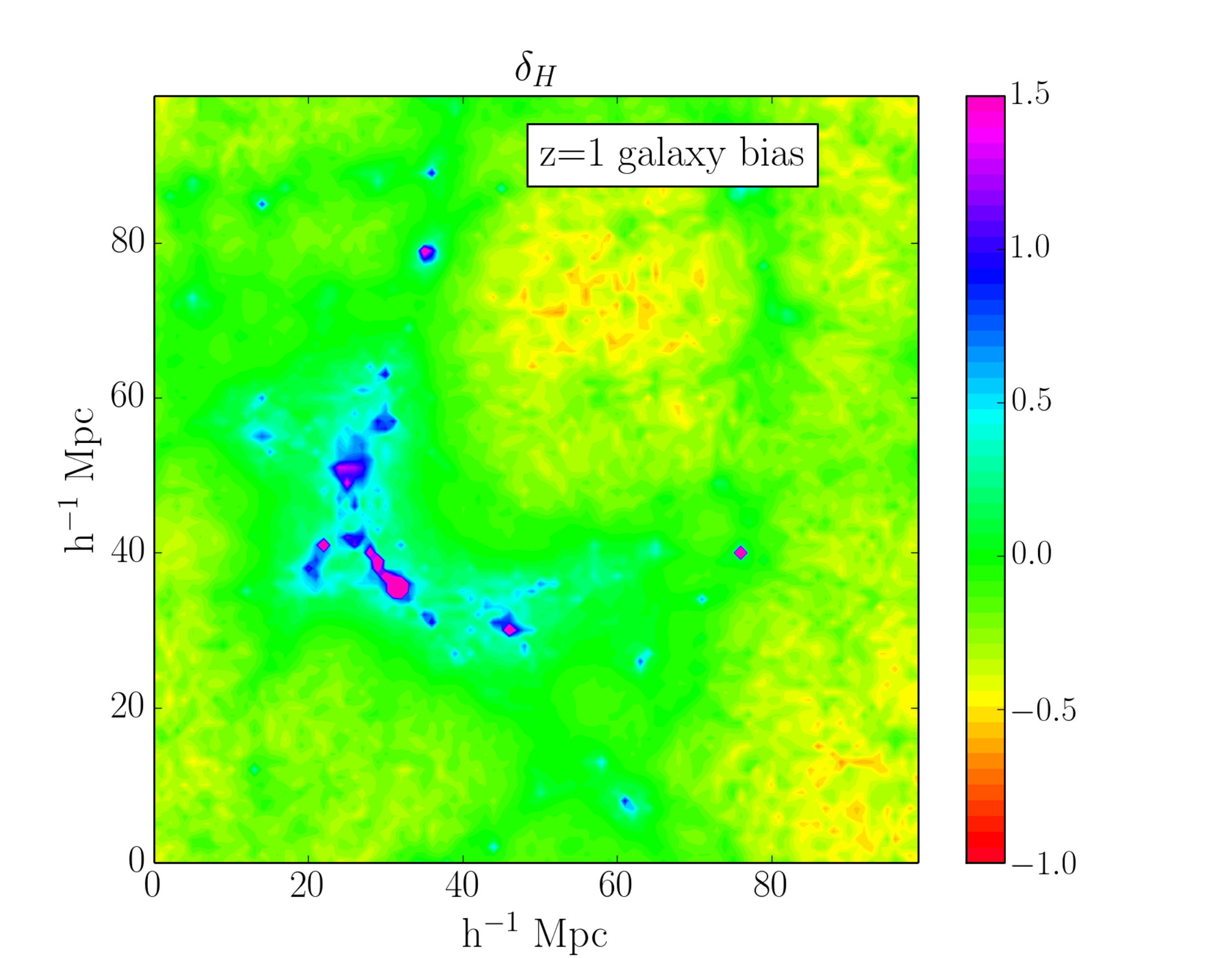}
\includegraphics[width = .45\textwidth ]{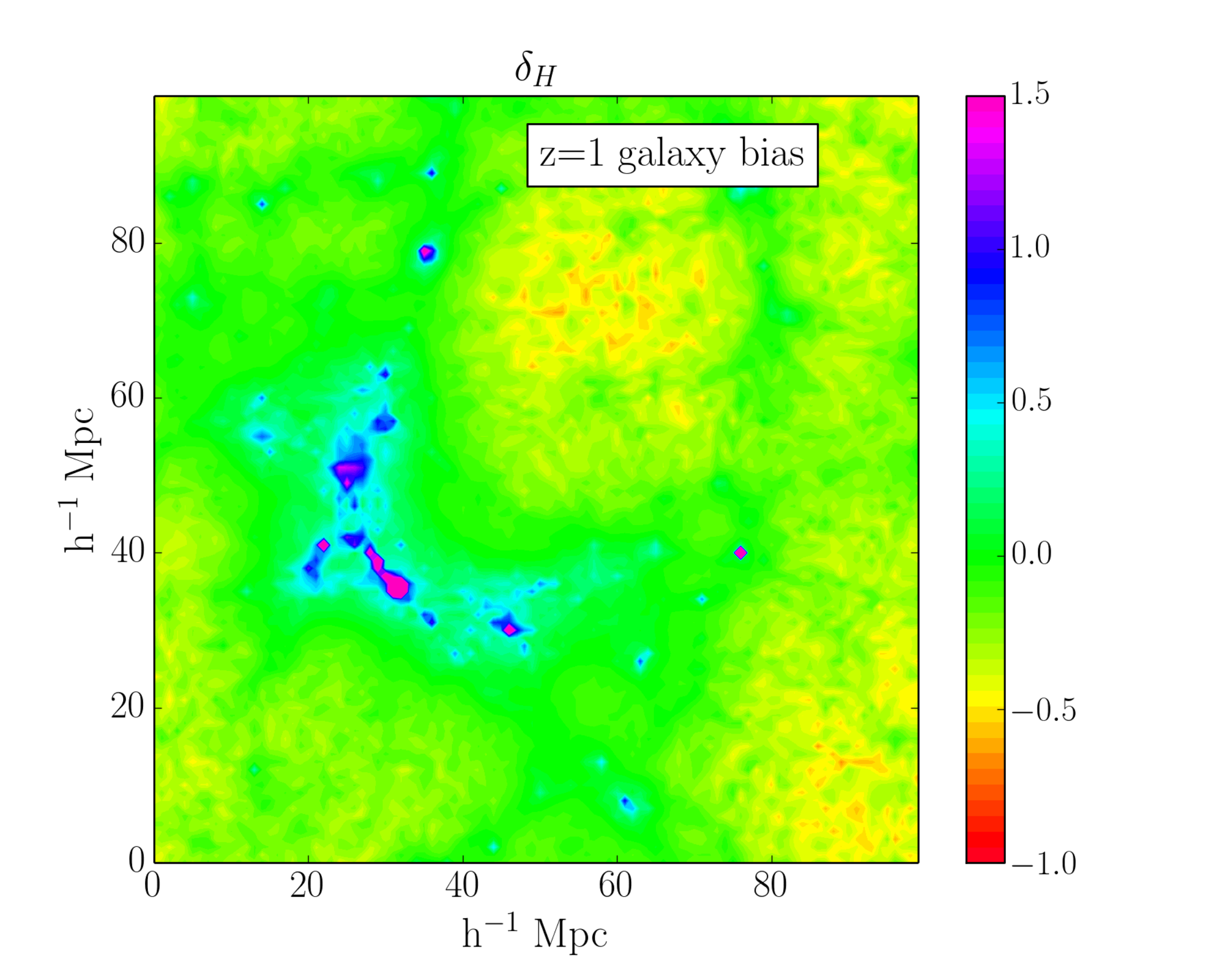}
\includegraphics[width = .45\textwidth ]{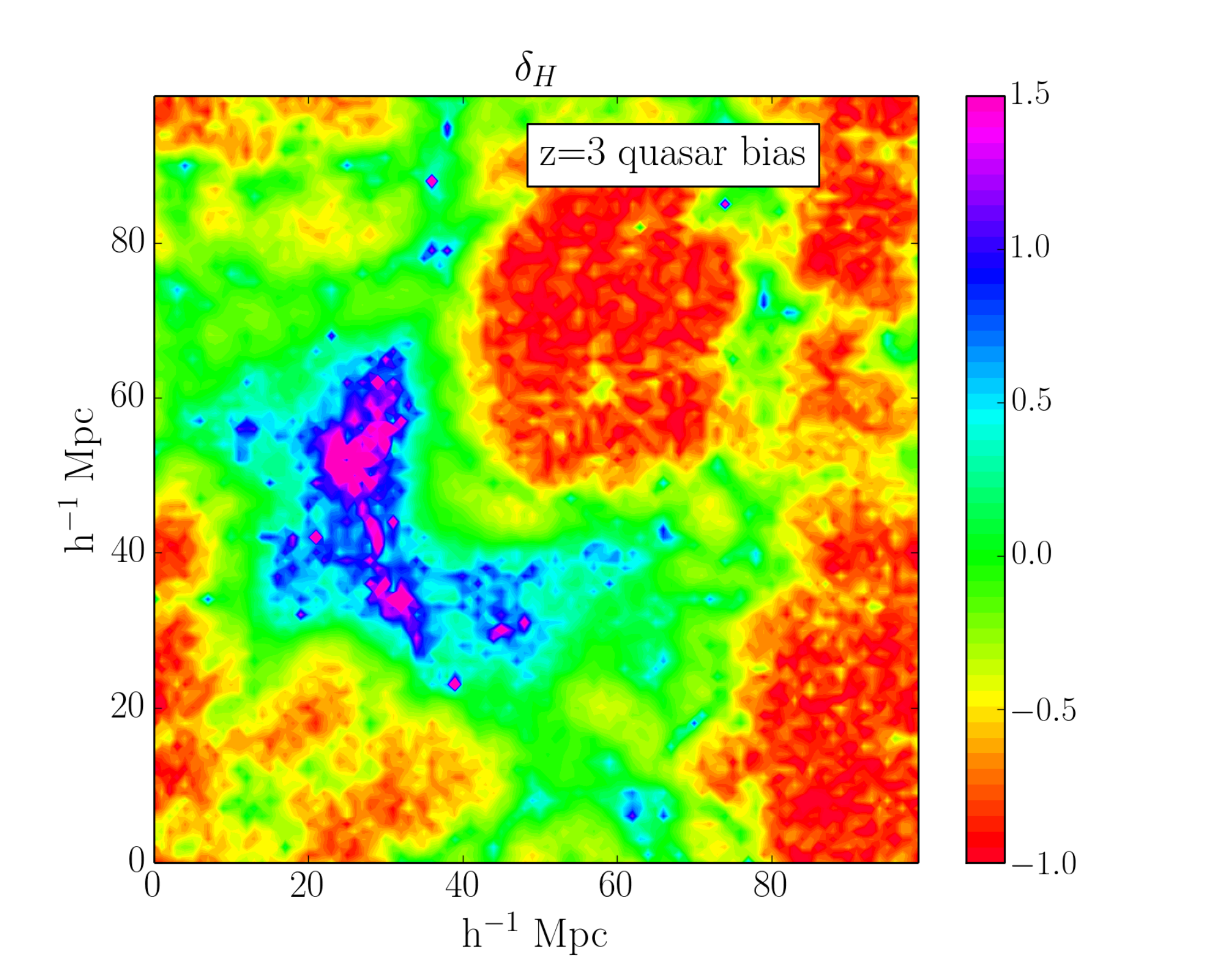}
\includegraphics[width = .45\textwidth ]{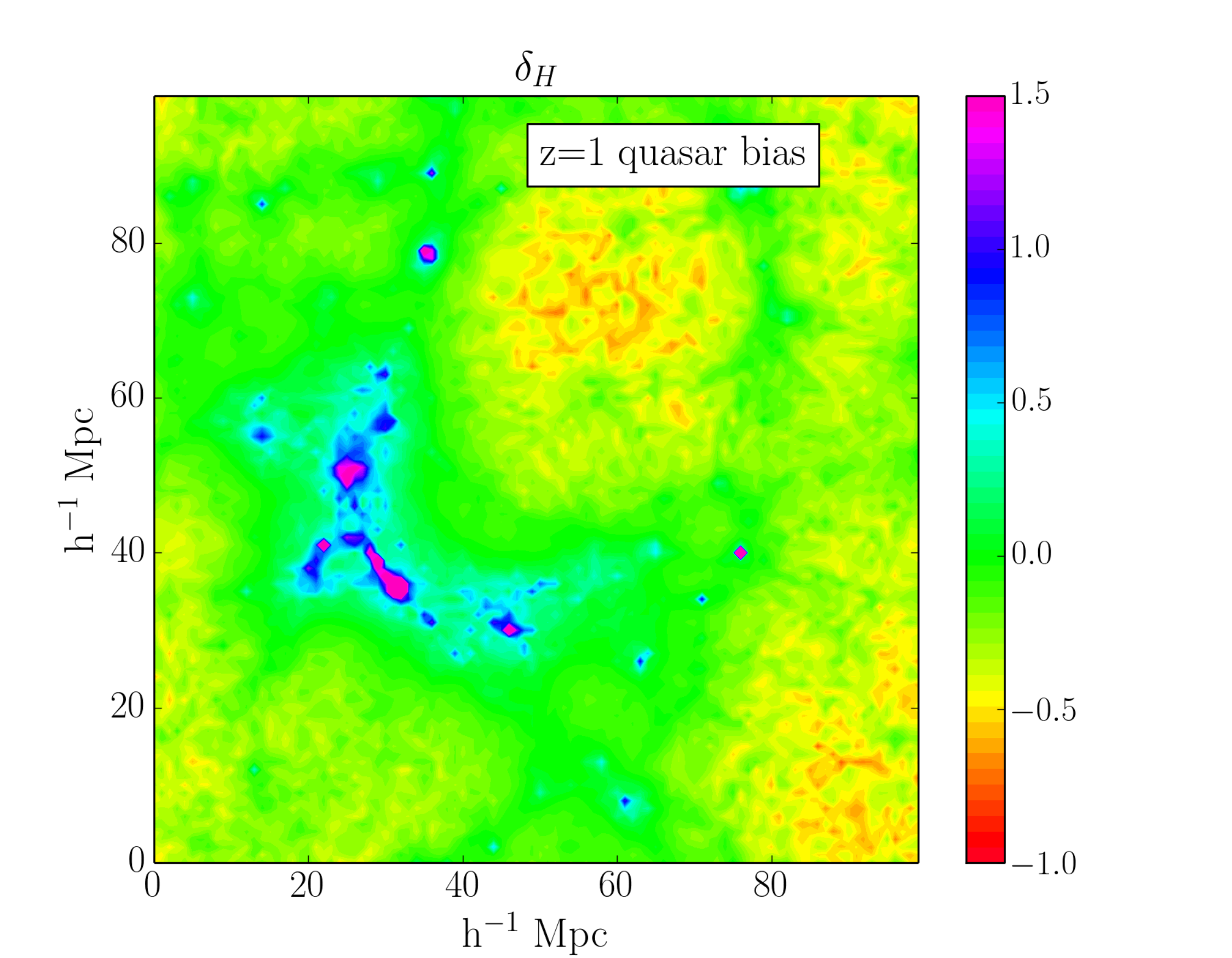}
\caption{ Slices through the midplane of the box, for $z=3$ (left column) and $z=1$ (right column): logarithm of the density with respect to the mean value (upper row), heating rate fluctuations with the galaxy (middle row) and quasar bias model (lower row). }
\label{fig:slice}
\end{figure*}
\begin{figure*}
\centering
\includegraphics[width = .45\textwidth ]{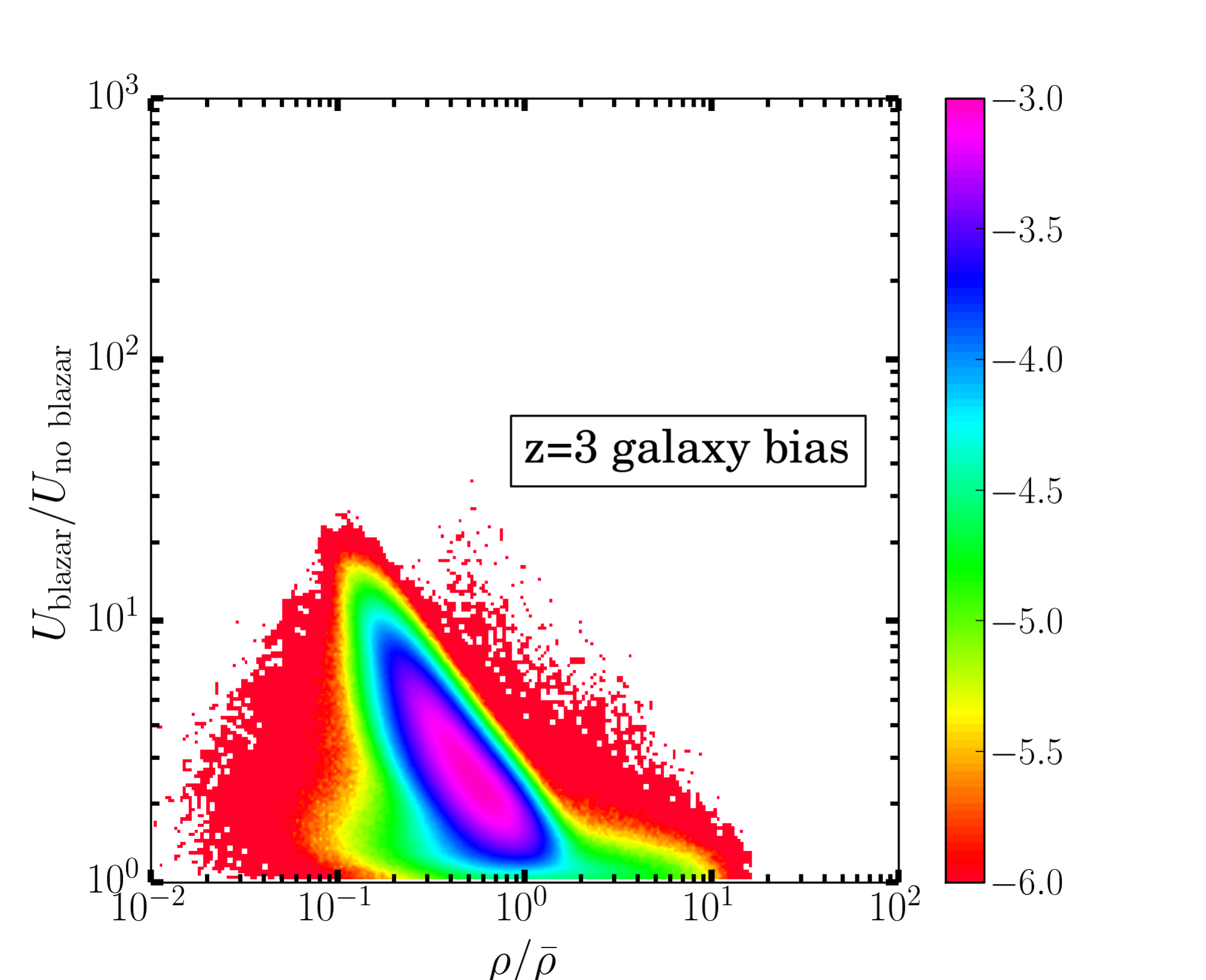}
\includegraphics[width = .45\textwidth ]{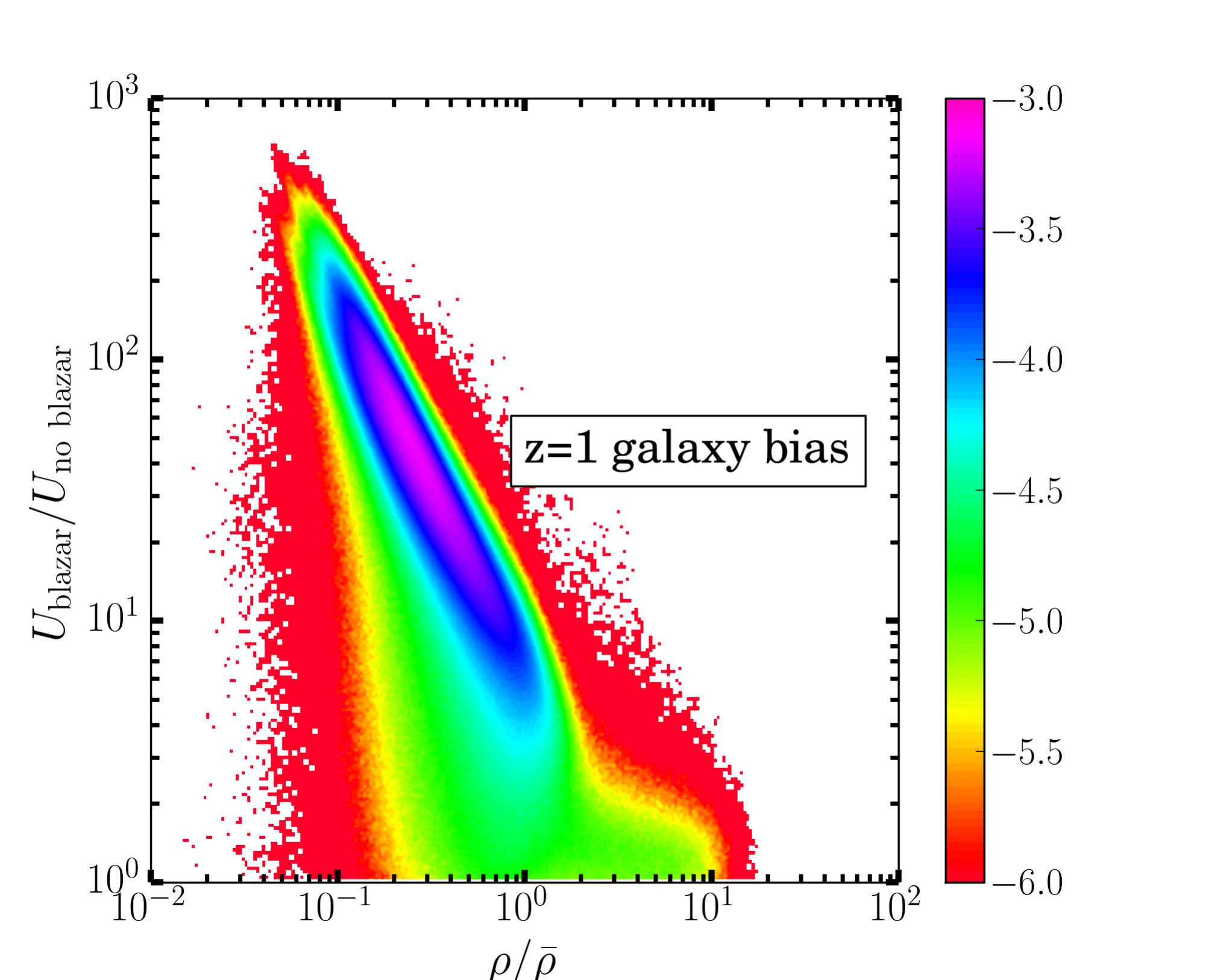}
\includegraphics[width = .45\textwidth ]{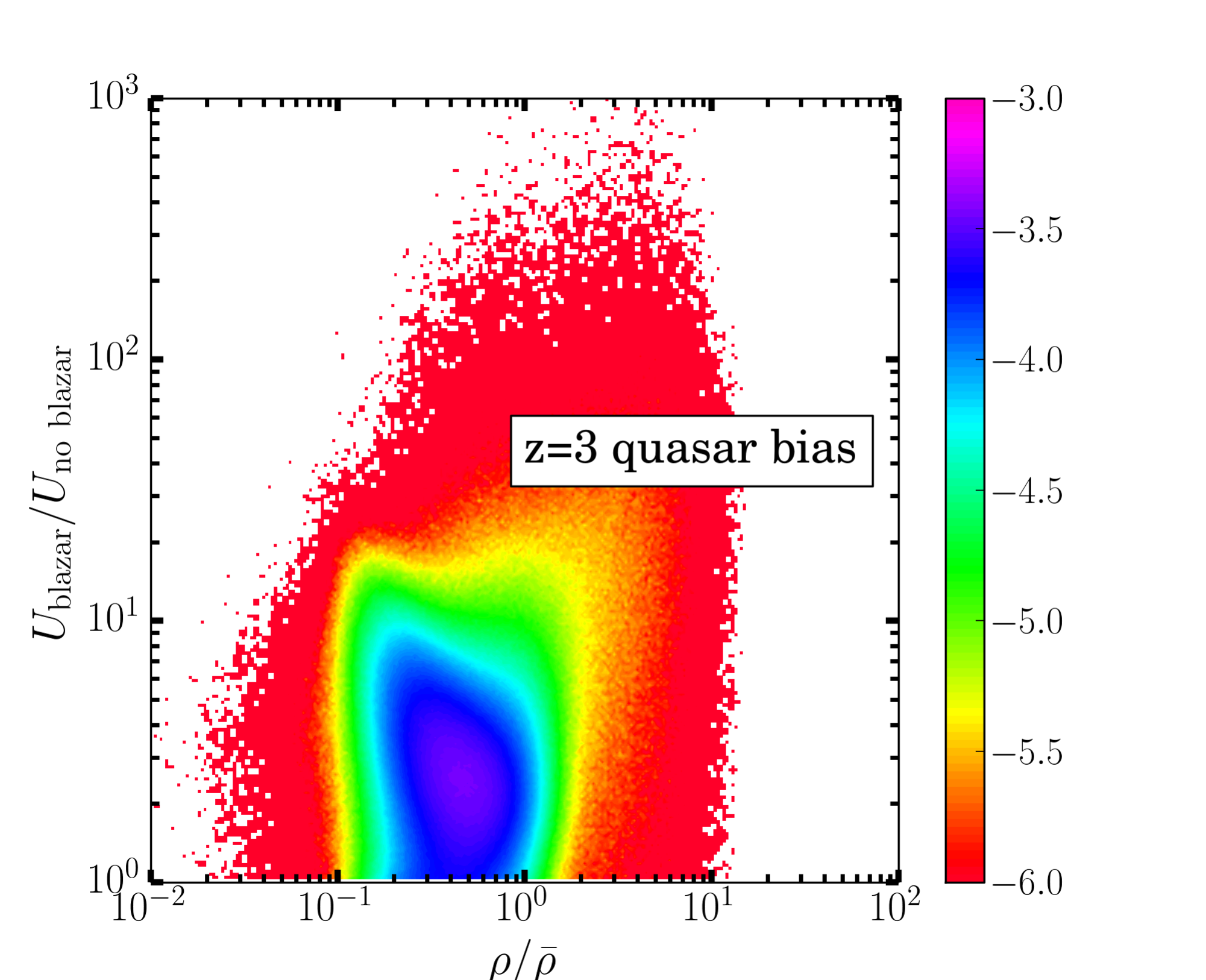}
\includegraphics[width = .45\textwidth ]{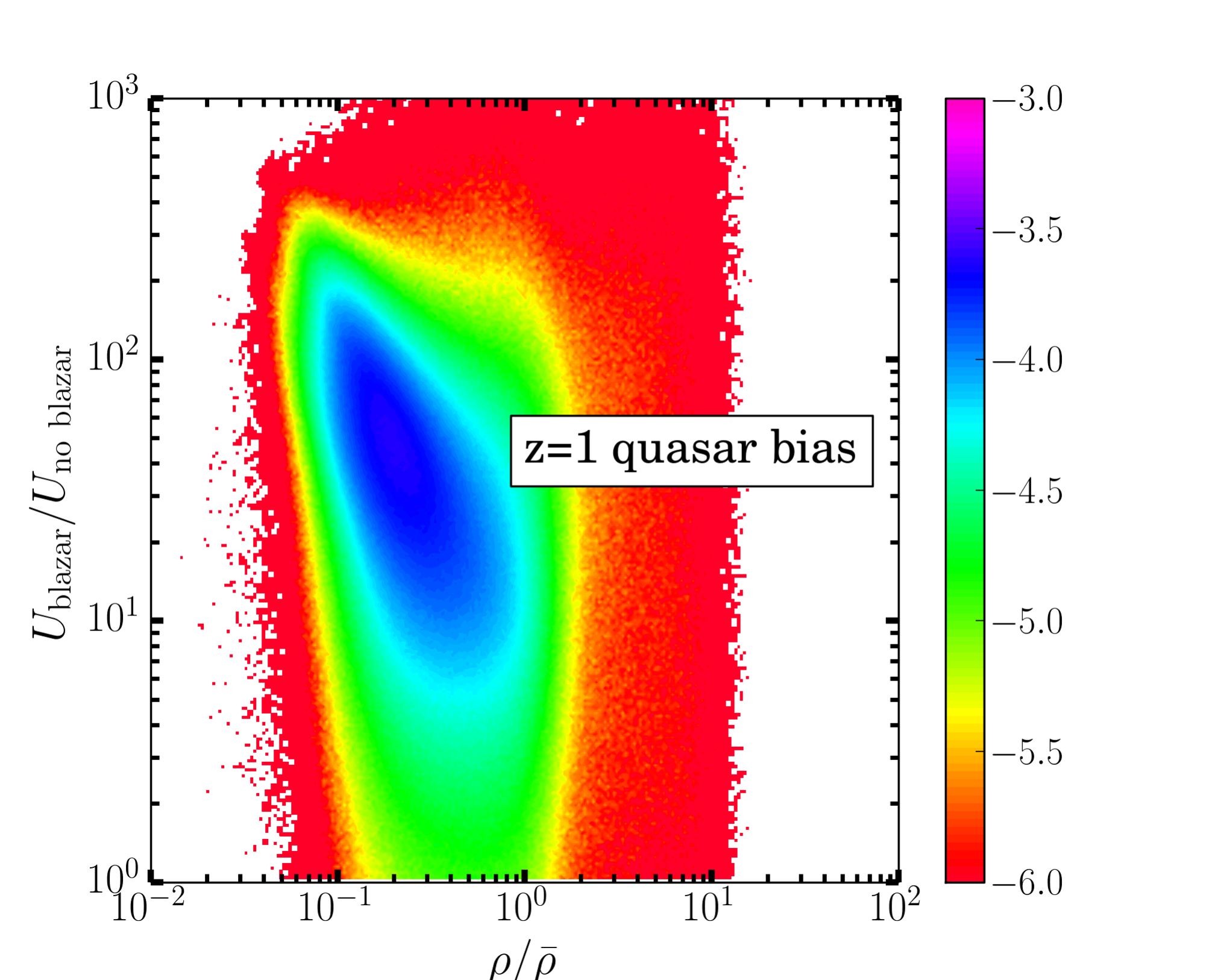}

\caption{Ratio between the internal energy ($U=3 k_B T / 2$) in simulations with blazar heating to those without, as a function of overdensity. The plots show the galaxy bias model (top) and the quasar bias model (bottom) for $z=3$ (left) and $z=1$ (right). The color scale is logarithmic.}
\label{fig:heating_ratio}
\end{figure*}
Figure~\ref{fig:slice} shows the heating rate fluctuations in the midplane of the $L_x=100h^{-1}$ Mpc simulation for $z=3$ and $1$ for both the galaxy and quasar bias model. Both simulation models are started from identical initial conditions such that any visible difference is solely due to the different bias assumptions in the blazar heating model. The corresponding density field is shown in the upper column and shows increasing structure formation as the redshift decreases. The heating map has a linear scale while the density scale is logarithmic. The heating rate fluctuations are, on average, much smaller and more spatially homogeneous than the density fluctuations. This is because the window function filters out small scales, which correspond to collapsed regions, where density fluctuations are the highest. To the zeroth order, one can thus consider TeV blazar heating to be uniform, as was assumed in \citet{2012ApJ...752...23C}.

Additional heating (i.e. $\delta_H>0$) occurs around clustered regions. This is expected, as the window function translates large scale density fluctuations into heating rate fluctuations. Conversely, underdense regions, such as the one around $\{x=60,y=70\}$ (for $z=3$) display heating below average as they are isolated from sources, and their flux decreases as $e^{-\tau} r^{-2}$. This is most obvious at high redshift, and for the quasar model, where bias is the strongest. As the redshift decreases, heating rate fluctuations decrease because of the decreasing bias.

Figure~\ref{fig:heating_ratio} shows the volume weighted ratio of the internal energy when blazar heating is included to the internal energy when blazar heating is not included, as a function of the density. For both simulation models, we divided the simulation volume into equally-sized small sub-volumes and computed the internal energy therein. Because both simulations have been started from identical initial conditions, the interference of the primordial density waves gives rise to a comparable morphology of the cosmic web so that we can compare thermodynamic quantities of (almost) identical sub-volumes of the different simulations. These maps clearly highlight that blazar heating has more impact in underdense regions, as the heating rate per baryon is higher. Even if these regions receive less heat than regions with higher density (see Figure~\ref{fig:slice}),  blazar heating increases the internal energy by a factor two at $z=3$ and about an order of magnitude at $z=1$. At low redshift, even in this inhomogeneous model, most of the gas is heated up and there is minimal difference between the two models we used for the bias.

\begin{figure*}
\centering

\includegraphics[trim=.5cm 0cm 3cm .5cm, clip ,width = .32\textwidth ]{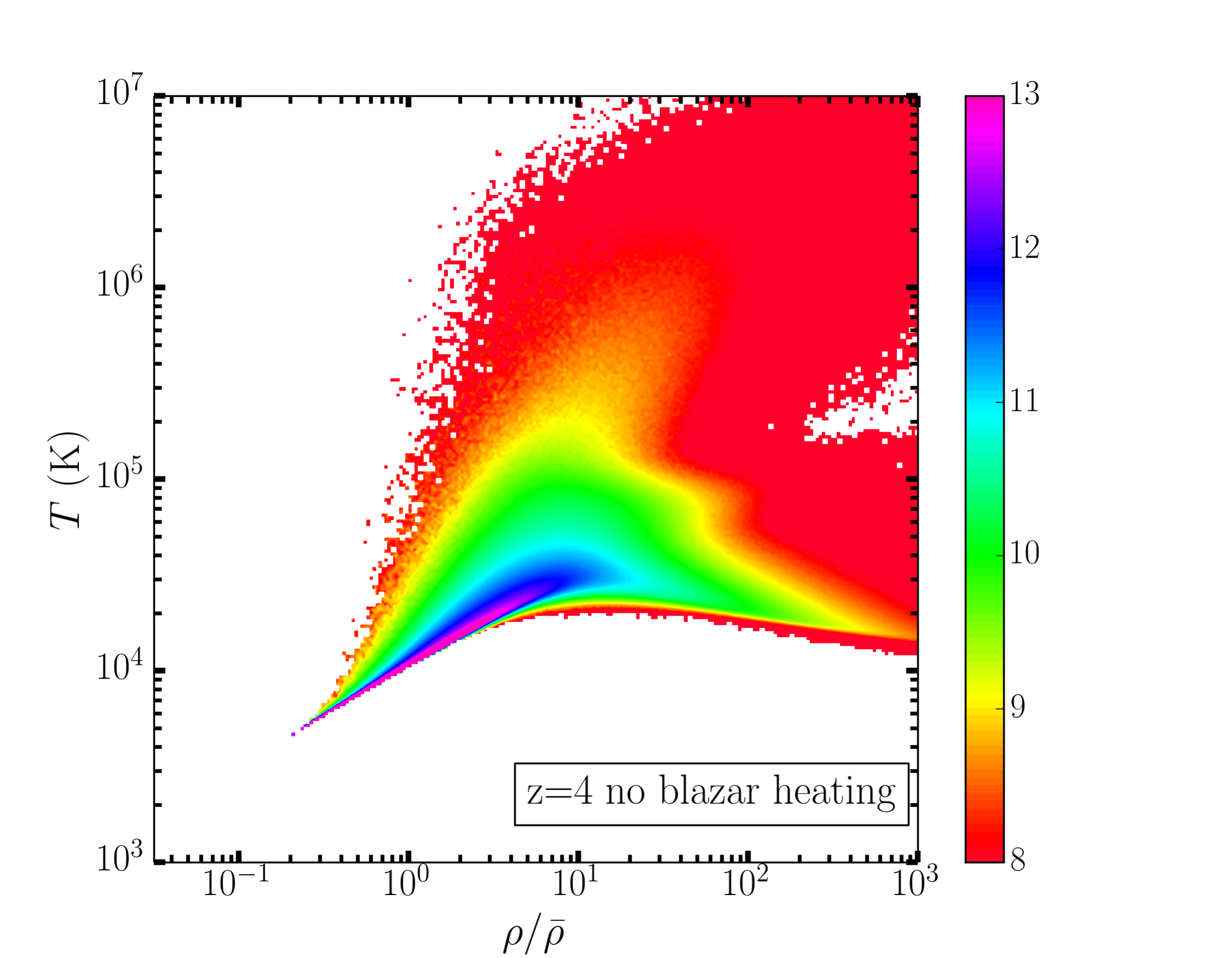}
\includegraphics[trim=.5cm 0cm 3cm .5cm, clip ,width = .32\textwidth ]{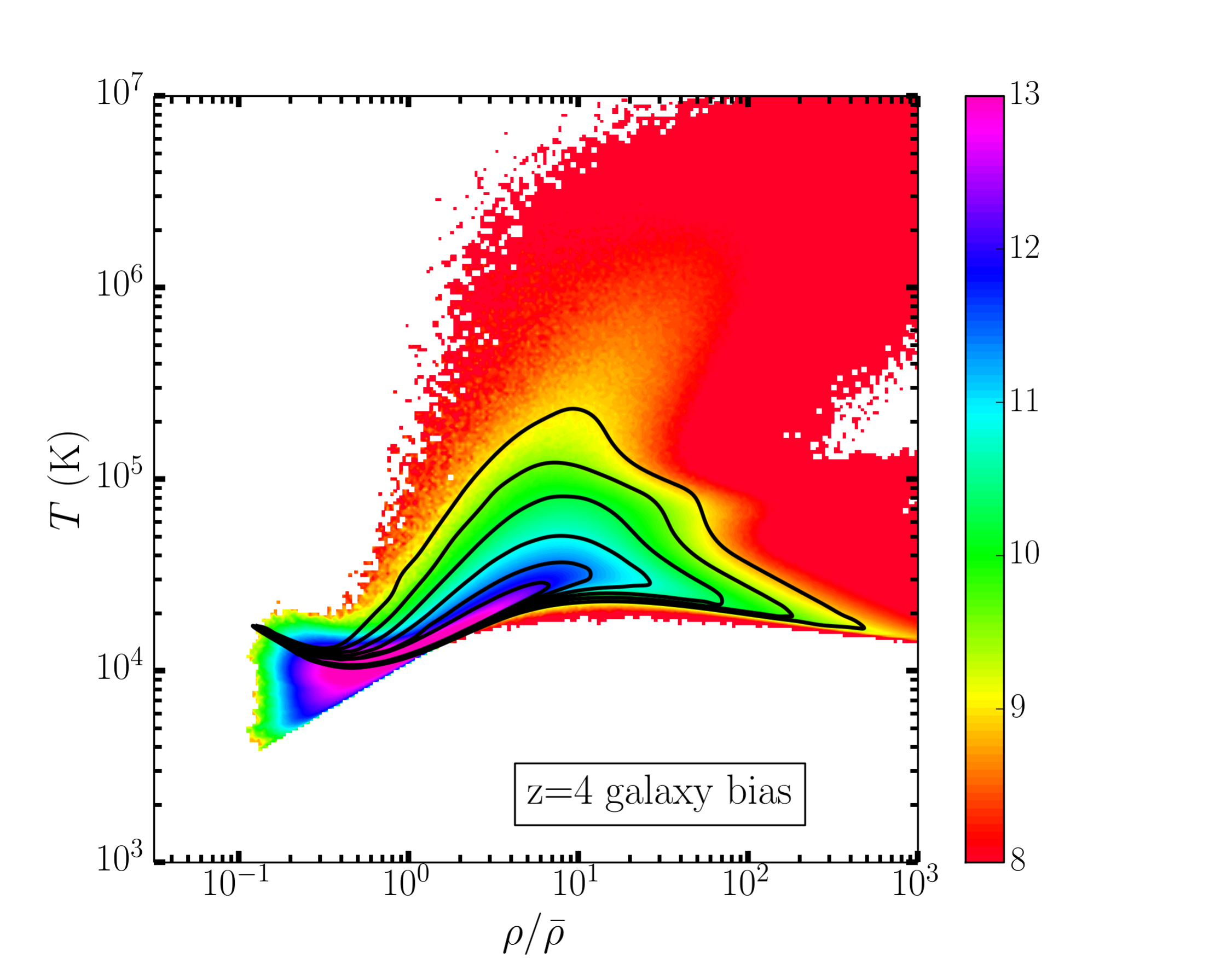}
\includegraphics[trim=.5cm 0cm 3cm .5cm, clip ,width = .32\textwidth ]{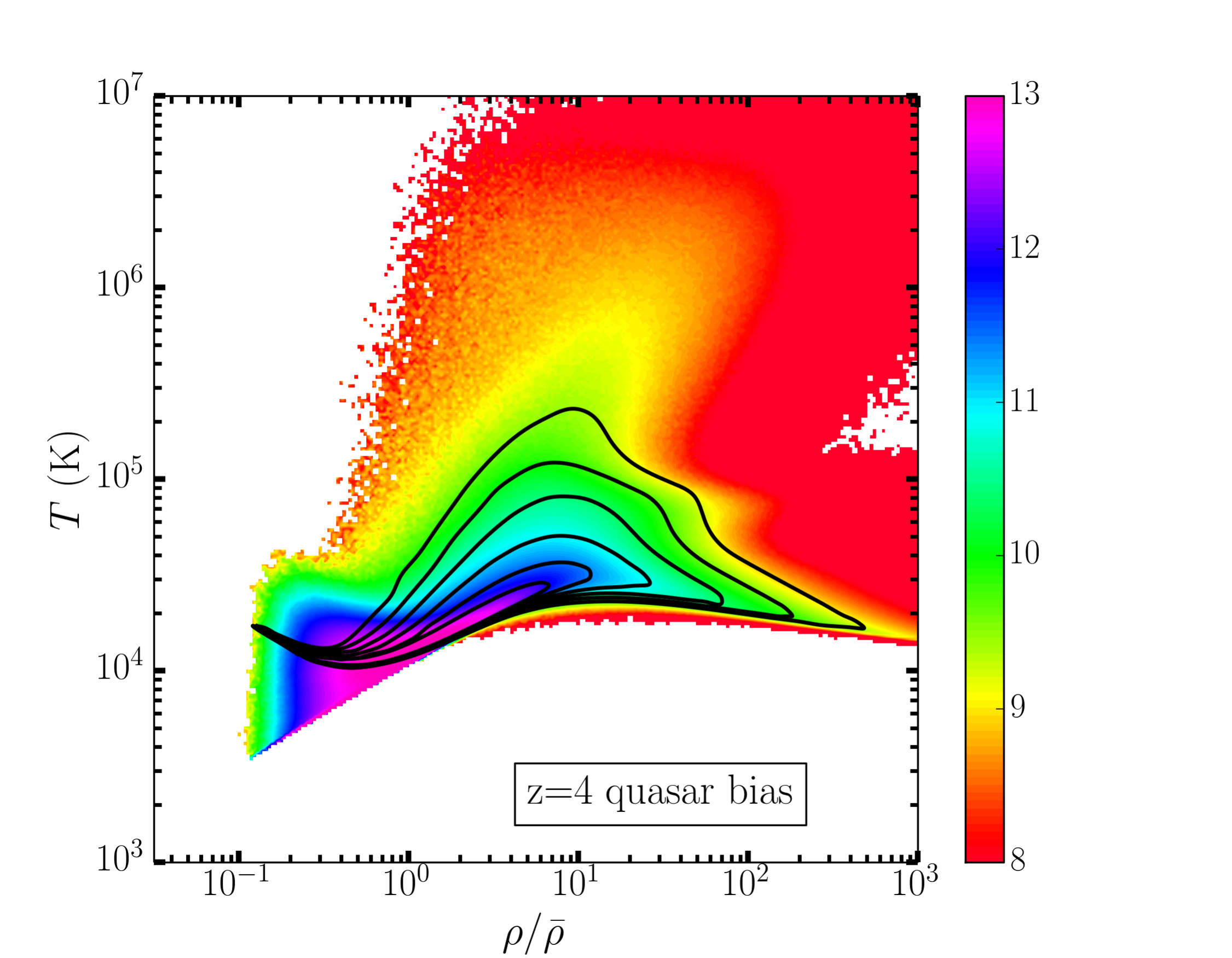}
\includegraphics[trim=.5cm 0cm 3cm .5cm, clip ,width = .32\textwidth ]{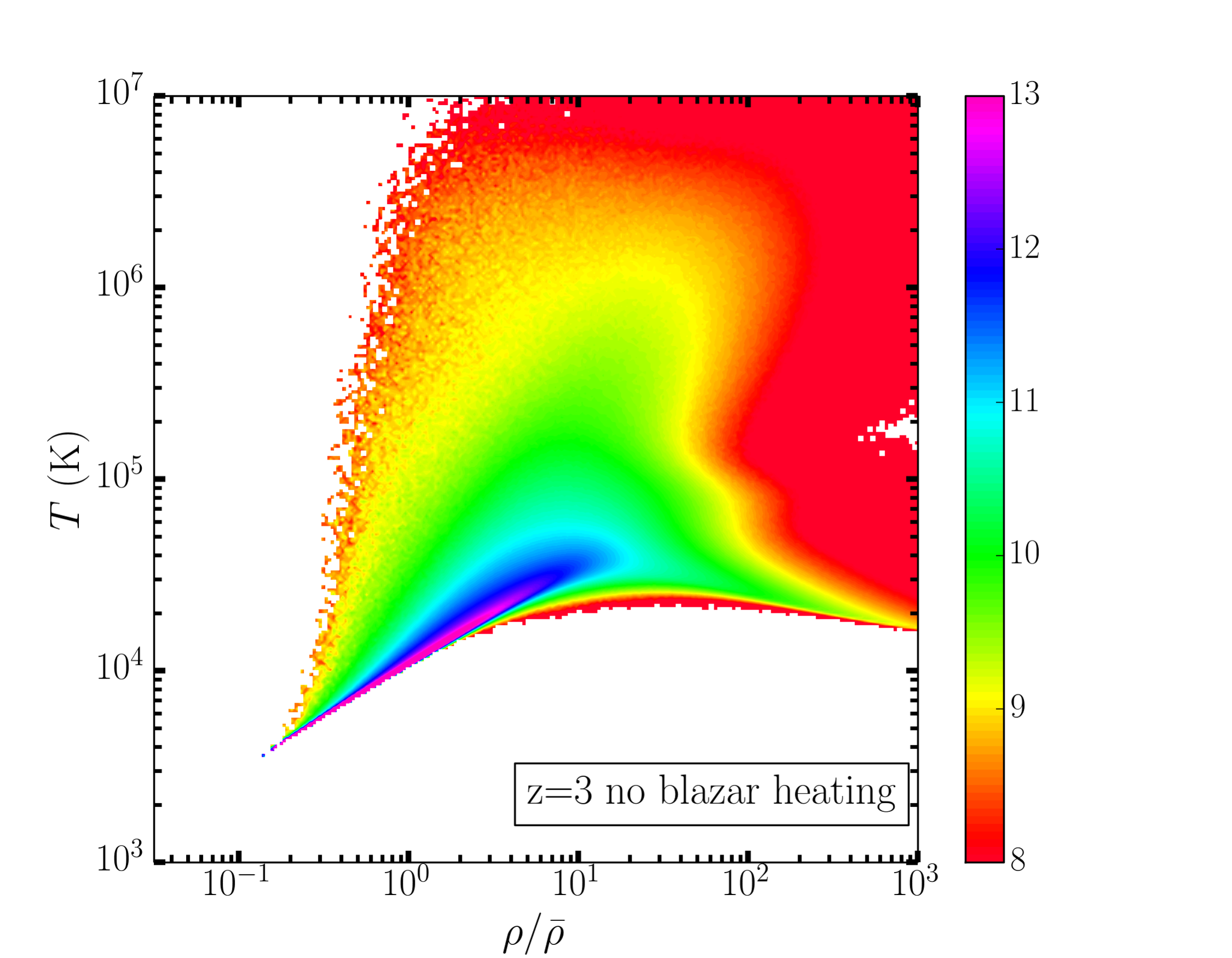}
\includegraphics[trim=.5cm 0cm 3cm .5cm, clip ,width = .32\textwidth ]{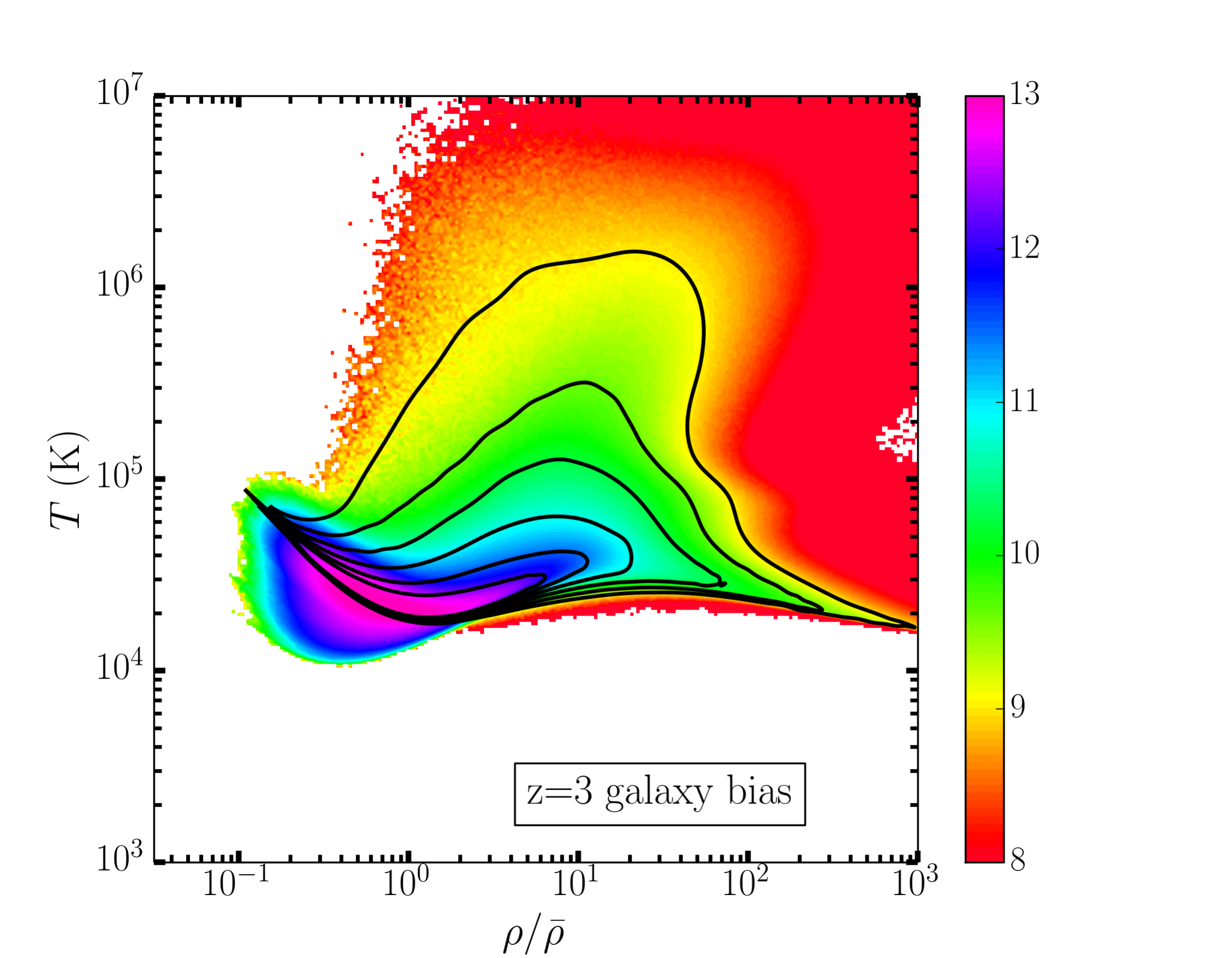}
\includegraphics[trim=.5cm 0cm 3cm .5cm, clip ,width = .32\textwidth ]{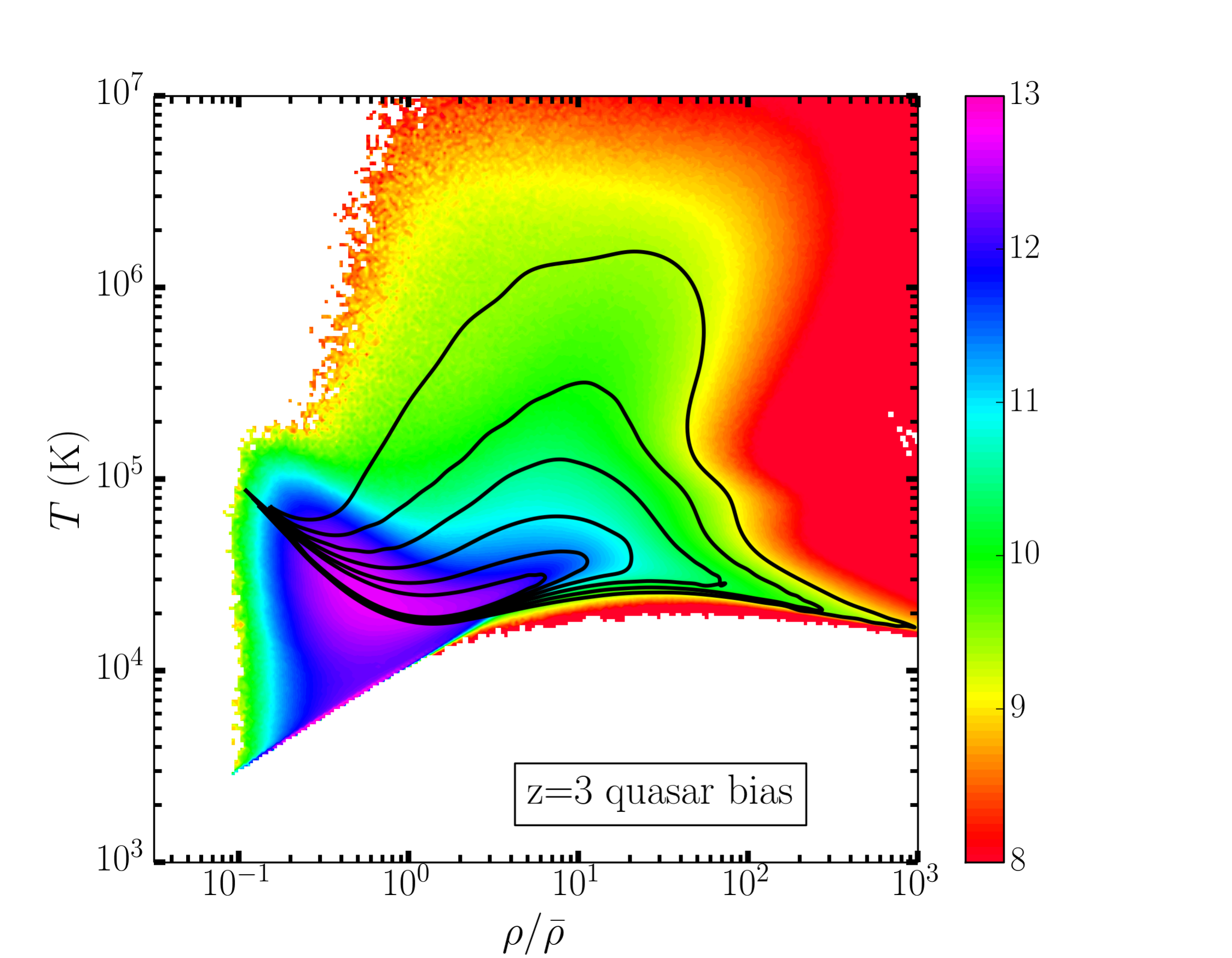}
\includegraphics[trim=.5cm 0cm 3cm .5cm, clip ,width = .32\textwidth ]{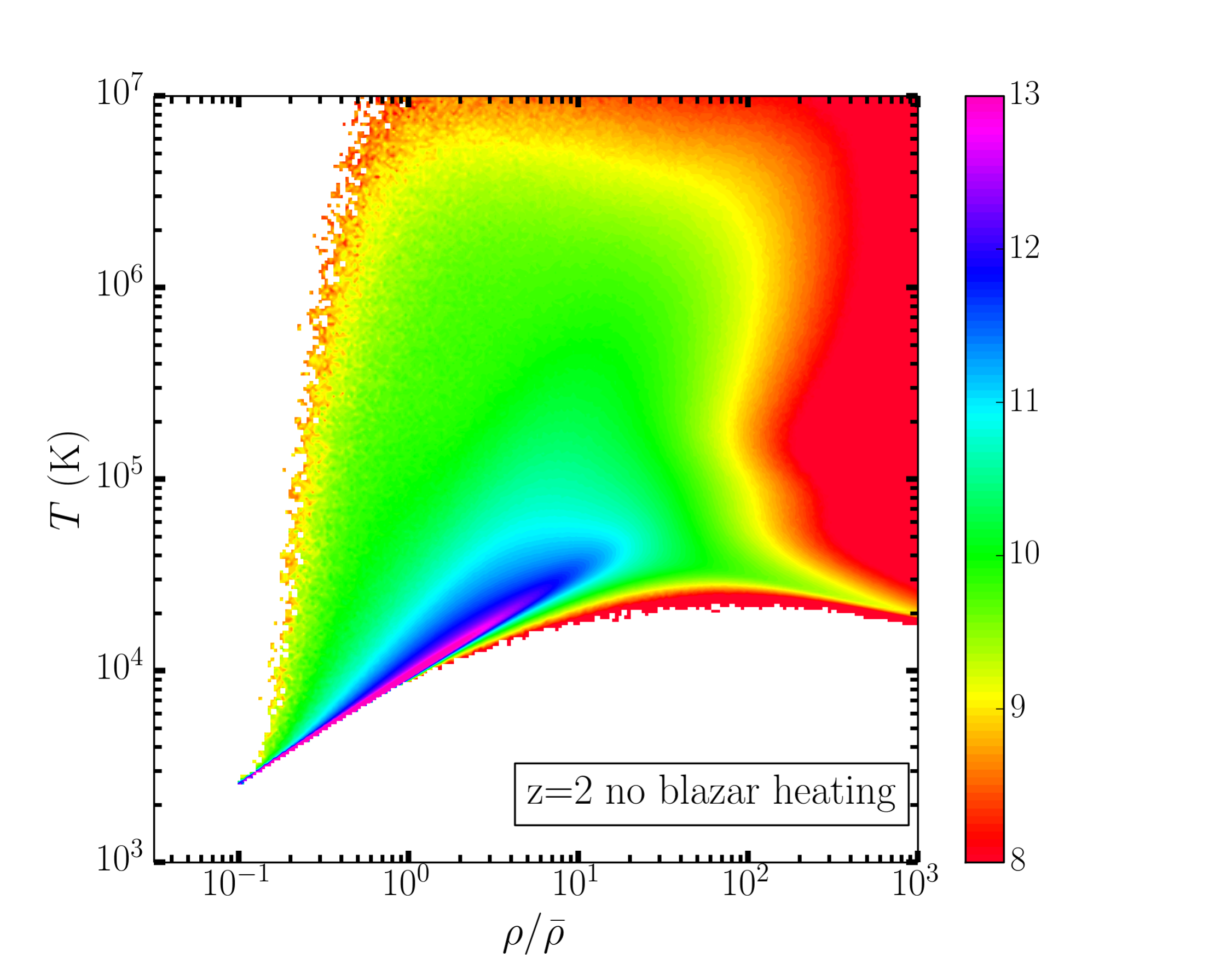}
\includegraphics[trim=.5cm 0cm 3cm .5cm, clip ,width = .32\textwidth ]{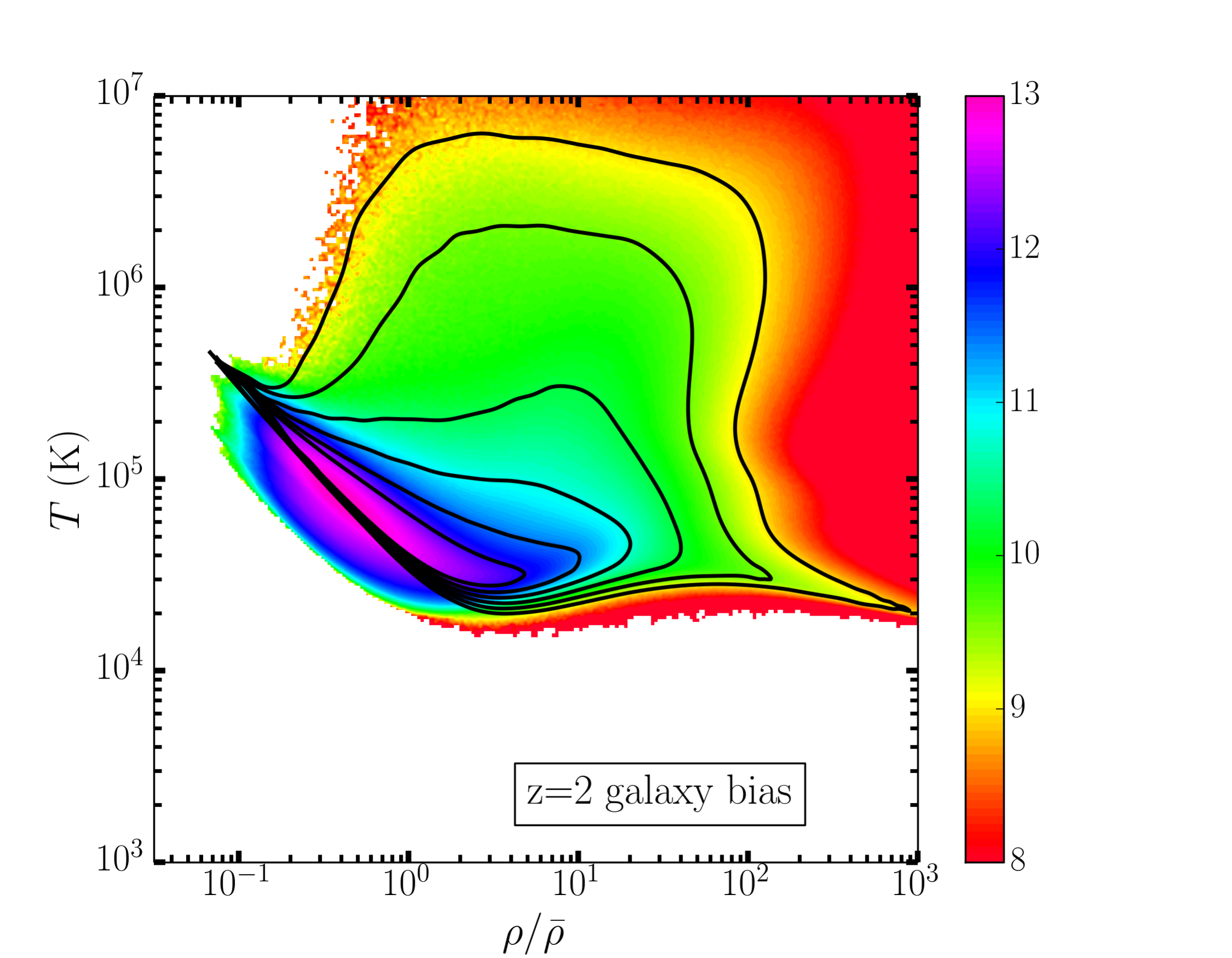}
\includegraphics[trim=.5cm 0cm 3cm .5cm, clip ,width = .32\textwidth ]{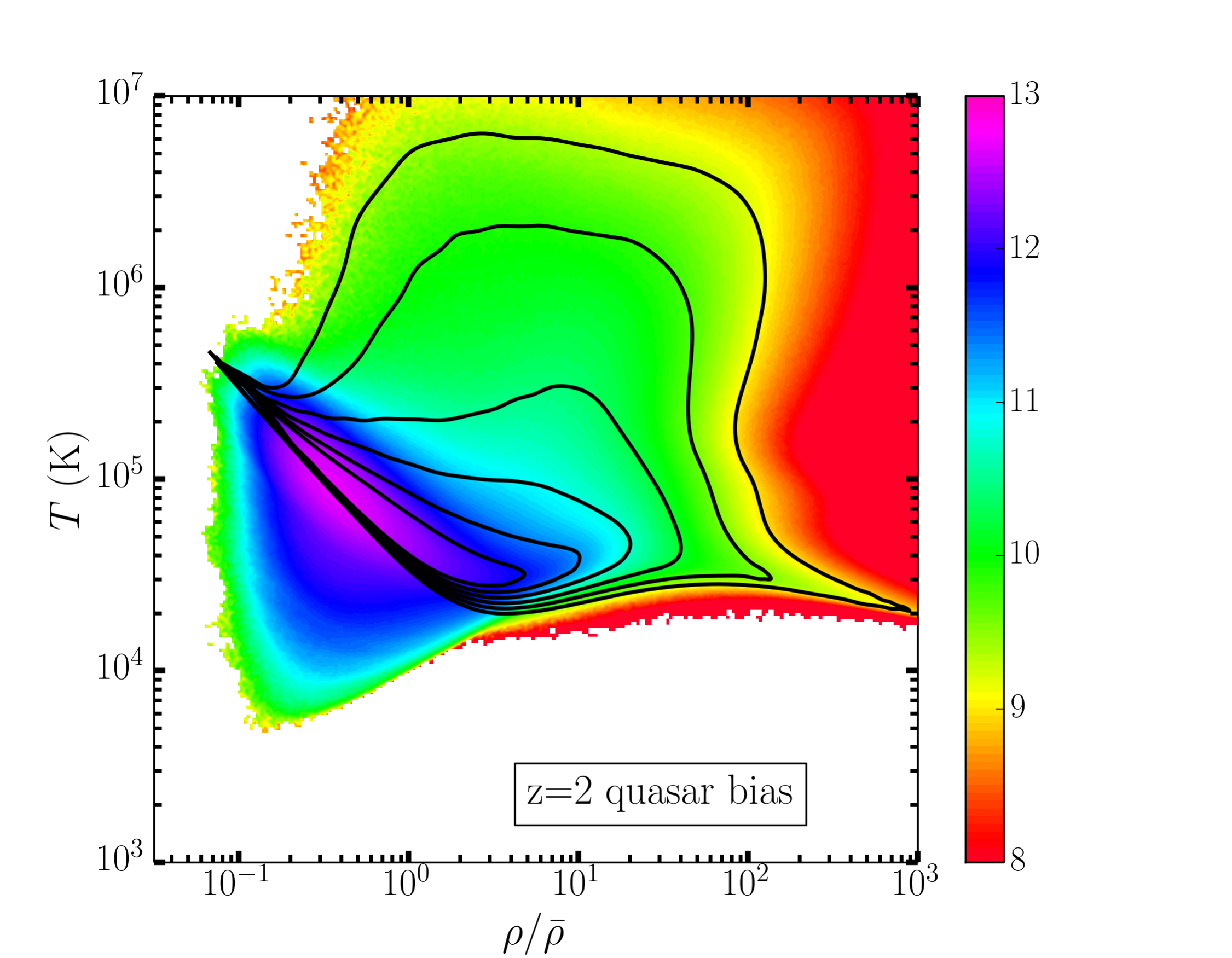}
\includegraphics[trim=.5cm 0cm 3cm .5cm, clip ,width = .32\textwidth ]{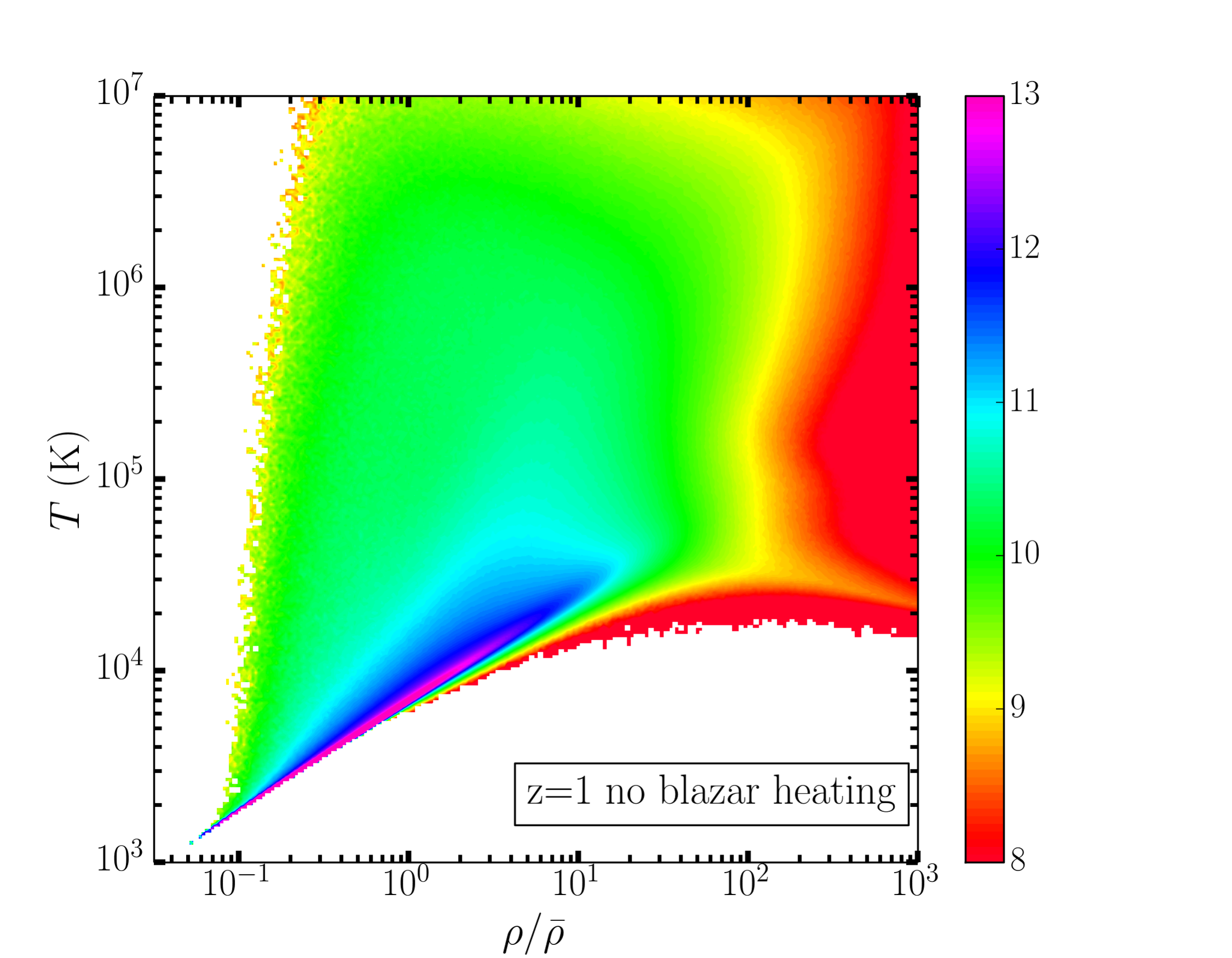}
\includegraphics[trim=.5cm 0cm 3cm .5cm, clip ,width = .32\textwidth ]{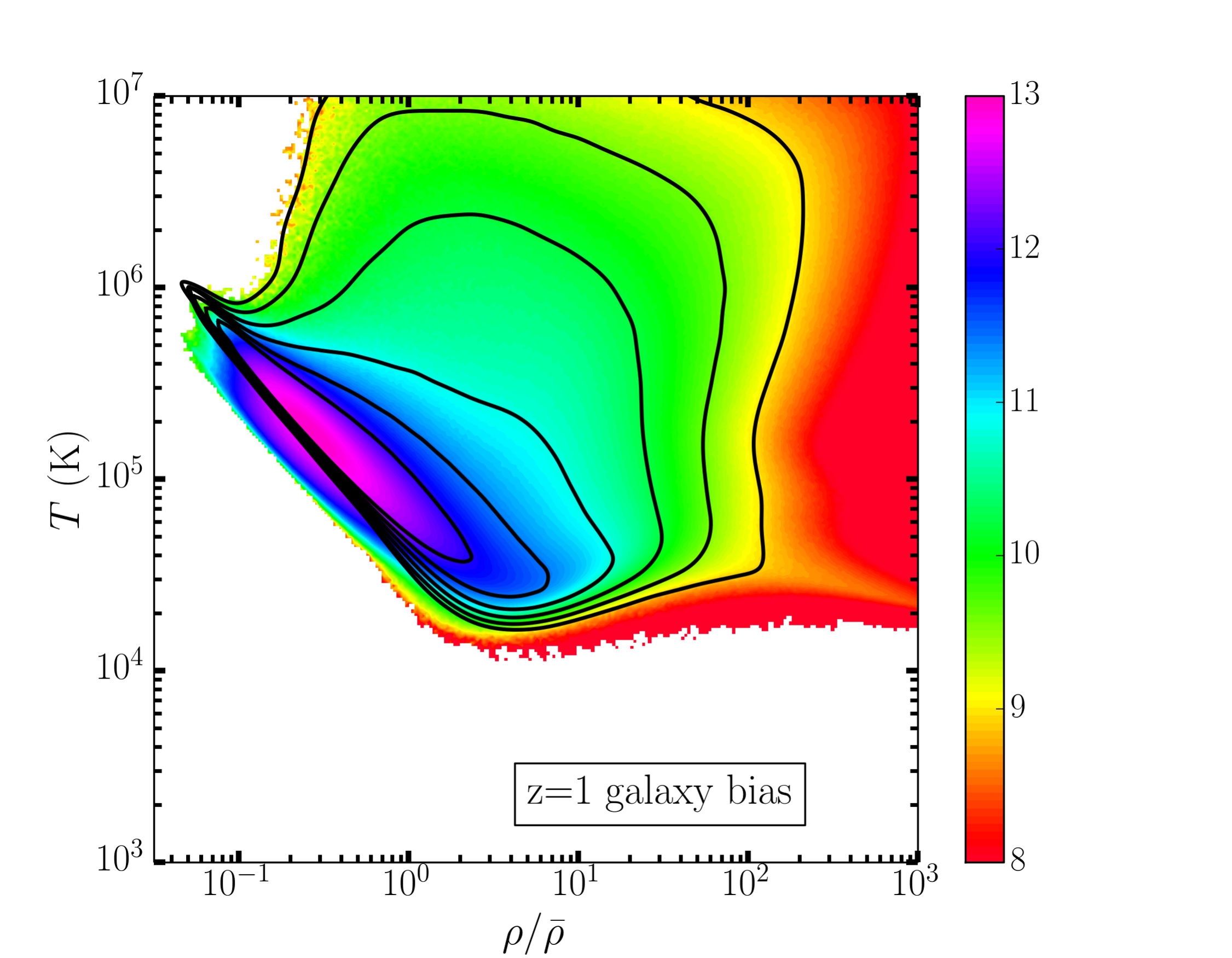}
\includegraphics[trim=.5cm 0cm 3cm .5cm, clip ,width = .32\textwidth ]{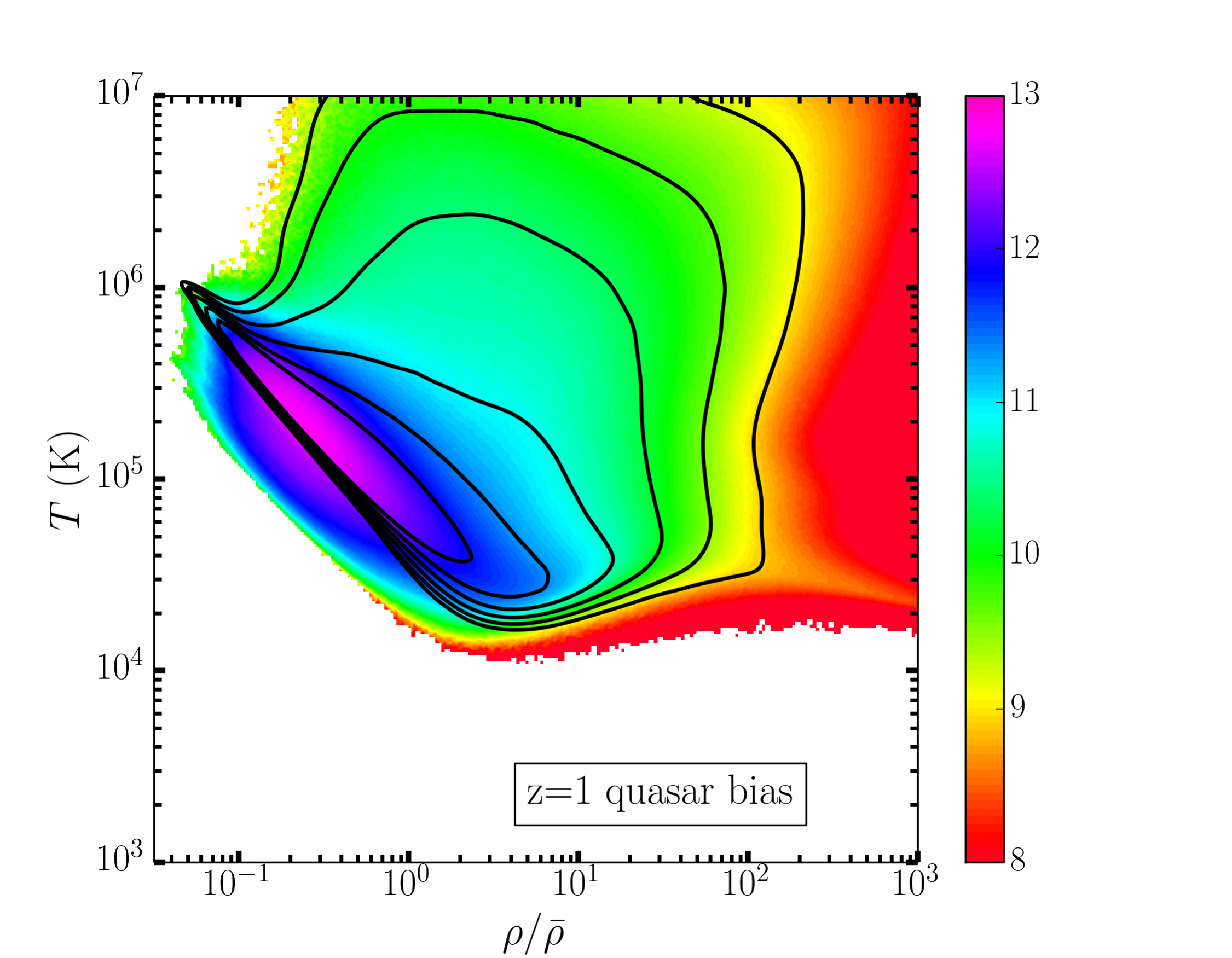}
\caption{ Volume-weighted temperature - density relation at $z=4,3,2,1$ (from top to bottom) for the simulations with no blazar heating (left), inhomogeneous heating following galaxy bias (middle), and the quasar bias model (right). The overlying black contours show the corresponding $T-\rho_{\mathrm{gas}}$ relation for uniform blazar heating \citep{2012MNRAS.423..149P} for the same redshift range. The color scale is logarithmic. None of the models includes radiative transfer effects of He\,\textsc{II} reionization}
\label{fig:T_rho}
\end{figure*}

The impact of inhomogeneous heating translates into a more complex temperature-density relation, as is shown on Figure~\ref{fig:T_rho}. The color map shows the mass weighted $T-\rho_{\mathrm{gas}}$ relation from our simulations and the grey contours show the case for uniform blazar heating with the same resolution \citep{2012MNRAS.423..149P}. Temperature measures the integrated impact of TeV blazar heating over time. The left column shows the $T-\rho_{\mathrm{gas}}$ relation in a model with no blazar heating. In the latter case, the underdense gas follows a very narrow distribution, where the lowest density gas is the coldest.  When blazar clustering is taken into account, the temperature-density relation has a significant scatter for underdense regions for $z\simeq 2-3$. For the quasar bias model, this scatter results in the lower envelope of the temperature-density that differs little from the case with no blazar heating. However, the mode of the temperature is very close to the uniform blazar heating case. At $z=1$  both models show a very similar behavior and blazar heating can be considered nearly homogeneous, though the lower envelope sits at a lower temperature.

\begin{figure}[h]
\centering
\includegraphics[width = .45\textwidth ]{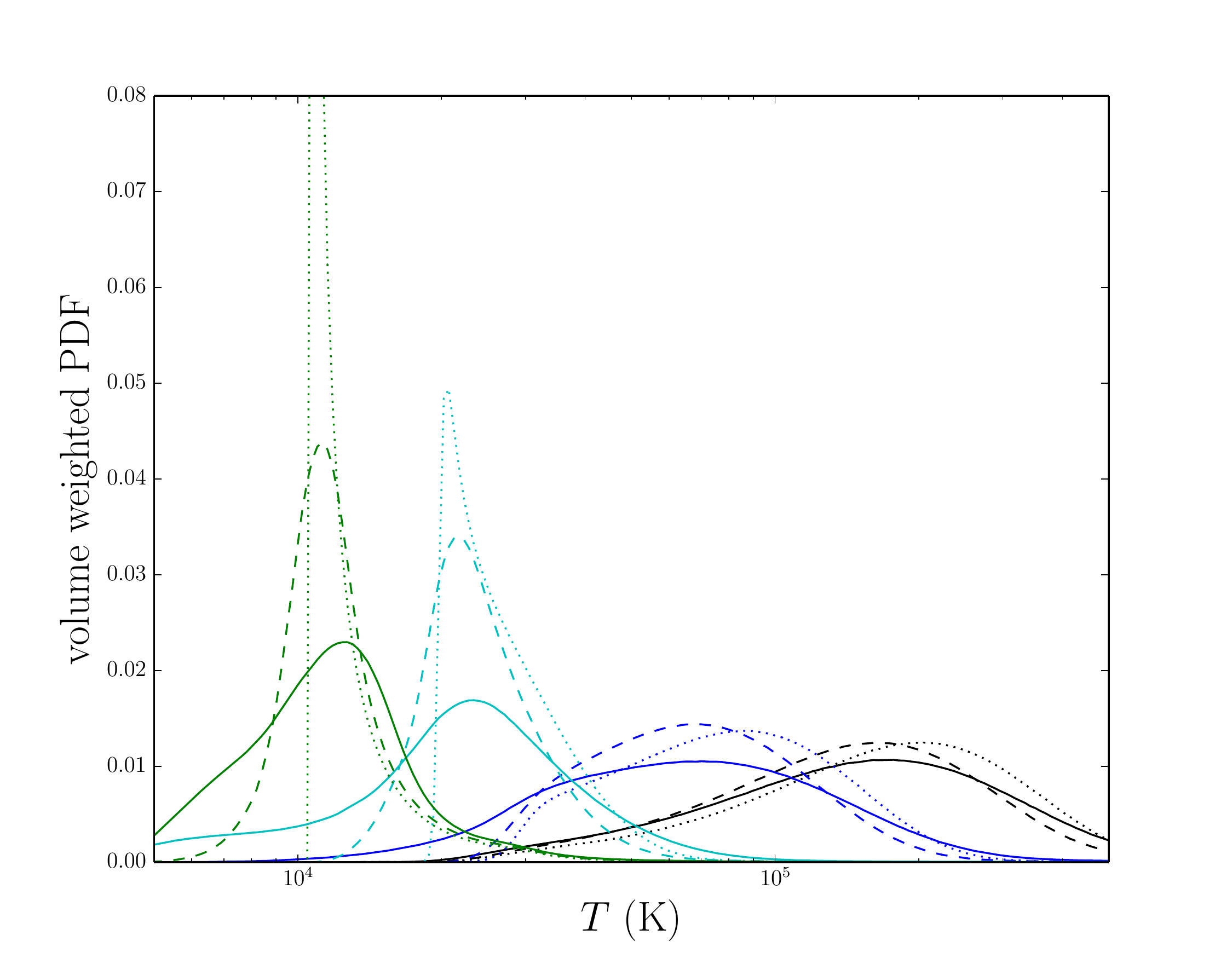}
\caption{Volume-weighted temperature probability distribution function for $z=1$ (black), 2 (blue), 3 (cyan) and 4 (green) for the quasar bias model (solid line), galaxy bias model (dashed line) and the uniform case (dotted line). The peaks in the temperature distribution move towards higher temperatures as the redshift decreases.}
\label{fig:PDF}
\end{figure}

Figure~\ref{fig:PDF} shows the volume-weighted probability distribution functions of the temperature for all the simulations. This provides a more quantitative view of the scatter in temperature and highlights the impact of clustering on TeV blazar heating. The simulations with  inhomogeneous blazar heating show significant deviation from the uniform case  for $z\geqslant 2$. The higher value of the quasar bias results in the presence of more unheated gas\ than in the galaxy bias model. Here the effect of the lower envelope is clear as there is a significant tail toward lower temperatures: the coldest zones have $T\leq 5 \times 10^3K$ while the warmest zones have $T\simeq   10^5K$ for $z=3$. Conversely, the warmest gas is only slightly warmer than the mode of the temperature. At lower redshift, the mode of the temperature is very similar in all models but the inhomogeneous models exhibit a larger amount of colder gas.

\begin{figure*}[h]
\centering
\includegraphics[width = .45\textwidth ]{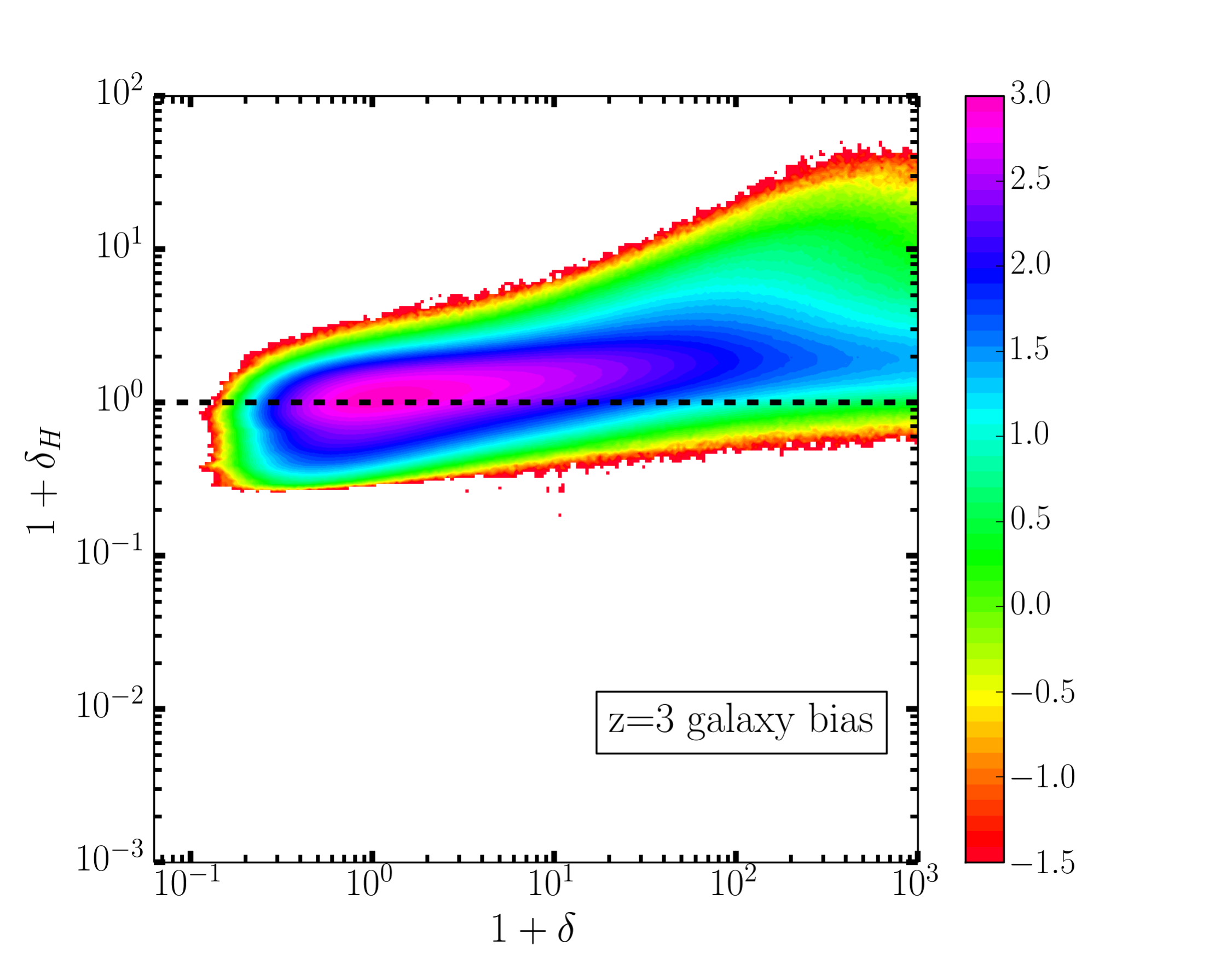}
\includegraphics[width = .45\textwidth ]{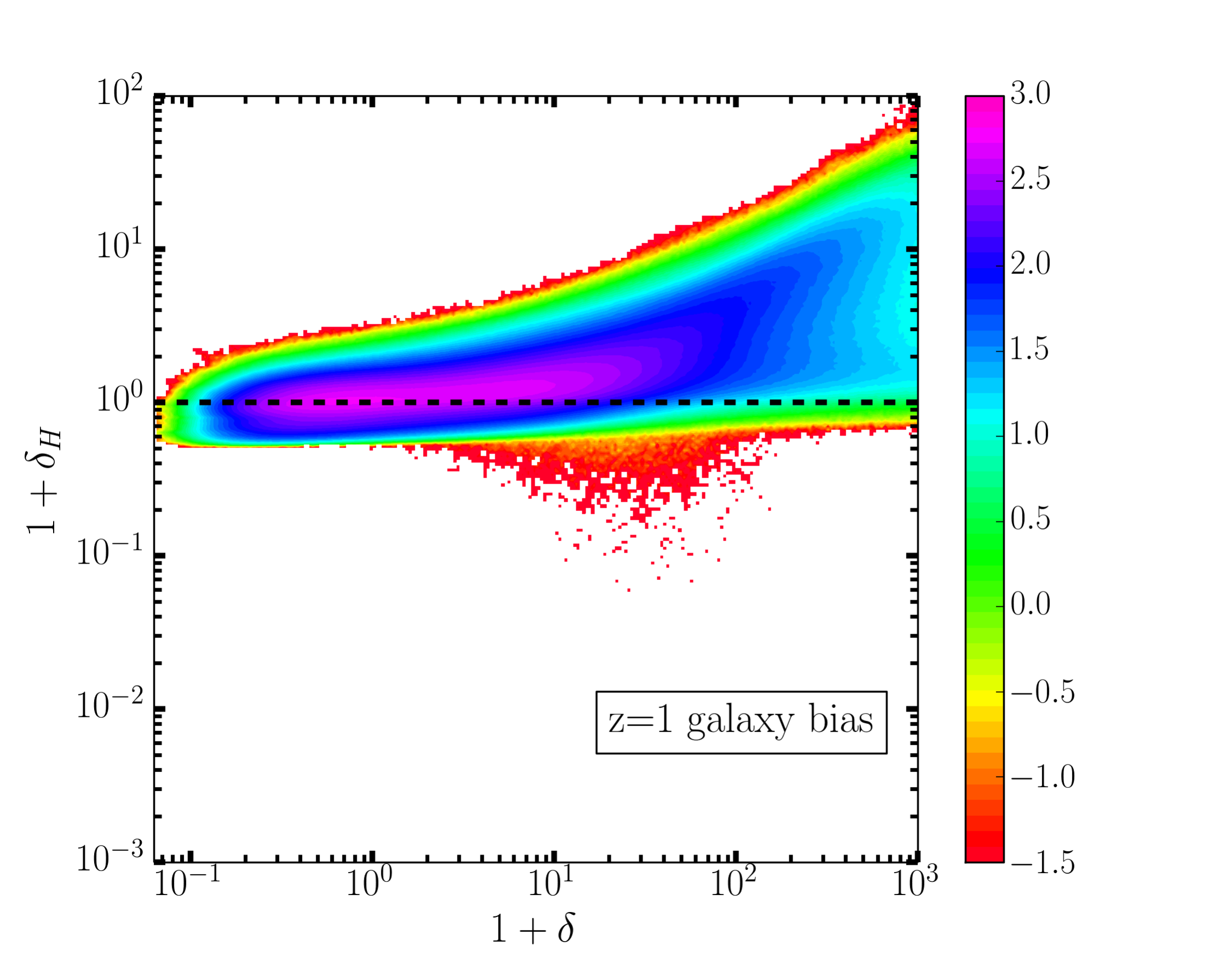}\\
\includegraphics[width = .45\textwidth ]{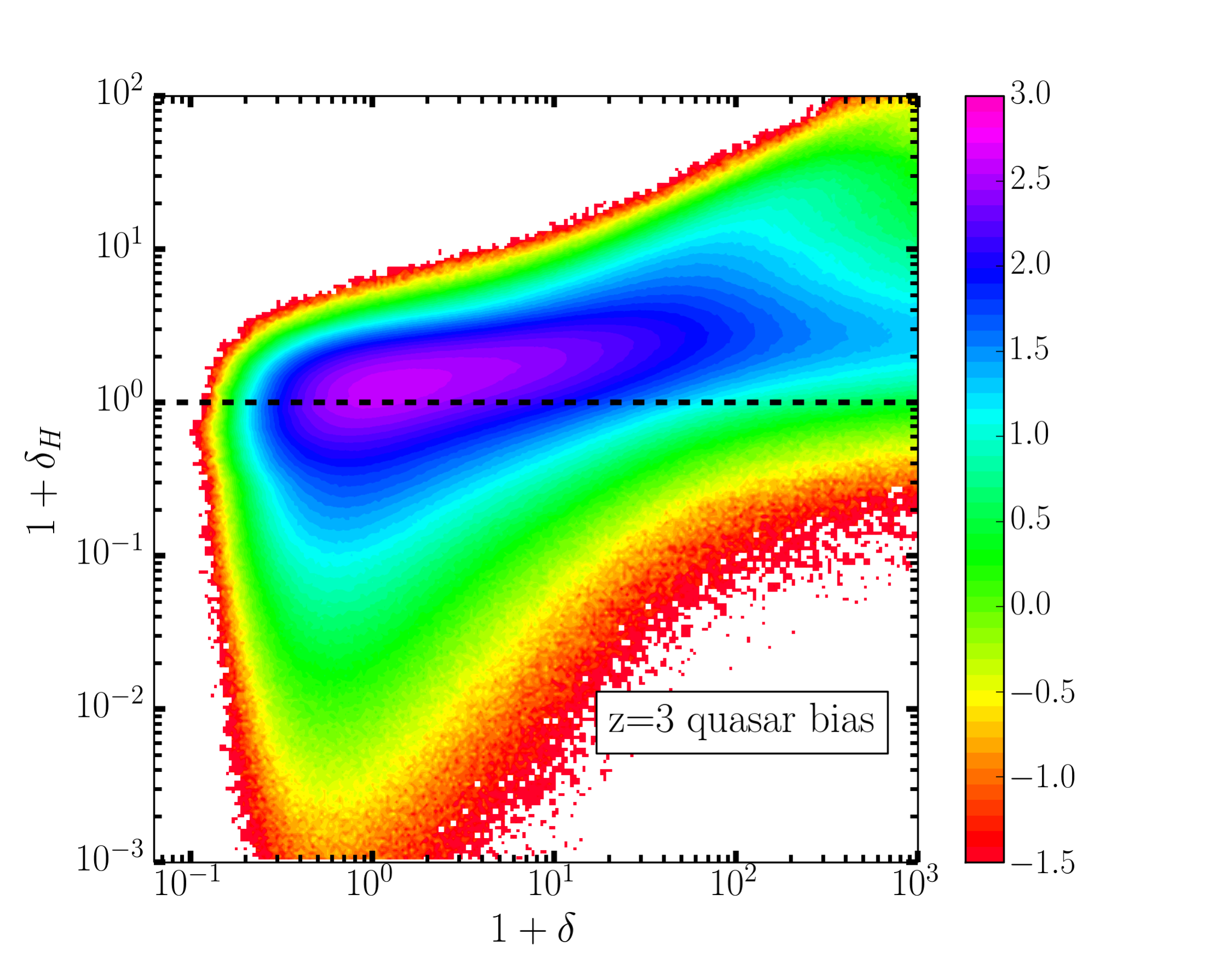}
\includegraphics[width = .45\textwidth ]{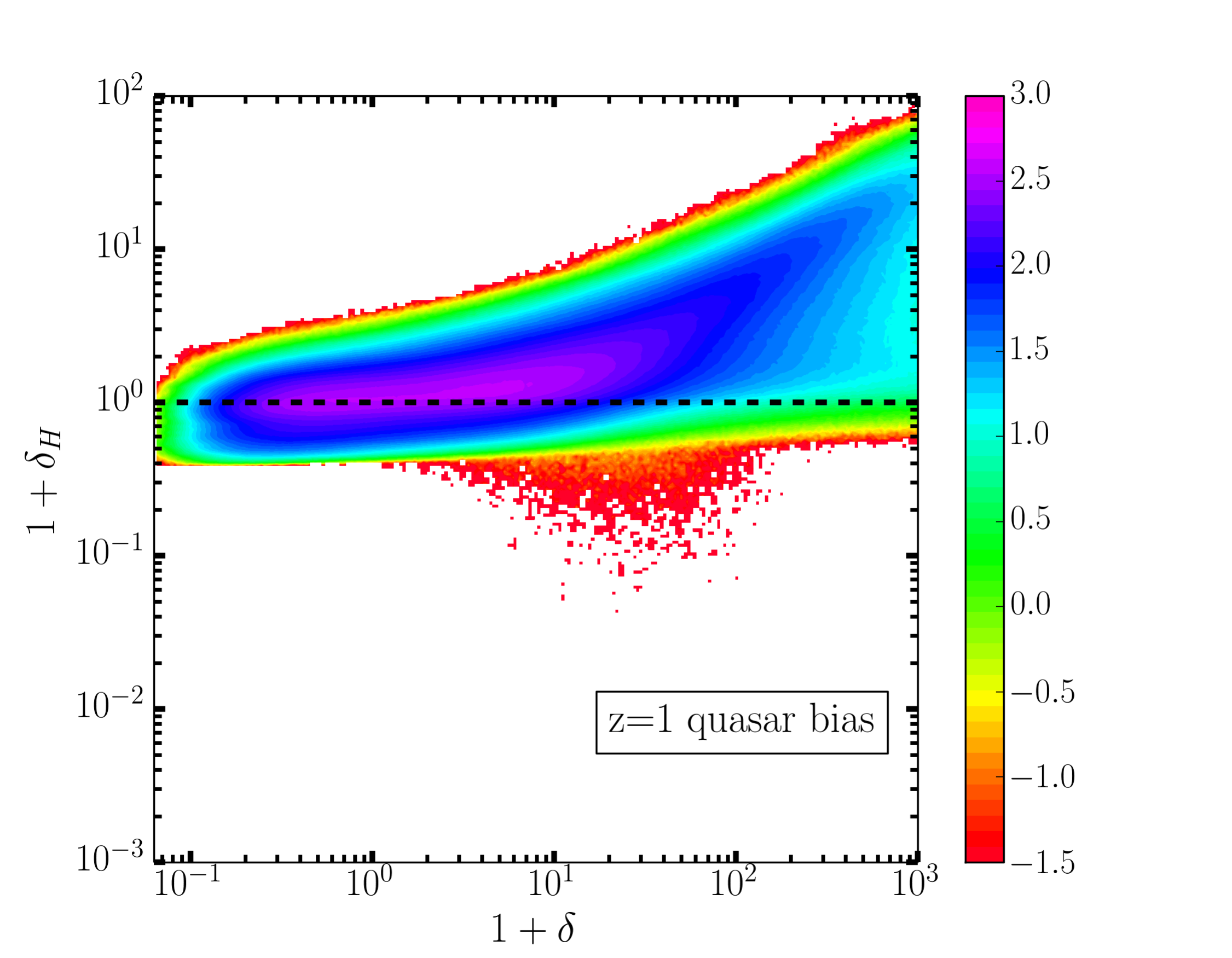}
\caption{Volume-weighted distribution of heating rate fluctuations $(1+\delta_H)$ with respect to the density fluctuations $(1+\delta_{\mathrm{gas}})$ for the galaxy bias model (upper row) and the quasar bias model (lower row) at $z=3$ and $z=1$. The black dashed line represents the case of uniform blazar heating. The color scale is logarithmic.}
\label{fig:deltas}
\end{figure*}

To have a better understanding of the heating fluctuations with respect to density fluctuations we show the mass-weighted $\delta_H-\delta_{\mathrm{gas}}$ distribution on Figure \ref{fig:deltas}. The heating rate represents an instantaneous view of the impact of TeV blazar heating as opposed to temperature or internal energy which probe the integrated injection history of non-gravitational heat. As in Figure \ref{fig:slice}, the quasar bias model stands out at high redshift (lower left panel).  In this model, most of the particles receive slightly more heat than in the uniform case. In contrast, the additional heat is only a few times more than the uniform case, while certain regions receive orders of magnitude less heat than the mean value. These areas suffer from the decrease of the TeV flux and isolation from massive structures. This is consistent with the temperature probability distribution function presenting a prominent low temperature tail and a low probability for high temperatures for $z\geqslant 2$. In the galaxy model, most of the gas is heated similarly to the uniform case, translating the lower bias for galaxies.
\section{Discussion}
Previous work on TeV blazar heating showed a significant increase in the temperature of the low density IGM \citep{2012ApJ...752...23C,2012MNRAS.423..149P}. These results, while in good agreement with observational data for the mean transmitted flux statistics as well as fits to individual lines, appear to be in potential tension with recent work by \citet{2012ApJ...757L..30R}. This observation suggest a significant presence of gas in the IGM has not been exposed to additional heating beyond photoheating. In the context of blazar heating, this suggests that the heating is not entirely uniform. In this work, we have studied the impact of inhomogeneous heating and find that when accounting for the clustering of sources, the variability in the heating rate is significant. In particular, we find that while most of the gas receives close to average heating, and has a temperature close to the uniform heating case, there is an important scatter in the temperature of the gas  for $z\gtrsim 2$. 

The lower envelope of the $T-\rho$ distribution we find is within the error bars of the powerlaw fit to the $T-\rho$ distribution derived in \citet{2014MNRAS.438.2499B}. Detailed comparison of the Lyman $\alpha$ forest properties will be the subject of a forthcoming paper. If confirmed with the analysis of the resulting Lyman $\alpha$ forest, our model would be able to account for both the observed increase in temperature in the low density IGM \citep{2014MNRAS.441.1916B,2011MNRAS.410.1096B,2009MNRAS.399L..39V} and the lower values of the line widths as suggested by \citet{2012ApJ...757L..30R}. Taken together, both observational diagnostics appear to point to a more complicated temperature-density relation with a possibly broad, multi-valued distribution as opposed to a simple power-law relation.

Around $2.7\lesssim z \lesssim 4$, the impact of TeV blazar heating is qualitatively similar to the impact of late reionization of He\,\textsc{II}.  Various physical models and assumptions for the quasar distribution can result in highly variable temperature-density distributions in models of He\,\textsc{II} reionization \citep{2004MNRAS.348L..43B,2009ApJ...694..842M}. However, comparing the temperature-density distributions  obtained from such models with our Figure ~\ref{fig:T_rho},  we notice two major differences. While our model and reionization models present a similar lower envelope of $\simeq 3\times 10^3$ K at $\rho/\bar{\rho}=0.1$ (at $z=3$), our model includes a significant fraction of gas above a few $10^4$ K, which is not present in reionization models. On top of this increased scatter, our model presents a warmer mean temperature than all the He\,\textsc{II} models. The same comparison holds at $z=4$.

Our model also predicts a warmer IGM at $z=2$, which would not be expected from He\,\textsc{II} reionization, as the dependence on the exact He\,\textsc{II} reinozation history will have largely faded at that time \citep{2013MNRAS.435.3169C}. An increased temperature, with a strong scatter in underdense regions could then be attributed to inhomogeneous blazar heating. Direct measurements of the low density IGM are hard to obtain as the Lyman $\alpha$ forest is sensitive to overdensities of at least a few. The Lyman $\alpha$ forest of He\,\textsc{II} traces lower density regions and is a promising observable, with more data becoming available \citep{2014arXiv1405.7405W}.
% Observations of the H Lyman alpha forest suggest an additional heating source, which could be related to He\,\textsc{II} reionization, provided it is sufficiently patchy and late. In this paper, we present a complementary (or alternate)  heating mechanism, which can reproduce the observed scatter and increased heating.  Improved measurements may clarify the respective contributions, although it will be hard, given the uncertainties on the current models. The strongest case for TeV blazar heating will come from low density, low redshift observations.}

Our model neglects important physical aspects of TeV blazars such as their duty cycle and beaming of the gamma-ray emission. The $\gamma$-ray duty cycle of BL Lac objects is of order of 10 $\%$ \citep{1996ApJ...464..600S}. As TeV blazars have low accretion rates, their lifetime is long $\simeq 5-7$ Gyr \citep{2002ApJ...571..226C}. The $\gamma$ ray sources are highly beamed, with observed opening angles peaking around 20$^{\circ}$ \citep{2009A&A...507L..33P}. Parsec scale VLBI observations of the inner parts of radio jets indicate variations of the position angle of around 20$^{\circ}$, up to 120$^{\circ}$ over more than a decade \citep{2013AJ....146..120L}. Assuming the gamma-ray emission follows the orientation of the radio jet, the rapid variability of the sources can potentially increase the scatter in the heating rate. Indeed, in the early universe, when few sources are present,  individual variability increases the inhomogeneity of blazar heating. In such case, our model is a lower limit for the true scatter. However, later on, as the number of sources at a  given point increases, the variability of the heating rate will be reduced.    On top of the intrinsic  variability  of the sources, shot noise, which becomes important when sources are rare, is not taken into account and will result in additional heating rate fluctuations. For instance, at $z \sim 1-2$, the number of sources that contributes half the heating is $\approx 10$ whereas at $z = 3$, it dwindles to $\approx 1$ \citep{2012ApJ...752...23C}.   While this would suggest that the shot noise is potentially large, this is mitigated somewhat by the fact that the next 25\% of the heating is provided by $\approx 100$ sources between $z=1-2$ and $\approx 10$ source at $z=3$. Hence, the fluctuations that we calculate in this work are likely a lower limit.

Our model is based on quasar bias, which is likely lower than TeV-blazar bias \citep{2014ApJ...797...96A}. Radio loud AGN are associated with red giant elliptical galaxies \citep{2007A&A...476..723H} and are often at the center of small clusters and groups. Quasars have a wider distribution of host galaxies and are rarely found at the center of clusters. Therefore, the scatter found in our simulations at low redshift is probably a lower limit to the exact scatter. In our simulations, any dense source emits TeV photons, regardless of the type of galaxy, and history of merger, accretion or star formation. However, based on the comparison of the galaxy and quasar bias models presented here with the uniform heating case we expect a more detailed physical model will produce comparable results. 
\section{Conclusions}
In this paper we have implemented inhomogeneous TeV blazar heating into cosmological simulations to study the impact of clustering on the thermal state of the intergalactic medium. This extends the work by \citet{2012ApJ...752...23C,2012MNRAS.423..149P}, based on uniform blazar heating.

We developed a filtering function relating heating rate fluctuations to the linear dark matter fluctuations. Using this window function, we are able to model the relevant length scales for the blazar heating fluctuations and include them in a statistical fashion. Our method is a cost effective alternative method to fully self consistent simulations of blazar heating where modeling black hole formation and growth and complete radiative transfer would have been prohibitive.

Our model for the blazar heating fluctuations yields a temperature-density relation which can potentially reconcile the observed increased mean temperature of the IGM \citep{2014MNRAS.441.1916B}, while maintaining a fraction of cold gas, responsible for the lower envelope of the linewidth distribution \citep{2012ApJ...757L..30R}. Therefore, detailed modeling of the Lyman $\alpha$ forest will be the subject of a forthcoming paper. If confirmed, this will clearly indicate a more complex thermal history of the IGM, with potentially an important impact on late forming structures.
\begin{acknowledgements}
AL and PC are supported by the UWM Research Growth Initiative, the NASA ATP
program through NASA grant NNX13AH43G, and NSF grant AST-1255469.
A.E.B.~and M.S.~receive financial support from the Perimeter
Institute for Theoretical Physics and the Natural Sciences and
Engineering Research Council of Canada through a Discovery Grant.
Research at Perimeter Institute is supported by the Government of
Canada through Industry Canada and by the Province of Ontario through
the Ministry of Research and Innovation.
C.P.~gratefully acknowledges
financial support of the Klaus Tschira Foundation. E.P. acknowledges support by the ERC grant ``The Emergence of Structure during the epoch of Reionization''.
The authors acknowledge the Texas Advanced Computing Center (TACC) at The University of Texas at Austin and the NASA Advanced Supercomputing Division for providing HPC resources that have contributed to the research results reported within this paper. The authors thank S. Furlanetto, A. Loeb and M. Heahnelt for fruitful discussions. 
\end{acknowledgements}

\appendix
We detail the derivation of the window function of Eq.~\eqref{eq:window}. To present a transparent derivation of the window function (and to clarify some confusion in the literature), we start with a simple toy model to which we add progressively more physics. A familiarized reader may directly switch to Appendix ~\ref{sec:window_complete}. In Appendix~\ref{sec:windon_newt}, we start from a purely Newtonian universe and assume that heating rate fluctuations perfectly trace density fluctuations and a monochromatic source of very high gamma-rays. Then in Appendix~\ref{sec:window_exp}, we include the impact of cosmological expansion and a spectral energy distribution of TeV blazars that evolve with redshift as quasars, albeit with a significantly smaller normalization in the luminosity density. This impacts on the energy-dependent mean free path of TeV photons. Finally, in Appendix~\ref{sec:window_complete} we also account for the clustering of sources (which we model through the linear bias factor) and redshift distortions as a result of peculiar infall velocities onto cluster potentials. \\

\section {Newtonian case}\label{sec:windon_newt}
\subsection {Fluctuations with respect to the mean heating rate}

The TeV flux received (in erg s$^{-1}$ cm$^{-2}$) at position $\mathbf{x}$, at a given energy, is given by the sum over all the sources 
\begin{equation}
  \label{eq:flux_recu0}
  J(\mathbf{x})=
  \int_{0}^{2\pi}d\phi\int_{0}^{\pi}\sin\theta d\theta\int_0^{\infty}|\mathbf{x}'-\mathbf{x}|^2 d|\mathbf{x}'-\mathbf{x}|
  \frac{\mathcal{E}(\mathbf{x}') }{4\pi |\mathbf{x}'-\mathbf{x}|^2} e^{-\tau}
  =\int_{\Omega} d\Omega\int_0^{\infty} dr \frac{\mathcal{E}(\mathbf{r}+\mathbf{x}) }{4\pi } e^{- r/D_{\mathrm{pp}}},
\end{equation}
where the emissivity $\mathcal{E}$ is an energy per unit time, per unit volume, $\tau=|\mathbf{x}'-\mathbf{x}|/D_{\mathrm{pp}}$ is the optical depth along the line of sight and $D_{\mathrm{pp}}$ is the mean free path for pair production, which is constant over the volume. In the last step, we introduced $\mathbf{r}=\mathbf{x}'-\mathbf{x}$, $d\Omega=\sin\theta d\theta d\phi$. We have further assumed that emissivity is monochromatic such that:
\begin{equation}
\mathcal{E}_E(\mathbf{x}) = \mathcal{E}(\mathbf{x}) \delta(E-E_0),
\end{equation}
where $\delta$ is the Dirac delta function and $E_0$ is the energy of the monochromatic photons. Similarly the spectral flux is 
\begin{equation}
J_E(\mathbf{x}) = J(\mathbf{x}) \delta(E-E_0).
\end{equation}
For simplicity, we have integrated equation (\ref{eq:flux_recu0}) over the monochromatic source.

 The differential heating rate is defined as $d\dot{Q}(\mathbf{x})\equiv dJ(\mathbf{x})/D_{\mathrm{pp}}$ such that the volume-integrated heating rate $\dot{Q}$ and its average (in units of erg cm$^{-3}$ s$^{-1}$) are given by 
\begin{equation}
  \label{eq:heating_rate0}
  \dot{Q}(\mathbf{x})=
  \frac{1}{4\pi}   \int_{\Omega}d\Omega\int_0^{\infty} dr \mathcal{E}(\mathbf{r}+\mathbf{x})\, \frac{e^{-r/D_{\mathrm{pp}}}}{D_{\mathrm{pp}}},\quad\mbox{and}\quad
  \bar{\dot{Q}}= \frac{1}{4\pi}   \int_{\Omega}d\Omega\int_0^{\infty} dr \bar{\mathcal{E}} \,\frac{e^{-r/D_{\mathrm{pp}}}}{D_{\mathrm{pp}}} = \bar{\mathcal{E}},
\end{equation}
The heating rate fluctuations at a given point $\mathbf{x}$ are then given by 
\begin{eqnarray}
  \label{eq:heat_fluc_newt0}
  \delta_H(\mathbf{x})&=&\frac{\dot{Q}(\mathbf{x})-\bar{\dot{Q}}}{\bar{\dot{Q}}}=
  \frac{1}{4\pi\bar{\dot{Q}}D_{\mathrm{pp}}} \int_{\Omega}d\Omega\int_0^{\infty} dr 
  [\mathcal{E}(\mathbf{r}+\mathbf{x})-\bar{\mathcal{E}}]e^{-r/D_{\mathrm{pp}}} \nonumber\\
  &=&\frac{1}{4\pi D_{\mathrm{pp}}}\int_{\Omega}d\Omega\int_0^{\infty} dr \delta_E(\mathbf{r}+\mathbf{x})\,e^{-r/D_{\mathrm{pp}}} ,
\end{eqnarray}

where $\delta_E$ are the fluctuations of the TeV photon emissivity.

\subsection{Window function}
As the universe is infinite and asymptotically flat, we can expand the fluctuations into planar waves, in order to get the length scale dependence of heating rate fluctuations,
\begin{eqnarray}
  \label{eq:FT_delta}
  \delta_H(\mathbf{x})&=&\frac{1}{(2\pi)^3}\int_{-\infty}^{\infty} d^3\mathbf{k'} \tilde{\delta}_H(\mathbf{k'}) e^{-i\mathbf{k'}\cdot\mathbf{x}},\\ \nonumber
  \delta_E(\mathbf{r}+\mathbf{x})&=&\frac{1}{(2\pi)^3}\int_{-\infty}^{\infty} d^3\mathbf{k'} \tilde{\delta}_E(\mathbf{k'}) e^{-i\mathbf{k'}\cdot(\mathbf{r}+\mathbf{x})}.
\end{eqnarray}
Performing a convolution of the density fluctuations with the kernel $C r^{-2} e^{-r/D_{\mathrm{pp}}}$ (with $C$ an arbitrary constant), and using the Fourier convolution theorem, Eq.~\eqref{eq:heat_fluc_newt0} rewrites
\begin{equation}
   \label{eq:1}
\tilde{\delta}_H(k) = \tilde{\delta}_E(k) \tilde{W}(k),
 \end{equation}
where the tilde denotes quantities in Fourier space and $\tilde{W}(k)$ is the Fourier transform of $C r^{-2} e^{-r/D_{\mathrm{pp}}}$. This yields

\begin{equation}
  \label{eq:heat_fluc_newt2}
  \tilde{\delta}_H(\mathbf{k})=\frac{1}{4\pi D_{\mathrm{pp}}} \int_{\Omega}d\Omega\int_0^{\infty}  dr  e^{- r/D_{\mathrm{pp}}}  \tilde{\delta}_E(\mathbf{k}) e^{-i\mathbf{k}\cdot{\mathbf{r}}}.
\end{equation}
Introducing $\mu=\cos\theta$, where $\theta$ is the angle between the wave vector and the line of sight, and assuming statistical isotropy and homogeneity, Eq.~\eqref{eq:heat_fluc_newt2} can be simplified as follows,
\begin{eqnarray}
  \label{eq:heat_fluc_newt3}
  \tilde{\delta}_H(k)&=&  
  \frac{1}{2D_{\mathrm{pp}}} \int_{-1}^{1} d\mu \int_0^{\infty}dr   e^{-r/D_{\mathrm{pp}}} \tilde{\delta}_E(k) e^{-ikr\mu}
  =\tilde{\delta}_E(k)\frac{1}{D_{\mathrm{pp}} }\int_0^{\infty} \frac{\sin(kr)}{kr} e^{-r/D_{\mathrm{pp}}}   dr\\ \nonumber
 &=&\frac{\tilde{\delta}_E(k)}{k D_{\mathrm{pp}}} \arctan\left(k D_{\mathrm{pp}}\right)
  = \tilde{\delta}(k)\frac{\arctan\left(k D_{\mathrm{pp}}\right)}{k D_{\mathrm{pp}}} .
\end{eqnarray}
In the last step, we have assumed for simplicity that the fluctuations in the TeV emissivity follow the DM density fluctuations, i.e., $\delta_E=\delta$. While we may justify this by the tight relation between the mass of the central black hole and the stellar mass of the bulge of the host galaxy \citep{2004ApJ...604L..89H}, we generalize the mapping between emissivity and DM fluctuations in Appendix~\ref{sec:window_complete}. Figure~\ref{fig:window_newt} shows  window functions with different  mean free paths in a purely Newtonian universe.  When absorption is present,  the impact of large scale structure (i.e. low wave numbers $k$) remains constant as distant sources are absorbed. For scales equal to or larger than the cutoff, heating fluctuations follow density fluctuations and overdense regions get more heat. At smaller scales, the density structure has less impact on the heating, unless a strong overdensity is present.  

\begin{figure}
\centering
\includegraphics[width = .45\textwidth ]{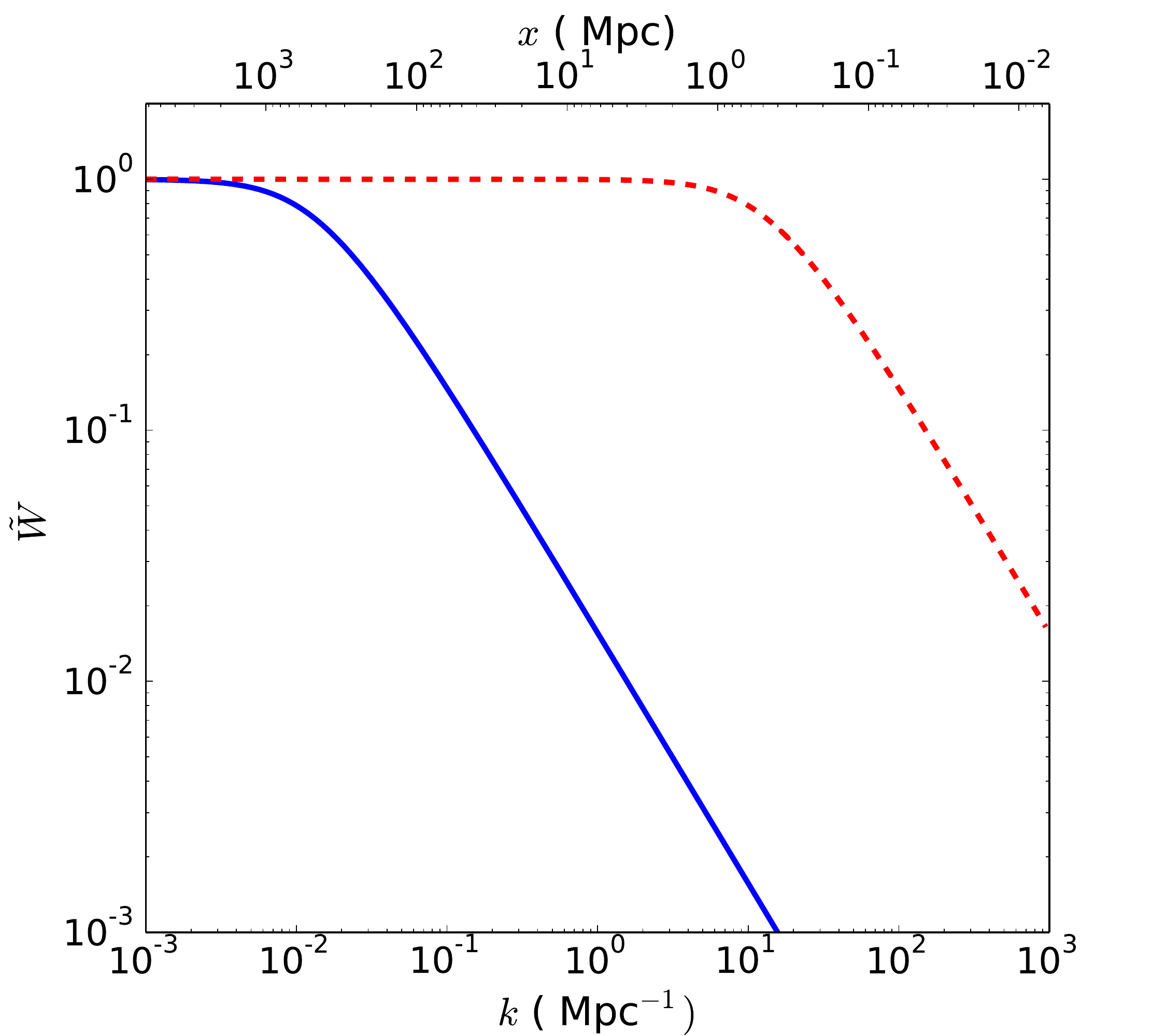}
\caption{Window function for a monochromatic source in a  non-expanding universe (Eq.~\eqref{eq:heat_fluc_newt3}) . The dashed red line
    shows a small mean free path
    ($D_{\mathrm{pp}}=10^{-1}$ Mpc), the solid blue line has
    $D_{\mathrm{pp}}=100 $ Mpc.}
\label{fig:window_newt}
\end{figure}

\section{Expanding universe}\label{sec:window_exp}

The heating rate in an expanding universe at a comoving point $\mathbf{\hat{x}}$ at redshift $z$ is given by
\begin{equation}
  \dot{Q}(\mathbf{\hat{x}})= \int_{E_{\rm min}}^{E_{\rm max}} dE \frac{J_E(\mathbf{\hat{x}})}{D_{\mathrm{pp}}(E, z)},
\end{equation}
where the interval $E_{\mathrm{min}}$ to $E_{\mathrm{max}}$ is the energy range over which pair
production is possible. Quantities indicated by a hat are measured in the
comoving frame.

$J_E$ is the specific intensity per unit energy (in erg s$^{-1}$ cm$^{-2}$ erg$^{-1}$). Assuming a flat geometry on spatial hypersurfaces of the Friedmann-Robertson-Walker metric, $J_E$ is given by
\begin{equation}\label{eq:expanding J}
  J_E(E, \mathbf{\hat{x}}) = \frac {(1 + z)^3} {4\pi} \int_{\Omega} d\Omega \int_z^{\infty} \frac{d z'}{1+z'}\,\frac{d\hat{r}}{dz'}\, {\hat{\mathcal{E}}_{E'}(\mathbf{\hat{r}}+\mathbf{\hat{x}}, z')}e^{-\tau(E,z,z')},
\end{equation}
where the vector $\mathbf{\hat{r}}(\hat{r}(z,z'), \theta, \phi)$, $z$ is the absorption redshift, $\hat{\mathcal{E}}_E = \epsilon_E/(1+z')^3$ is the comoving specific emissivity defined in Eq.~\eqref{eq:spec_lum_dens}, $E' = E (1+z')/(1+z)$ and $\tau$ is the approximate optical depth given by
\begin{equation}
\label{eq:tau}
\tau(E,z,z')\equiv \tau=\int_z^{z'}dz''\frac{1}{\hat D_{\mathrm{pp}}(E,z'')}\frac{d\hat r}{dz''}.
\end{equation}
An alert reader will recognize that the true optical depth must include the effect of a changing $E'' = E(1+z'')/(1+z)$ as we integrate along $z''$ from $z$ to $z'$.  However, here we assume the typical mean free paths for high energy photons are much smaller than the Hubble length so that $E'' \approx E$ over the redshift interval, which  is small compared to $z$.  
We then find that the energy-integrated and volume-integrated heating rate is then given by 
\begin{equation}
  \label{eq:int_exp_heat}
  \dot{Q}[\mathbf{\hat{x}}(z)]=\frac{(1+z)^3}{4\pi }\int_{E_{\mathrm{min}}}^{E_{\mathrm{max}}} \frac{dE}{D_{\mathrm{pp}}(E,z)}\int_{\Omega}d\Omega\int_z^{\infty} \frac{d z'}{1+z'}\,\frac{d\hat r}{dz'}
\mathcal{\hat E}_{E'}(\calR, z') e^{-\tau(E,z,z')},
\end{equation}
where $\calR = \mathbf{\hat{x}} + \mathbf{\hat{r}}$. Similarly to Eq.~\eqref{eq:heating_rate0}, the mean heating rate is given by
\begin{equation}
\label{eq:mean_exp_heat}
\bar{\dot{Q}} (z)=\frac{(1+z)^3}{4\pi}\int_{E_{\mathrm{min}}}^{E_{\mathrm{max}}} \frac{dE}{D_{\mathrm{pp}}(E, z)}\int_{\Omega}d\Omega\int_z^{\infty} \frac{d z'}{1+z'}\, \frac{d\hat r}{dz'}\bar{\hat{\mathcal{E}}}_{E'}(z') e^{-\tau(E,z,z')}.
\end{equation}

This enables us to define the heating rate fluctuations
\begin{eqnarray}
\label{eq:fluc_exp0}
\delta_H[\mathbf{\hat{x}}(z)]&=&\frac{\dot{Q}[\mathbf{\hat{x}}(z)]-\bar{\dot{Q}}(z)}{\bar{\dot{Q}}(z)}=\frac{(1+z)^3c}{4\pi\bar{\dot{Q}}} \int_{E_{\mathrm{min}}}^{E_{\mathrm{max}}}\frac{dE}{D_{\mathrm{pp}}(E,z)}\int_{\Omega}d\Omega\int_z^{\infty} d z'\,\frac{ \left[ \mathcal{\hat E}_{E'}(\calR, z')-\bar{\hat{\mathcal{E}}}_{E'}(z')\right] e^{-\tau(E,z,z')}}{(1+z')\,H(z')} \nonumber\\ 
&=&\frac{(1+z)^3 c}{4\pi\bar{\dot{Q}}} \int_{E_{\mathrm{min}}}^{E_{\mathrm{max}}} \frac{dE}{D_{\mathrm{pp}}(E,z)}\int_{\Omega}d\Omega\int_z^{\infty} d z'\, \frac{\bar{\hat{\mathcal{E}}}_{E'}(z')\delta_E(\calR, z') e^{-\tau(E,z,z')}}{(1+z')\,H(z')},
\end{eqnarray}
The TeV emission is related to the presence of supermassive black holes at the center of galaxies, which are located in collapsed dark matter halos. We can thus connect the fluctuations of the TeV emission, within a certain radius $\hat r$, to the underlying dark matter fluctuations $\delta$.
At this stage we assume that TeV fluctuations exactly match the dark matter fluctuations $\delta_E=\delta$. We explain in the next section that this is not exactly true and will take into account various corrections. The initial density fluctuations represent a Gaussian random field, whose exact properties depend on the earliest stages of the universe prior to recombination \citep{1986ApJ...304...15B,Peebles}. They grow linearly between $z'$ and $z$ following  \citep{ 1977MNRAS.179..351H}
\begin{equation}
  \label{eq:growth}
\delta(\calR, z)=\delta(\calR, z')D(z)/D(z'),
\end{equation}
where the linear growth factor is given by
\begin{equation}
\label{eq:growth_1}
D(z)=D_0H(z)\int_z^{\infty}dz'\frac{1+z'}{H^3(z')}.
\end{equation}
The linear approximation breaks down when the amplitude of the root mean square of the perturbations approaches unity. The evolution of the density field is then determined by the spherical collapse \citep{1972ApJ...176....1G} and the virialization of halos. In the linear regime, the growth of the modes is independent of the wavenumber and we have
\begin{equation}
\label{eq:FT_delta}
\delta_E(\mathbf{\hat{r}}, z')=\delta(\mathbf{\hat{r}}, z')
=\delta(\mathbf{\hat{r}}, z)\frac{D(z')}{D(z)}
=\frac{D(z')}{D(z)}\frac{1}{(2\pi)^3}\int d^3\mathbf{\hat k'} \tilde{\delta}(\mathbf{\hat k'}) e^{-i\mathbf{\hat k'}\cdot\mathbf{\hat r}}.
\end{equation}
And Eq.~\eqref{eq:fluc_exp0} rewrites to
\begin{equation}
\label{eq:heat_fluc_exp0}
\delta_H[\mathbf{\hat{x}}(z)]=\frac{(1+z)^3  c}{4\pi\bar{\dot{Q}}}\int_{E_{\mathrm{min}}}^{E_{\mathrm{max}}} \frac{dE}{D_{\mathrm{pp}}(E,z)} \int_{\Omega}d\Omega\int_z^{\infty}dz' \frac{D(z')}{D(z)} \frac{\bar{\hat{\mathcal{E}}}_{E'}(z')\delta(\calR, z) e^{-\tau(E,z,z')}}{(1+z')\,H(z')}.
\end{equation}
Fourier transforming yields $\tilde{\delta}(\mathbf{\hat k})$ on the left hand side while the right hand side transforms in a similar fashion to Eq.~\eqref{eq:heat_fluc_newt3}. As the power spectrum of density fluctuations is statistically isotropic and homogeneous, this simplifies to
\begin{equation}
\label{eq:heat_fluc_exp1}
\tilde{\delta}_H(\hat k, z)=\tilde{\delta}(\hat k, z) \frac{(1+z)^3c}{\bar{\dot{Q}}} \int_{E_{\mathrm{min}}}^{E_{\mathrm{max}}} \frac{dE}{D_{\mathrm{pp}}(E,z)}\int_z^{\infty} dz' \frac{D(z')}{D(z)}\frac{\sin\left(\hat k\hat r(z,z')\right)}{\hat k \hat r(z, z')} \frac{\bar{\hat{\mathcal{E}}}_{E'}(z') e^{-\tau(E,z,z')}} {(1+z')\,H(z')}.
\end{equation}
\begin{figure}[h]
\centering
\includegraphics[width = .45\textwidth ]{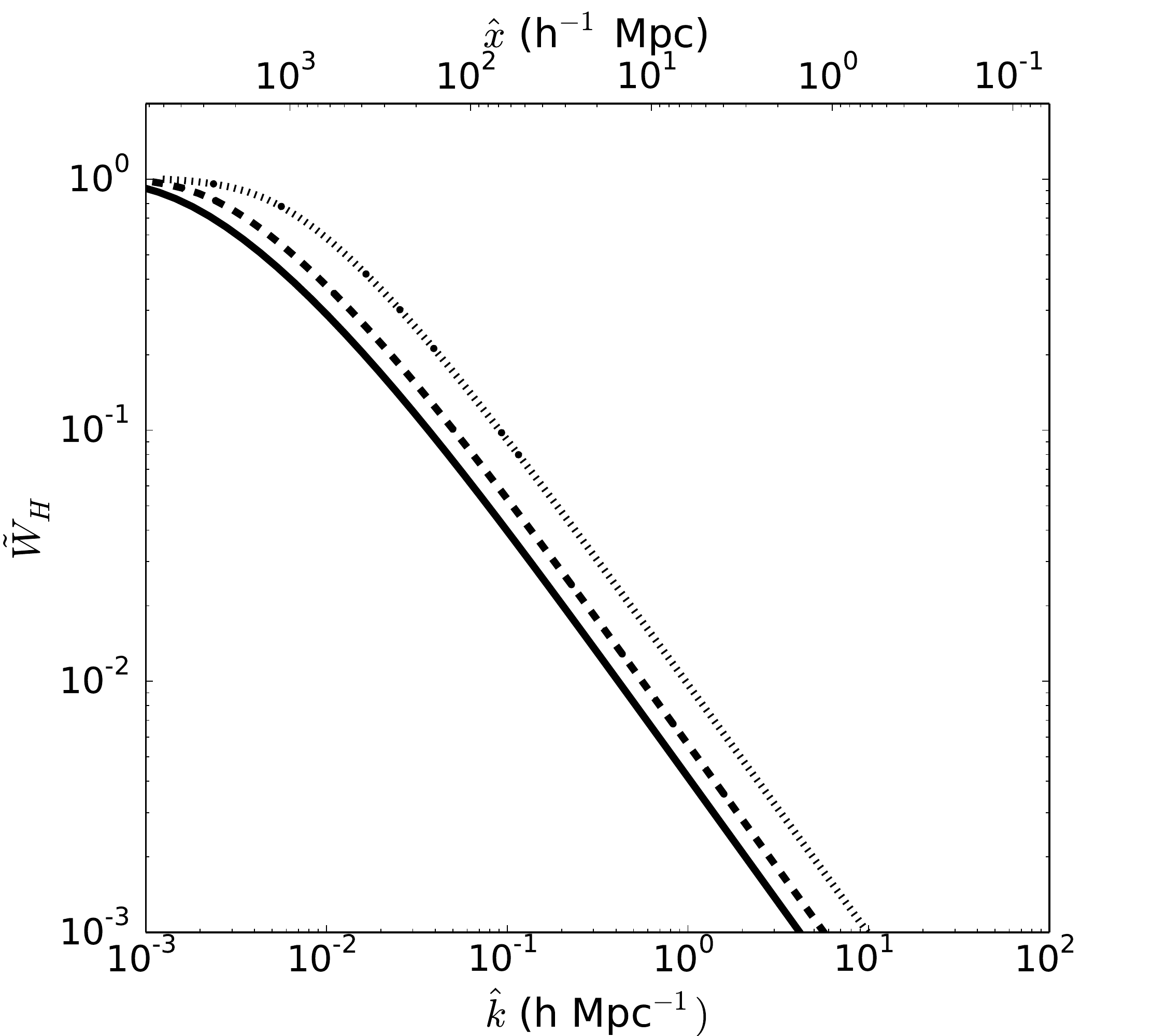}
\caption{Window function for blazar heating in an expanding universe without account for clustering bias (Eq.~\eqref{eq:heat_fluc_exp1}) for $z=1$ (solid line), 2 (dashed line) and 4 (dotted line).}
\label{fig:window_nobiases}
\end{figure}
Figure~\ref{fig:window_nobiases} shows the window function in an expanding universe at different redshifts.  The position of the breaks results from a combination between the value of the mean free path and the redshift evolution of the blazar luminosity density. 
As discussed in the main text, the rapidly decreasing blazar luminosity density with increasing redshift for $z\gtrsim 2$, which moves the break toward larger wave vectors, dominates over the increasing comoving mean free path, which moves the break toward smaller wave vectors.  This is due to the fact that the integration over the energy of the photons, which includes lower energy photons with larger mean free paths, allows the former effect to dominate and moves the break toward smaller wave vectors with increasing redshift.
The TeV emission fluctuations are not exactly equal to the dark matter density fluctuations. In the next section we will account for the various corrections that have to be taken into account to determine a more accurate window function.

\section{Complete window function}
\label{sec:window_complete}
The TeV fluctuations are biased with respect to the dark matter fluctuations, as we detail in $\S$\ref{sec:window}, yielding
\begin{equation}
\delta_E(\mathbf{\hat{x}},z)=b(z)\delta(\mathbf{\hat{x}},z),
\label{eq:bias}
\end{equation}
with $b$ the Eulerian bias. If galaxies were moving exactly with the Hubble flow, their redshift would yield their exact distance to an observer. However, structures falling towards a central potential of a cluster have an infall velocity  towards the central overdensity, yielding different overdensities in redshift space $\delta_E^{(\hat{s})}$ and real space. In our case, the blazar luminosity density evolution, the mean free path, and the bias as a function of redshift are all subject to this redshift space distortions as measured by the $z=0$ observer. We follow the derivation of the redshift space distortions of density fluctuations of \citet{MFW}. The key difference that we encounter is that redshift space distortions impact on $\calR=\bm\hat{\mathbf{x}}+\bm\hat{\mathbf{r}}$, whereas our computation involves the displacements $\bm\hat{\mathbf{r}}$ between the heated point and a distant source. To this end, we define a comoving redshift space distance
\begin{equation}
\mathbf{\hat{s}} \equiv\frac{\dot{\calR}}{a H(z)} = 
\calR + \frac{\hat{v}_{\mathcal{R}}}{H(z)} \mathbf{e}_\mathcal{R},
\label{eq:s_redshift}
\end{equation}
which is given in units of length, $a=1/(1+z)$ is the cosmological scale factor, $\hat{v}_\mathcal{R}$ is the radial component of the comoving peculiar velocity ($\mathbf{\hat{v}} = \mathbf{v}/a$ with $\mathbf{v}$ denoting the physical peculiar velocity), and $\mathbf{e}_\mathcal{R}$ is the unit vector along the line of sight. The conservation of TeV blazar number then implies
\begin{equation}
\hat n_E^{(\hat s)}(\mathbf{\hat s}) d^3\mathbf{\hat s} = \hat n_E(\calR) d^3\calR,
\qquad\mbox{or}\qquad
\left[1+\delta_E^{(\hat s)}(\mathbf{\hat s})\right] d^3\mathbf{\hat s} = 
\left[1+\delta_E(\calR)\right] d^3\calR,
\label{eq:conservation}
\end{equation}
where we employed the fact that the mean number density is the same in both spaces. Assuming statistical isotropy, we find
\begin{equation}
1+\delta_E^{(\hat s)}(\mathbf{\hat s}) = 
\left[1+\delta_E(\calR)\right] \frac{\magcalR^2}{\left(\magcalR +\hat{v}_{\mathcal{R}}/H(z)\right)^2}
\left(1+\frac{1}{H(z)}\frac{\partial \hat{v}_{\mathcal{R}}}{\partial\magcalR}\right)^{-1},
\label{eq:conservation2}
\end{equation}
Assuming $\left|v_{\mathcal{R}}\right|\ll \magcalR H(z)$ and keeping only terms up to lowest non-trivial order, we obtain
\begin{equation}
  \label{eq:real_vs_z}
  \delta_E^{(\hat s)}\left[\mathbf{\hat s}(\calR) \right]=
  \delta_E(\calR)-\frac{1}{H(z)}\left(\frac{\partial}{\partial \magcalR} + 
    \frac{2}{\magcalR}\right)\hat{v}_{\mathcal{R}}\approx
  \delta_E(\calR)-\frac{1}{H(z)}\frac{\partial \hat{v}_{\mathcal{R}}}{\partial \magcalR},
\end{equation}
where we assumed in the second step that the scale of the fluctuations is small compared to the distance to the sources (i.e., we used the plane-parallel approximation according to \citealt{1987MNRAS.227....1K}). In the linear regime ($\delta_E(\calR)\ll 1$), the peculiar velocity field is related to the density field via the Zel'dovich approximation \citep{MFW},
\begin{equation}
  \label{eq:Zeldovich}
  \delta_E(\calR) = -\frac{b}{f H(z)}\frac{\partial \bm\hat{\mathbf{v}}}{\partial \calR},
\end{equation}
where $f\equiv d\log\delta/d\log a$. It can be shown in the linear regime that non-trivial solutions of the velocity field must be irrotational so that $\mathbf{v} = \boldsymbol{\nabla}_{\calR} \Phi_v$. Using this property, we can rewrite Eq.~\eqref{eq:Zeldovich} to yield
\begin{equation}
  \label{eq:Zeldovich2}
  \hat{v}_{\mathcal{R}} = - \frac{f H(z)}{b} \frac{\partial}{\partial \magcalR} \boldsymbol{\nabla}_{\calR}^{-2}\delta_E(\calR),
\end{equation}
where $\boldsymbol{\nabla}_{\calR}^{-2}$ represents the inverse Laplacian. It follows that
\begin{equation}
  \label{eq:Zeldovich3}
  \delta_E^{(\hat s)}[\calR(\bm\hat{\mathbf{r}})] = 
  \left[1+\frac{f}{b}\frac{\partial^2}{\partial \magcalR^2}
    \boldsymbol{\nabla}_{\calR}^{-2}
\right]\delta_E[\calR(\bm\hat{\mathbf{r}})] =
  \left[1+\frac{f}{b}\frac{\partial^2}{\partial \hat{r}^2}
    \boldsymbol{\nabla}_{\bm\hat{\mathbf{r}}}^{-2}
\right]\delta_E[\calR(\bm\hat{\mathbf{r}})].
\end{equation}
This demonstrates that redshift space distortions do not depend on the line-of-sight distance from the observer to the source but {\em only} on the shape of the local gravitational potential of the source! This property enables us to treat redshift space distortions locally by performing a Fourier decomposition of the density field into plane waves, which yields
\begin{equation}
  \label{eq:real_vs_z2}
  \tilde{\delta}_E^{(\hat s)}\left[\mathbf{\hat{k}}^{(\hat s)}(\mathbf{\hat{k}})\right]=
  \left(1+\frac{f}{b}\mu^2\right) \tilde{\delta}_E(\mathbf{\hat{k}})=
  \left(b+f\mu^2\right) \tilde{\delta}(\mathbf{\hat{k}}),
\end{equation}
where $\mu\equiv k_z/k$. Keeping only first order corrections for density fluctuations, Eq.~\eqref{eq:int_exp_heat} yields
\begin{equation}
\label{eq:mean_heat0}
\dot{Q}(\mathbf{\hat{x}},z)=
\frac{(1+z)^3c}{4\pi }\int_{E_{\mathrm{min}}}^{E_{\mathrm{max}}}\frac{dE}{D_{\mathrm{pp}}(E,z)}
\int_{\Omega}d\Omega\int_z^{\infty} dz'\frac{\mathcal{\bar{\hat E}}_{E'}(z')e^{-\tau(E,z,z')}}{(1+z')\,H(z')}
\left[1+\delta_E^{(\hat s)}(\calR,z')\right].
\end{equation}
Subtracting the mean heating rate (Eq.~\eqref{eq:mean_exp_heat}) then yields the fluctuations
\begin{equation}
\label{eq:heat_fluc0}
\delta_H(\mathbf{\hat{x}}, z)=\frac{1}{4\pi\bar{\dot{Q}}}\int_{E_{\mathrm{min}}}^{E_{\mathrm{max}}} dE
\int_{\Omega}d\Omega\int_z^{\infty}dz'X(E,z,z')\delta_E^{(\hat s)}(\calR,z'),
\end{equation}
where we introduced
\begin{equation}
\label{eq:def_X}
X(E,z,z')=\frac{(1+z)^3c}{D_{\mathrm{pp}}(E,z)}\frac{\mathcal{\bar{\hat E}}_{E'}(z')e^{-\tau(E,z,z')}}{(1+z')\,H(z')}
\end{equation}
for convenience. Transforming to Fourier space yields
\begin{eqnarray}
\label{eq:heat_fluc0}
\tilde{\delta}_H(\mathbf{\hat k}, z)&=&\frac{1}{4\pi\bar{\dot{Q}}}\int_{E_{\mathrm{min}}}^{E_{\mathrm{max}}} dE\int_{\Omega}
\!d\Omega\int_z^{\infty}\! dz'
\frac{X(E,z,z')}{(2\pi)^3}\!\int d^3\mathbf{\hat x}e^{i\mathbf{\hat k \cdot \hat x}}
\int\! d^3 \mathbf{\hat k'}e^{-i\mathbf {\hat k'} \cdot (\mathbf{\hat r} +\mathbf{ \hat x})} 
\tilde\delta(\mathbf {\hat k'}, z)\frac{D(z')}{D(z)}\left[b(z') + f \mu^2\right].
\end{eqnarray}
Assuming statistical isotropy and homogeneity allows us to simplify this expression, and we arrive at
\begin{eqnarray}
\tilde{\delta}_H(z,\hat k)&=&\frac{\tilde{\delta}(z,\hat k)}{2\bar{\dot{Q}}}\int_{E_{\mathrm{min}}}^{E_{\mathrm{max}}} dE \int_{\Omega}d\Omega\int_z^{\infty} dz'X(E,z,z')\frac{D(z')}{D(z)}\int_{-1}^{1}d\mu\left[e^{-i\hat k \hat r\mu} (b +f \mu^2)\right]\\ \nonumber
&=&\frac{\tilde{\delta}(z,\hat k)}{\bar{\dot{Q}}}\int_{E_{\mathrm{min}}}^{E_{\mathrm{max}}} dE\int_z^{\infty}dz'X(E,z,z')\frac{D(z')}{D(z)}\left[\left(b(z')+\frac{f}{3}\right)j_0(\hat k \hat r)-\frac{2}{3}f j_2(\hat k \hat r)\right],
\end{eqnarray}
where we used the spherical Bessel functions
\begin{eqnarray}
\label{eq:bessel}
j_0(x)&=& \frac{\sin(x)}{x},\\
j_2(x)&=& \left(\frac{3}{x^2}-1\right)\frac{\sin(x)}{x}-\frac{3 \cos(x)}{x^2} .
\end{eqnarray}
We have further used
\begin{equation}
\label{eq:bes2}
\int_{-1}^{1}\mu^2 e^{-i k r \mu} d\mu=\frac{2 \sin(kr)}{kr}+4\frac{\cos(kr)}{(kr)^2}-4\frac{\sin(kr)}{(kr)^3}.
\end{equation}
The window function for the heating rate fluctuations is then given by
\begin{eqnarray}
\label{eq:heat_fluc}
\tilde{W}_H(z,\hat k)&=&\frac{(1+z)^3c }{\bar{\dot{Q}}}\int_{E_{\mathrm{min}}}^{E_{\mathrm{max}}} \frac{dE}{D_{\mathrm{pp}}(E,z)}\int_z^{\infty} dz' \frac{\mathcal{\bar{\hat E}}_{E'}(z')e^{-\tau(E,z,z')}}{(1+z')\,H(z')}\frac{D(z')}{D(z)}\left[\left(b(z')+\frac{f}{3}\right)j_0(\hat k \hat r)-\frac{2}{3}f j_2(\hat k \hat r)\right]\\
&=& \frac{\displaystyle \int_{E_{\mathrm{min}}}^{E_{\mathrm{max}}} \frac{dE}{{D_{\mathrm{pp}}}(E,z)}\int_z^{\infty} dz'   \frac{\mathcal{\bar{\hat E}}_{E'}(z')e^{-\tau(E,z,z')}}{(1+z')\,H(z')}\frac{D(z')}{D(z)}\left[\left(b(z')+\frac{f}{3}\right)j_0(\hat k \hat r)-\frac{2f}{3}j_2(\hat k \hat r)\right]}{\displaystyle \int_{E_{\mathrm{min}}}^{E_{\mathrm{max}}} \frac{dE}{{D_{\mathrm{pp}}}(E,z)}\int_z^{\infty} dz'  \mathcal{\bar{\hat E}}_{E'}(z') \frac{e^{-\tau(E,z,z')}}{(1+z')\,H(z')}}.
\end{eqnarray}
This equation is similar to \citet{2007MNRAS.376.1680P} and \citet{2005ApJ...626....1B} but does not contain the unnecessary area correction factor, which would be present only for a Lagrangian description of bias.
\bibliographystyle{apj}
\bibliography{biblio_total}

\begin{thebibliography}{}
\expandafter\ifx\csname natexlab\endcsname\relax\def\natexlab#1{#1}\fi

\bibitem[{{Abazajian} {et~al.}(2011){Abazajian}, {Blanchet}, \&
  {Harding}}]{Abazajian:2011}
{Abazajian}, K.~N., {Blanchet}, S., \& {Harding}, J.~P. 2011, \prd, 84, 103007

\bibitem[{{Aleksi{\'c}} {et~al.}(2010){Aleksi{\'c}}, {Antonelli}, {Antoranz},
  {Backes}, {Baixeras}, {Barrio}, {Bastieri}, {Becerra Gonz{\'a}lez},
  {Bednarek}, {Berdyugin}, {Berger}, {Bernardini}, {Biland}, {Blanch}, {Bock},
  {Bonnoli}, {Bordas}, {Borla Tridon}, {Bosch-Ramon}, {Bose}, {Braun}, {Bretz},
  {Britzger}, {Camara}, {Carmona}, {Carosi}, {Colin}, {Commichau}, {Contreras},
  {Cortina}, {Costado}, {Covino}, {Dazzi}, {de Angelis}, {de Cea Del Pozo}, {de
  Los Reyes}, {de Lotto}, {de Maria}, {de Sabata}, {Delgado Mendez}, {Doert},
  {Dom{\'{\i}}nguez}, {Dominis Prester}, {Dorner}, {Doro}, {Elsaesser},
  {Errando}, {Ferenc}, {Fonseca}, {Font}, {Garc{\'{\i}}a L{\'o}pez},
  {Garczarczyk}, {Gaug}, {Godinovic}, {Hadasch}, {Herrero}, {Hildebrand},
  {H{\"o}hne-M{\"o}nch}, {Hose}, {Hrupec}, {Hsu}, {Jogler}, {Klepser},
  {Kr{\"a}henb{\"u}hl}, {Kranich}, {La Barbera}, {Laille}, {Leonardo},
  {Lindfors}, {Lombardi}, {Longo}, {L{\'o}pez}, {Lorenz}, {Majumdar}, {Maneva},
  {Mankuzhiyil}, {Mannheim}, {Maraschi}, {Mariotti}, {Mart{\'{\i}}nez},
  {Mazin}, {Meucci}, {Miranda}, {Mirzoyan}, {Miyamoto}, {Mold{\'o}n}, {Moles},
  {Moralejo}, {Nieto}, {Nilsson}, {Ninkovic}, {Orito}, {Oya}, {Paiano},
  {Paoletti}, {Paredes}, {Partini}, {Pasanen}, {Pascoli}, {Pauss}, {Pegna},
  {Perez-Torres}, {Persic}, {Peruzzo}, {Prada}, {Prandini}, {Puchades},
  {Puljak}, {Reichardt}, {Rhode}, {Rib{\'o}}, {Rico}, {Rissi}, {R{\"u}gamer},
  {Saggion}, {Saito}, {Salvati}, {S{\'a}nchez-Conde}, {Satalecka}, {Scalzotto},
  {Scapin}, {Schultz}, {Schweizer}, {Shayduk}, {Shore}, {Sierpowska-Bartosik},
  {Sillanp{\"a}{\"a}}, {Sitarek}, {Sobczynska}, {Spanier}, {Spiro}, {Stamerra},
  {Steinke}, {Struebig}, {Suric}, {Takalo}, {Tavecchio}, {Temnikov}, {Terzic},
  {Tescaro}, {Teshima}, {Torres}, {Vankov}, {Wagner}, {Weitzel}, {Zabalza},
  {Zandanel}, {Zanin}, {Neronov}, \& {Semikoz}}]{2010A&A...524A..77A}
{Aleksi{\'c}}, J., {Antonelli}, L.~A., {Antoranz}, P., {et~al.} 2010, \aap,
  524, A77

\bibitem[{{Allevato} {et~al.}(2014){Allevato}, {Finoguenov}, \&
  {Cappelluti}}]{2014ApJ...797...96A}
{Allevato}, V., {Finoguenov}, A., \& {Cappelluti}, N. 2014, \apj, 797, 96

\bibitem[{{Bardeen} {et~al.}(1986){Bardeen}, {Bond}, {Kaiser}, \&
  {Szalay}}]{1986ApJ...304...15B}
{Bardeen}, J.~M., {Bond}, J.~R., {Kaiser}, N., \& {Szalay}, A.~S. 1986, \apj,
  304, 15

\bibitem[{{Barkana} \& {Loeb}(2005)}]{2005ApJ...626....1B}
{Barkana}, R., \& {Loeb}, A. 2005, \apj, 626, 1

\bibitem[{{Basilakos} {et~al.}(2008){Basilakos}, {Plionis}, \&
  {Ragone-Figueroa}}]{2008ApJ...678..627B}
{Basilakos}, S., {Plionis}, M., \& {Ragone-Figueroa}, C. 2008, \apj, 678, 627

\bibitem[{{Becker} {et~al.}(2011){Becker}, {Bolton}, {Haehnelt}, \&
  {Sargent}}]{2011MNRAS.410.1096B}
{Becker}, G.~D., {Bolton}, J.~S., {Haehnelt}, M.~G., \& {Sargent}, W.~L.~W.
  2011, \mnras, 410, 1096

\bibitem[{{Boera} {et~al.}(2014){Boera}, {Murphy}, {Becker}, \&
  {Bolton}}]{2014MNRAS.441.1916B}
{Boera}, E., {Murphy}, M.~T., {Becker}, G.~D., \& {Bolton}, J.~S. 2014, \mnras,
  441, 1916

\bibitem[{{Bolton} {et~al.}(2004){Bolton}, {Meiksin}, \&
  {White}}]{2004MNRAS.348L..43B}
{Bolton}, J., {Meiksin}, A., \& {White}, M. 2004, \mnras, 348, L43

\bibitem[{{Bolton} \& {Becker}(2009)}]{2009MNRAS.398L..26B}
{Bolton}, J.~S., \& {Becker}, G.~D. 2009, \mnras, 398, L26

\bibitem[{{Bolton} {et~al.}(2014){Bolton}, {Becker}, {Haehnelt}, \&
  {Viel}}]{2014MNRAS.438.2499B}
{Bolton}, J.~S., {Becker}, G.~D., {Haehnelt}, M.~G., \& {Viel}, M. 2014,
  \mnras, 438, 2499

\bibitem[{{Bolton} {et~al.}(2008){Bolton}, {Viel}, {Kim}, {Haehnelt}, \&
  {Carswell}}]{2008MNRAS.386.1131B}
{Bolton}, J.~S., {Viel}, M., {Kim}, T.-S., {Haehnelt}, M.~G., \& {Carswell},
  R.~F. 2008, \mnras, 386, 1131

\bibitem[{{Bond} {et~al.}(1996){Bond}, {Kofman}, \&
  {Pogosyan}}]{1996Natur.380..603B}
{Bond}, J.~R., {Kofman}, L., \& {Pogosyan}, D. 1996, \nat, 380, 603

\bibitem[{Bret {et~al.}(2004)Bret, Firpo, \& Deutsch}]{PhysRevE.70.046401}
Bret, A., Firpo, M.-C., \& Deutsch, C. 2004, Phys. Rev. E, 70, 046401

\bibitem[{{Broderick} {et~al.}(2012){Broderick}, {Chang}, \&
  {Pfrommer}}]{2012ApJ...752...22B}
{Broderick}, A.~E., {Chang}, P., \& {Pfrommer}, C. 2012, \apj, 752, 22

\bibitem[{{Broderick} {et~al.}(2014{\natexlab{a}}){Broderick}, {Pfrommer},
  {Puchwein}, \& {Chang}}]{2014ApJ...790..137B}
{Broderick}, A.~E., {Pfrommer}, C., {Puchwein}, E., \& {Chang}, P.
  2014{\natexlab{a}}, \apj, 790, 137

\bibitem[{{Broderick} {et~al.}(2014{\natexlab{b}}){Broderick}, {Pfrommer},
  {Puchwein}, {Chang}, \& {Smith}}]{2014ApJ...796...12B}
{Broderick}, A.~E., {Pfrommer}, C., {Puchwein}, E., {Chang}, P., \& {Smith},
  K.~M. 2014{\natexlab{b}}, \apj, 796, 12

\bibitem[{{Cavaliere} \& {D'Elia}(2002)}]{2002ApJ...571..226C}
{Cavaliere}, A., \& {D'Elia}, V. 2002, \apj, 571, 226

\bibitem[{{Chang} {et~al.}(2012){Chang}, {Broderick}, \&
  {Pfrommer}}]{2012ApJ...752...23C}
{Chang}, P., {Broderick}, A.~E., \& {Pfrommer}, C. 2012, \apj, 752, 23

\bibitem[{{Chang} {et~al.}(2014){Chang}, {Broderick}, {Pfrommer}, {Puchwein},
  {Lamberts}, \& {Shalaby}}]{2014ApJ...797..110C}
{Chang}, P., {Broderick}, A.~E., {Pfrommer}, C., {et~al.} 2014, \apj, 797, 110

\bibitem[{{Clowes} {et~al.}(2013){Clowes}, {Harris}, {Raghunathan},
  {Campusano}, {S{\"o}chting}, \& {Graham}}]{2013MNRAS.429.2910C}
{Clowes}, R.~G., {Harris}, K.~A., {Raghunathan}, S., {et~al.} 2013, \mnras,
  429, 2910

\bibitem[{{Compostella} {et~al.}(2013){Compostella}, {Cantalupo}, \&
  {Porciani}}]{2013MNRAS.435.3169C}
{Compostella}, M., {Cantalupo}, S., \& {Porciani}, C. 2013, \mnras, 435, 3169

\bibitem[{{Cooray} \& {Sheth}(2002)}]{2002PhR...372....1C}
{Cooray}, A., \& {Sheth}, R. 2002, \physrep, 372, 1

\bibitem[{{Croom} {et~al.}(2005){Croom}, {Boyle}, {Shanks}, {Smith}, {Miller},
  {Outram}, {Loaring}, {Hoyle}, \& {da {\^A}ngela}}]{2005MNRAS.356..415C}
{Croom}, S.~M., {Boyle}, B.~J., {Shanks}, T., {et~al.} 2005, \mnras, 356, 415

\bibitem[{{Dermer} {et~al.}(2011){Dermer}, {Cavadini}, {Razzaque}, {Finke},
  {Chiang}, \& {Lott}}]{2011ApJ...733L..21D}
{Dermer}, C.~D., {Cavadini}, M., {Razzaque}, S., {et~al.} 2011, \apjl, 733, L21

\bibitem[{{Durrer} \& {Neronov}(2013)}]{2013A&ARv..21...62D}
{Durrer}, R., \& {Neronov}, A. 2013, \aapr, 21, 62

\bibitem[{{Faucher-Gigu{\`e}re} {et~al.}(2009){Faucher-Gigu{\`e}re}, {Lidz},
  {Zaldarriaga}, \& {Hernquist}}]{2009ApJ...703.1416F}
{Faucher-Gigu{\`e}re}, C.-A., {Lidz}, A., {Zaldarriaga}, M., \& {Hernquist}, L.
  2009, \apj, 703, 1416

\bibitem[{{Franceschini} {et~al.}(2008){Franceschini}, {Rodighiero}, \&
  {Vaccari}}]{2008A&A...487..837F}
{Franceschini}, A., {Rodighiero}, G., \& {Vaccari}, M. 2008, \aap, 487, 837

\bibitem[{{Furniss} {et~al.}(2015){Furniss}, {Sutter}, {Primack}, \&
  {Dom{\'{\i}}nguez}}]{2015MNRAS.446.2267F}
{Furniss}, A., {Sutter}, P.~M., {Primack}, J.~R., \& {Dom{\'{\i}}nguez}, A.
  2015, \mnras, 446, 2267

\bibitem[{{Giommi} {et~al.}(2013){Giommi}, {Padovani}, \&
  {Polenta}}]{Giommi:2013}
{Giommi}, P., {Padovani}, P., \& {Polenta}, G. 2013, \mnras, 431, 1914

\bibitem[{{Giommi} {et~al.}(2012){Giommi}, {Padovani}, {Polenta}, {Turriziani},
  {D'Elia}, \& {Piranomonte}}]{Giommi:2012}
{Giommi}, P., {Padovani}, P., {Polenta}, G., {et~al.} 2012, \mnras, 420, 2899

\bibitem[{{Gould} \& {Schr{\'e}der}(1967)}]{1967PhRv..155.1408G}
{Gould}, R.~J., \& {Schr{\'e}der}, G.~P. 1967, Physical Review, 155, 1408

\bibitem[{{Gunn} \& {Gott}(1972)}]{1972ApJ...176....1G}
{Gunn}, J.~E., \& {Gott}, III, J.~R. 1972, \apj, 176, 1

\bibitem[{{H.~E.~S.~S.~Collaboration}
  {et~al.}(2014){H.~E.~S.~S.~Collaboration}, {Abramowski}, {Aharonian}, {Ait
  Benkhali}, {Akhperjanian}, {Ang{\"u}ner}, {Anton}, {Backes}, {Balenderan},
  {Balzer}, \& et~al.}]{2014A&A...562A.145H}
{H.~E.~S.~S.~Collaboration}, {Abramowski}, A., {Aharonian}, F., {et~al.} 2014,
  \aap, 562, A145

\bibitem[{{H{\"a}ring} \& {Rix}(2004)}]{2004ApJ...604L..89H}
{H{\"a}ring}, N., \& {Rix}, H.-W. 2004, \apjl, 604, L89

\bibitem[{{Heath}(1977)}]{1977MNRAS.179..351H}
{Heath}, D.~J. 1977, \mnras, 179, 351

\bibitem[{{Hockney} \& {Eastwood}(1981)}]{1981csup.book.....H}
{Hockney}, R.~W., \& {Eastwood}, J.~W. 1981, {Computer Simulation Using
  Particles}

\bibitem[{{Hopkins} {et~al.}(2007){Hopkins}, {Richards}, \&
  {Hernquist}}]{2007ApJ...654..731H}
{Hopkins}, P.~F., {Richards}, G.~T., \& {Hernquist}, L. 2007, \apj, 654, 731

\bibitem[{{Hyv{\"o}nen} {et~al.}(2007){Hyv{\"o}nen}, {Kotilainen}, {Falomo},
  {{\"O}rndahl}, \& {Pursimo}}]{2007A&A...476..723H}
{Hyv{\"o}nen}, T., {Kotilainen}, J.~K., {Falomo}, R., {{\"O}rndahl}, E., \&
  {Pursimo}, T. 2007, \aap, 476, 723

\bibitem[{{Inoue} \& {Ioka}(2012)}]{Inoue:2012}
{Inoue}, Y., \& {Ioka}, K. 2012, \prd, 86, 023003

\bibitem[{{Kaiser}(1987)}]{1987MNRAS.227....1K}
{Kaiser}, N. 1987, \mnras, 227, 1

\bibitem[{{Kashikawa} {et~al.}(2006){Kashikawa}, {Yoshida}, {Shimasaku},
  {Nagashima}, {Yahagi}, {Ouchi}, {Matsuda}, {Malkan}, {Doi}, {Iye}, {Ajiki},
  {Akiyama}, {Ando}, {Aoki}, {Furusawa}, {Hayashino}, {Iwamuro}, {Karoji},
  {Kobayashi}, {Kodaira}, {Kodama}, {Komiyama}, {Miyazaki}, {Mizumoto},
  {Morokuma}, {Motohara}, {Murayama}, {Nagao}, {Nariai}, {Ohta}, {Okamura},
  {Sasaki}, {Sato}, {Sekiguchi}, {Shioya}, {Tamura}, {Taniguchi}, {Umemura},
  {Yamada}, \& {Yasuda}}]{2006ApJ...637..631K}
{Kashikawa}, N., {Yoshida}, M., {Shimasaku}, K., {et~al.} 2006, \apj, 637, 631

\bibitem[{{Kirkman} \& {Tytler}(1997)}]{1997ApJ...484..672K}
{Kirkman}, D., \& {Tytler}, D. 1997, \apj, 484, 672

\bibitem[{{Kneiske} {et~al.}(2004){Kneiske}, {Bretz}, {Mannheim}, \&
  {Hartmann}}]{2004A&A...413..807K}
{Kneiske}, T.~M., {Bretz}, T., {Mannheim}, K., \& {Hartmann}, D.~H. 2004, \aap,
  413, 807

\bibitem[{{Kneiske} \& {Mannheim}(2008)}]{Knei-Mann:08}
{Kneiske}, T.~M., \& {Mannheim}, K. 2008, \aap, 479, 41

\bibitem[{{Komatsu} {et~al.}(2011){Komatsu}, {Smith}, {Dunkley}, {Bennett},
  {Gold}, {Hinshaw}, {Jarosik}, {Larson}, {Nolta}, {Page}, {Spergel},
  {Halpern}, {Hill}, {Kogut}, {Limon}, {Meyer}, {Odegard}, {Tucker}, {Weiland},
  {Wollack}, \& {Wright}}]{2011ApJS..192...18K}
{Komatsu}, E., {Smith}, K.~M., {Dunkley}, J., {et~al.} 2011, \apjs, 192, 18

\bibitem[{{Kravtsov}(2010)}]{2010AdAst2010E...8K}
{Kravtsov}, A. 2010, Advances in Astronomy, 2010, arXiv:0906.3295

\bibitem[{{Kriss} {et~al.}(2001){Kriss}, {Shull}, {Oegerle}, {Zheng},
  {Davidsen}, {Songaila}, {Tumlinson}, {Cowie}, {Deharveng}, {Friedman},
  {Giroux}, {Green}, {Hutchings}, {Jenkins}, {Kruk}, {Moos}, {Morton},
  {Sembach}, \& {Tripp}}]{2001Sci...293.1112K}
{Kriss}, G.~A., {Shull}, J.~M., {Oegerle}, W., {et~al.} 2001, Science, 293,
  1112

\bibitem[{{Lister} {et~al.}(2013){Lister}, {Aller}, {Aller}, {Homan},
  {Kellermann}, {Kovalev}, {Pushkarev}, {Richards}, {Ros}, \&
  {Savolainen}}]{2013AJ....146..120L}
{Lister}, M.~L., {Aller}, M.~F., {Aller}, H.~D., {et~al.} 2013, \aj, 146, 120

\bibitem[{{Mandelbaum} {et~al.}(2009){Mandelbaum}, {Li}, {Kauffmann}, \&
  {White}}]{2009MNRAS.393..377M}
{Mandelbaum}, R., {Li}, C., {Kauffmann}, G., \& {White}, S.~D.~M. 2009, \mnras,
  393, 377

\bibitem[{{Marinoni} {et~al.}(2005){Marinoni}, {Le F{\`e}vre}, {Meneux},
  {Iovino}, {Pollo}, {Ilbert}, {Zamorani}, {Guzzo}, {Mazure}, {Scaramella},
  {Cappi}, {McCracken}, {Bottini}, {Garilli}, {Le Brun}, {Maccagni}, {Picat},
  {Scodeggio}, {Tresse}, {Vettolani}, {Zanichelli}, {Adami}, {Arnouts},
  {Bardelli}, {Blaizot}, {Bolzonella}, {Charlot}, {Ciliegi}, {Contini},
  {Foucaud}, {Franzetti}, {Gavignaud}, {Marano}, {Mathez}, {Merighi},
  {Paltani}, {Pell{\`o}}, {Pozzetti}, {Radovich}, {Zucca}, {Bondi},
  {Bongiorno}, {Busarello}, {Colombi}, {Cucciati}, {Lamareille}, {Mellier},
  {Merluzzi}, {Ripepi}, \& {Rizzo}}]{2005A&A...442..801M}
{Marinoni}, C., {Le F{\`e}vre}, O., {Meneux}, B., {et~al.} 2005, \aap, 442, 801

\bibitem[{{McQuinn} {et~al.}(2009){McQuinn}, {Lidz}, {Zaldarriaga},
  {Hernquist}, {Hopkins}, {Dutta}, \&
  {Faucher-Gigu{\`e}re}}]{2009ApJ...694..842M}
{McQuinn}, M., {Lidz}, A., {Zaldarriaga}, M., {et~al.} 2009, \apj, 694, 842

\bibitem[{{Meiksin} \& {Tittley}(2012)}]{2012MNRAS.423....7M}
{Meiksin}, A., \& {Tittley}, E.~R. 2012, \mnras, 423, 7

\bibitem[{{Miniati} \& {Elyiv}(2013)}]{2013ApJ...770...54M}
{Miniati}, F., \& {Elyiv}, A. 2013, \apj, 770, 54

\bibitem[{{Mo} {et~al.}(2011){Mo}, {Van den Bosch}, \& {White}}]{MFW}
{Mo}, H., {Van den Bosch}, F., \& {White}, S. 2011, {Galaxy formation and
  evolution}, ed. {Cambridge University Press}

\bibitem[{{Mo} \& {White}(1996)}]{1996MNRAS.282..347M}
{Mo}, H.~J., \& {White}, S.~D.~M. 1996, \mnras, 282, 347

\bibitem[{{Murase} {et~al.}(2012){Murase}, {Beacom}, \& {Takami}}]{Murase:2012}
{Murase}, K., {Beacom}, J.~F., \& {Takami}, H. 2012, JCAP, 8, 30

\bibitem[{{Myers} {et~al.}(2007){Myers}, {Brunner}, {Nichol}, {Richards},
  {Schneider}, \& {Bahcall}}]{2007ApJ...658...85M}
{Myers}, A.~D., {Brunner}, R.~J., {Nichol}, R.~C., {et~al.} 2007, \apj, 658, 85

\bibitem[{{Neronov} \& {Semikoz}(2009)}]{2009PhRvD..80l3012N}
{Neronov}, A., \& {Semikoz}, D.~V. 2009, \prd, 80, 123012

\bibitem[{{Papageorgiou} {et~al.}(2012){Papageorgiou}, {Plionis}, {Basilakos},
  \& {Ragone-Figueroa}}]{2012MNRAS.422..106P}
{Papageorgiou}, A., {Plionis}, M., {Basilakos}, S., \& {Ragone-Figueroa}, C.
  2012, \mnras, 422, 106

\bibitem[{{Peebles}(1982)}]{Peebles}
{Peebles}, P. 1982, {Principles of Physical Cosmology}, ed. T.~Anderson,
  Wightman (Princeton University Press)

\bibitem[{{Pfrommer} {et~al.}(2012){Pfrommer}, {Chang}, \&
  {Broderick}}]{2012ApJ...752...24P}
{Pfrommer}, C., {Chang}, P., \& {Broderick}, A.~E. 2012, \apj, 752, 24

\bibitem[{{Pontzen}(2014)}]{2014PhRvD..89h3010P}
{Pontzen}, A. 2014, \prd, 89, 083010

\bibitem[{{Pritchard} \& {Furlanetto}(2007)}]{2007MNRAS.376.1680P}
{Pritchard}, J.~R., \& {Furlanetto}, S.~R. 2007, \mnras, 376, 1680

\bibitem[{{Puchwein} {et~al.}(2014){Puchwein}, {Bolton}, {Haehnelt}, {Madau},
  \& {Becker}}]{2014arXiv1410.1531P}
{Puchwein}, E., {Bolton}, J.~S., {Haehnelt}, M.~G., {Madau}, P., \& {Becker},
  G.~D. 2014, ArXiv e-prints, arXiv:1410.1531

\bibitem[{{Puchwein} {et~al.}(2012){Puchwein}, {Pfrommer}, {Springel},
  {Broderick}, \& {Chang}}]{2012MNRAS.423..149P}
{Puchwein}, E., {Pfrommer}, C., {Springel}, V., {Broderick}, A.~E., \& {Chang},
  P. 2012, \mnras, 423, 149

\bibitem[{{Pushkarev} {et~al.}(2009){Pushkarev}, {Kovalev}, {Lister}, \&
  {Savolainen}}]{2009A&A...507L..33P}
{Pushkarev}, A.~B., {Kovalev}, Y.~Y., {Lister}, M.~L., \& {Savolainen}, T.
  2009, \aap, 507, L33

\bibitem[{{Rauch} {et~al.}(1997){Rauch}, {Miralda-Escude}, {Sargent}, {Barlow},
  {Weinberg}, {Hernquist}, {Katz}, {Cen}, \& {Ostriker}}]{1997ApJ...489....7R}
{Rauch}, M., {Miralda-Escude}, J., {Sargent}, W.~L.~W., {et~al.} 1997, \apj,
  489, 7

\bibitem[{{Regan} {et~al.}(2007){Regan}, {Haehnelt}, \&
  {Viel}}]{2007MNRAS.374..196R}
{Regan}, J.~A., {Haehnelt}, M.~G., \& {Viel}, M. 2007, \mnras, 374, 196

\bibitem[{{Ricotti} {et~al.}(2000){Ricotti}, {Gnedin}, \&
  {Shull}}]{2000ApJ...534...41R}
{Ricotti}, M., {Gnedin}, N.~Y., \& {Shull}, J.~M. 2000, \apj, 534, 41

\bibitem[{{Rudie} {et~al.}(2012){Rudie}, {Steidel}, \&
  {Pettini}}]{2012ApJ...757L..30R}
{Rudie}, G.~C., {Steidel}, C.~C., \& {Pettini}, M. 2012, \apjl, 757, L30

\bibitem[{{Schaye} {et~al.}(2000){Schaye}, {Theuns}, {Rauch}, {Efstathiou}, \&
  {Sargent}}]{2000MNRAS.318..817S}
{Schaye}, J., {Theuns}, T., {Rauch}, M., {Efstathiou}, G., \& {Sargent},
  W.~L.~W. 2000, \mnras, 318, 817

\bibitem[{{Schlickeiser} {et~al.}(2012){Schlickeiser}, {Ibscher}, \&
  {Supsar}}]{2012ApJ...758..102S}
{Schlickeiser}, R., {Ibscher}, D., \& {Supsar}, M. 2012, \apj, 758, 102

\bibitem[{{Schlickeiser} {et~al.}(2013){Schlickeiser}, {Krakau}, \&
  {Supsar}}]{2013ApJ...777...49S}
{Schlickeiser}, R., {Krakau}, S., \& {Supsar}, M. 2013, \apj, 777, 49

\bibitem[{{Shen} {et~al.}(2007){Shen}, {Strauss}, {Oguri}, {Hennawi}, {Fan},
  {Richards}, {Hall}, {Gunn}, {Schneider}, {Szalay}, {Thakar}, {Vanden Berk},
  {Anderson}, {Bahcall}, {Connolly}, \& {Knapp}}]{2007AJ....133.2222S}
{Shen}, Y., {Strauss}, M.~A., {Oguri}, M., {et~al.} 2007, \aj, 133, 2222

\bibitem[{{Simionescu} {et~al.}(2009){Simionescu}, {Werner}, {B{\"o}hringer},
  {Kaastra}, {Finoguenov}, {Br{\"u}ggen}, \& {Nulsen}}]{2009A&A...493..409S}
{Simionescu}, A., {Werner}, N., {B{\"o}hringer}, H., {et~al.} 2009, \aap, 493,
  409

\bibitem[{{Simpson} {et~al.}(2012){Simpson}, {Rawlings}, {Ivison}, {Akiyama},
  {Almaini}, {Bradshaw}, {Chapman}, {Chuter}, {Croom}, {Dunlop}, {Foucaud}, \&
  {Hartley}}]{2012MNRAS.421.3060S}
{Simpson}, C., {Rawlings}, S., {Ivison}, R., {et~al.} 2012, \mnras, 421, 3060

\bibitem[{{Sironi} \& {Giannios}(2014)}]{2014ApJ...787...49S}
{Sironi}, L., \& {Giannios}, D. 2014, \apj, 787, 49

\bibitem[{{Springel}(2005)}]{2005MNRAS.364.1105S}
{Springel}, V. 2005, \mnras, 364, 1105

\bibitem[{{Springel} \& {Hernquist}(2002)}]{2002MNRAS.333..649S}
{Springel}, V., \& {Hernquist}, L. 2002, \mnras, 333, 649

\bibitem[{{Stecker} {et~al.}(1992){Stecker}, {de Jager}, \&
  {Salamon}}]{1992ApJ...390L..49S}
{Stecker}, F.~W., {de Jager}, O.~C., \& {Salamon}, M.~H. 1992, \apjl, 390, L49

\bibitem[{{Stecker} {et~al.}(2006){Stecker}, {Malkan}, \&
  {Scully}}]{2006ApJ...648..774S}
{Stecker}, F.~W., {Malkan}, M.~A., \& {Scully}, S.~T. 2006, \apj, 648, 774

\bibitem[{{Stecker} \& {Salamon}(1996)}]{1996ApJ...464..600S}
{Stecker}, F.~W., \& {Salamon}, M.~H. 1996, \apj, 464, 600

\bibitem[{{Steidel} {et~al.}(1998){Steidel}, {Adelberger}, {Dickinson},
  {Giavalisco}, {Pettini}, \& {Kellogg}}]{1998ApJ...492..428S}
{Steidel}, C.~C., {Adelberger}, K.~L., {Dickinson}, M., {et~al.} 1998, \apj,
  492, 428

\bibitem[{{Syphers} \& {Shull}(2014)}]{2014ApJ...784...42S}
{Syphers}, D., \& {Shull}, J.~M. 2014, \apj, 784, 42

\bibitem[{{Tittley} \& {Meiksin}(2007)}]{2007MNRAS.380.1369T}
{Tittley}, E.~R., \& {Meiksin}, A. 2007, \mnras, 380, 1369

\bibitem[{{Venters}(2010)}]{Vent:10}
{Venters}, T.~M. 2010, \apj, 710, 1530

\bibitem[{{Viel} {et~al.}(2009){Viel}, {Bolton}, \&
  {Haehnelt}}]{2009MNRAS.399L..39V}
{Viel}, M., {Bolton}, J.~S., \& {Haehnelt}, M.~G. 2009, \mnras, 399, L39

\bibitem[{{Viel} {et~al.}(2004){Viel}, {Haehnelt}, \&
  {Springel}}]{2004MNRAS.354..684V}
{Viel}, M., {Haehnelt}, M.~G., \& {Springel}, V. 2004, \mnras, 354, 684

\bibitem[{{Vovk} {et~al.}(2012){Vovk}, {Taylor}, {Semikoz}, \&
  {Neronov}}]{2012ApJ...747L..14V}
{Vovk}, I., {Taylor}, A.~M., {Semikoz}, D., \& {Neronov}, A. 2012, \apjl, 747,
  L14

\bibitem[{{Werner} {et~al.}(2010){Werner}, {Simionescu}, {Million}, {Allen},
  {Nulsen}, {von der Linden}, {Hansen}, {B{\"o}hringer}, {Churazov}, {Fabian},
  {Forman}, {Jones}, {Sanders}, \& {Taylor}}]{2010MNRAS.407.2063W}
{Werner}, N., {Simionescu}, A., {Million}, E.~T., {et~al.} 2010, \mnras, 407,
  2063

\bibitem[{{Worseck} {et~al.}(2014){Worseck}, {Prochaska}, {Hennawi}, \&
  {McQuinn}}]{2014arXiv1405.7405W}
{Worseck}, G., {Prochaska}, J.~X., {Hennawi}, J.~F., \& {McQuinn}, M. 2014,
  ArXiv e-prints, arXiv:1405.7405

\end{thebibliography}
\end{document}